\documentclass[12pt,preprint]{aastex}
\usepackage{graphicx}
\usepackage[intlimits]{amsmath}
\usepackage{txfonts}
\usepackage{natbib}
\usepackage{textcomp}
\usepackage{color}
\usepackage{fixltx2e}

\begin{document}
\title{Calculating the Habitable Zone of Multiple Star Systems\\
(http://astro.twam.info/hz)}

\author{Tobias~W.~A.~M\"uller\altaffilmark{1} and Nader~Haghighipour\altaffilmark{1,2}}
\altaffiltext{1}{Institute for Astronomy and Astrophysics, University of T\"ubingen, Auf der Morgenstelle 10, 72076 T\"ubingen,	Germany}
\altaffiltext{2}{Institute for Astronomy and NASA Astrobiology Institute, University of Hawaii-Manoa, Honolulu, HI 96822, USA}

\begin{abstract}
We have developed a comprehensive methodology and an interactive website for calculating the habitable zone (HZ) of multiple star systems. 
Using the concept of spectral weight factor, as introduced in our previous studies of the calculations of HZ in and around
binary star systems, we calculate the contribution of each star (based on its spectral energy distribution) to the total 
flux received at the top of the atmosphere of an Earth-like planet, and use the models of the HZ of the Sun to determine the
boundaries of the HZ in multiple star systems. Our interactive website for carrying out these calculations is 
publicly available at {\url http://astro.twam.info/hz}\,. We discuss the details of our methodology and present its application to some 
of the multiple star systems detected by the {\it Kepler} space telescope. We also present the instructions for using our interactive 
website, and demonstrate its capabilities by calculating the HZ for two interesting analytical solutions of the three-body
problem. 
\end{abstract}


\section{Introduction}

It is widely accepted that stars form in clusters. Surveys of star-forming regions have indicated that
approximately 70\% of all stars in our galaxy are in binary or multiple star systems \citep{Batten89}.
An examination of 164 nearest G-dwarfs by \citet{Duquennoy91}, for instance, has shown that 62 of these stars are in binaries, 
7 are in triplets, and 2 are members of two quadruple star systems. In the past few years, the {\it Kepler} space telescope has also 
detected many binary and multiple star systems. The {\it Kepler} Eclipsing Binary Catalog lists more than 2100 eclipsing binaries 
\citep{Slawson11} among which $\sim 20\%$ are within triple star systems.

A survey of the currently known planet-hosting stars indicates that slightly more than 8\% of these stars have stellar companions 
\citep{Rein12}. While the majority of these stars are in binaries, some are also in triple and quadruple systems. 
For instance, Kepler 64, which has a binary stellar companion
and is host to a circumbinary planet \citep{Schwamb13}, is in fact a member of a quadruple stellar system. The M star 
Gliese\,667 \citep{Anglada12,Delfosse13}, which is known to host at least seven planets, is part of a triple stellar
system. The star 16\,Cyg \citep{Cochran97}, the first multiple star system discovered to host a planet,
consists of three stellar components.

The discovery of planets in multiple star systems has raised the question that whether such planetary systems can be habitable. 
As the habitability of a planet (and, therefore, the system's HZ), in addition to the size, atmospheric composition, 
and orbital dynamics of the planet, depends also on the total flux received at the top of the planet's atmosphere, the stellar multiplicity
plays an important role in determining the range and location of the system's HZ. Depending on their surface temperatures and orbital 
characteristics, each star of the system will have a different contribution to the total flux at the location of the planet. 
Within the context of binary stars during the past few years this topic has been addressed by \citet{Mason13,Liu13,Kane13,Eggl13,Eggl12,Quarles12}.

Recently, \citet[][hereafter, HK13]{Haghighipour13} and \citet[][hereafter, KH13]{Kaltenegger13} studied this concept within the context of
binary star systems. These authors have shown that as the atmosphere of a planet interacts differently with the incident radiation
from stars with different spectral energy distributions (SED), the contribution of each star of the binary to the total flux received by
the planet will be different. In other words, considering the direct summation of the fluxes of the two stars as the total flux received
at the top of the planet's atmosphere, and using that quantity to calculate the total insulation on the planet's surface will not be a 
correct approach and will result in an inaccurate value for the planet's equilibrium temperature. The fact that the planet's atmosphere 
responds differently to stellar radiations with different incident energy indicates that the contribution of each star to the total 
flux received at the top of the planet's atmosphere has to be weighted according to the star's SED. As shown by HK13 and KH13, 
such a weight factor will be
a function of the star's effective temperature and will have different forms for different models of the Sun's HZ. 
Considering the latest models of the habitability of Earth as presented by \citet{Kopparapu13a,Kopparapu13b}, HK13 and KH13 
derived a formula for the {\it spectral weight factor} of a star based on the star's effective temperature,
and presented an analytical formalism for calculating the HZ in S-type and P-type binary stars systems.

In this paper, we follow the same approach as presented by these authors
and generalize their methodology to calculate the HZ in systems with $N \geq 2$ stars. 
Although in order to maintain stability, most multi-star systems have evolved into hierarchical configurations and have
developed large stellar separations (which depending on the effective temperatures of their stars may imply minimal contribution 
from one star in the extent of the HZ around others), as we will explain in next sections, it proves useful to develop a self-consistent 
and comprehensive methodology that can be used to calculate the HZ of any system with more than one stellar component. The latter constitutes
the main goal of this paper. We consider the HZ to
be a region where an Earth-like planet (that is, a rocky planet with a ${{\rm CO}_2}/{\rm {H_2}O}/{\rm {N_2}}$ atmosphere and 
sufficiently large water content)
can permanently maintain liquid water on its solid surface. This definition of the HZ assumes that similar to Earth, the planet has
a dynamic interior and a geophysical cycle similar to Earth's carbonate silicate cycle that naturally regulates the abundance of ${\rm CO}_2$
and ${\rm {H_2}O}$ in its  atmosphere. The boundaries of the HZ are then associated with an ${\rm {H_2}O}$-dominated atmosphere for its outer
boundary and a ${\rm CO}_2$-dominated atmosphere for its inner limit. Between those limits on a geologically active planet, climate 
stability is established by a feedback mechanism through which the concentration of ${\rm CO}_2$ in the atmosphere varies inversely 
with planetary surface temperature.

It is important to note that in a multiple star system, the close approach of each star of the system to a planet can substantially 
affect the contribution of that star to the overall flux received by the planet. Also, the interaction between the star and the planet
can affect the orbital motion of the planet, and therefore, its habitability. This all implies that the (instantaneous) shape of the HZ 
in a multiple star system will vary during the motion of the stars. In the next sections, we discuss this in more detail and explain  
how our methodology treats this matter properly.

In Section \ref{sec:model}, we present the generalization of the calculation of a Binary HZ to a system of $N$ stars. In Section \ref{sec:examples}, 
we demonstrate the time-variation of the HZ in a multiple star system by applying our methodology to some of the systems from
the {\it Kepler} catalog. Motivated by the \emph{Habitable Zone Gallery} \citep[hzgallery.org,][]{Kane12} 
which provides information about the HZ around
single stars, we present in section \ref{sec:website} our fully interactive website for calculating the HZ in binary 
and multiple star systems. We also demonstrate in this section how to use our website by calculating
the HZ of two analytical solutions of
the three-body problem. In Section \ref{sec:summary}, we conclude this study by summarizing the results.

\section{Calculation of the Habitable Zone}
\label{sec:model}

As mentioned in the previous section, we consider the habitable zone to be the region where a fictitious Earth-like planet with  
a CO\textsubscript{2}/H\textsubscript{2}O/N\textsubscript{2} atmosphere, and similar geophysical and geodynamical properties as
those of Earth can maintain liquid water on its solid surface. 
As the capability of retaining liquid water depends on the planet's equilibrium temperature, and because this
temperature depends on the total flux received by the planet, the statement
above is equivalent to considering the HZ to be the region where the total flux received at the top of the atmosphere of a
fictitious Earth-like planet is equal to that of Earth received from the Sun. To calculate this flux for a star in a multiple star system, 
we generalize the methodology presented by HK13 and KH13 (which has been developed for binary star systems) 
to systems with $N$ number of stars. Using equation (1) in KH13 and equation (3) in HK13, that implies,

\begin{align}
\label{eq:totalflux}
F_\mathrm{total}\,=\, \sum_{i = 1}^N {W_i}\,({T_{\rm star}})\,\frac{L_i/L_{\sun}}{d_i^2} \,.
\end{align}

\noindent
In this equation, $F_{\rm total}$ is the total flux received by the planet, $ L_i $ is the luminosity of star $i$ in units of 
the solar luminosity $(L_{\sun})$, $d_i$ is the distance of the planet to the $i$-th star in astronomical units (AU), and 
${W_i}({T_{\rm star}})$ is the spectral weight factor which accounts for the different spectral energy distribution of the 
$i$-th star compared to the Sun. The quantity $T_{\rm star}$ is the effective stellar temperature.

The value of the spectral weight factor $W(T)$, in addition to the star's effective temperature, depends also on the models of the Sun's HZ.
We consider the models recently developed by \citet{Kopparapu13a,Kopparapu13b} for which the spectral weight factor of a star 
with an effective temperature in the range of $2600\,\mathrm{K} \leq T_\mathrm{star} \leq 7200\,\mathrm{K}$,
is given by (HK13 and KH13)

\begin{align}
\label{eq:weightfactor}
W({T_{\rm star}}) = \left[ 1 + \alpha\,({T_S})\,{d_{\rm Sun}^2} \right]^{-1} \,.
\end{align}

\noindent
In this equation, ${T_S}= {T_\mathrm{star}} - 5780$ K is the star's temperature-difference compared to the Sun, and

\begin{equation}
\label{eq:alpha}
\alpha ({T_S})\,=\, {a\, {T_S}} + {b\, {T_S^2}} + {c\, {T_S^3}} + {d\, {T_S^4}}\,.
\end{equation}

\noindent
The quantities $a, b, c,$ and $d$ in equation (\ref{eq:alpha}) are constant coefficients with values that depend on the conditions that
determine the inner and outer boundaries of the Sun's HZ. Table \ref{table1} shows these values. 

As mentioned earlier, to calculate the HZ of a multiple star system, we compare the total flux received by a fictions Earth-like
planet with that received by Earth from the Sun. From equation (\ref{eq:totalflux}), that means,

\begin{align}
\label{eq:4}
\sum_{i = 1}^N {W_i}\,({T_{\rm star}})\,\frac{L_i/L_{\sun}}{d_i^2}\,=\,\frac{L_{\Sun}}{{l_{\rm x-Sun}^2}}\,,
\end{align}

\noindent
where ${\rm x= (In,Out)}$, and $l_{\rm x-Sun}$ denotes the boundaries of the Sun's HZ. To determine the location
and range of the HZ of a multiple star system, the distances $d_i$ have to be calculated for different values of 
$l_{\rm x-Sun}$ by solving a set of differential equations corresponding to the motion of the stellar $N$-body
system along with the algebraic equation (\ref{eq:4}). 

To determine the values of $l_{\rm x-Sun}$, we follow HK13 and KH13, and consider a {\it narrow} HZ corresponding to the 
region between the runaway greenhouse and the maximum greenhouse limits in the most recent Sun's HZ mode by \citet{Kopparapu13a,Kopparapu13b}. 
As this model does not include cloud feedback, the boundaries of the narrow HZ in our definition do not include the feedback
from clouds as well. To mimic the effects of clouds (and have a more realistic definition for the HZ), HK13 and KH13 introduced
the {\it empirical} (nominal) HZ as the region with an inner boundary at the Recent Venus limit and an outer boundary at the 
limit of Early Mars in the model by \citet{Kopparapu13a,Kopparapu13b}. 
These boundaries have been derived using the fluxes received by Mars and Venus at 3.5 and 1.0 Gyr ago, respectively. 
At these times, the two planets did not show indications for liquid water on their surfaces \citep{Kasting93}. 
We follow HK13 and KH13, and consider also the empirical HZ, as the second range for determining the values of $l_{\rm x-Sun}$.
In these definitions, the locations of the HZs are determined based on the flux received by the planet
\citep{Kasting93,Selsis07,Kaltenegger11,Kopparapu13a}.

It is important to note that, except for some special orbital configurations of the stars (e.g., when the stars orbit the center
of mass of the system so close to one another that the system's HZ can only exist around the entire stellar system, as in HK13), 
it may not be possible to determine a distinct inner or outer edge for the HZ of a multiple star system.
The motions of the stars cause the HZ of the system to be dynamic and change boundaries and locations as the stars move over the time.
For this reason, in order to determine the location of the HZ of the system, at any given time, we consider a large grid of points
over the entire system and calculate the flux at each grid point using the equation (\ref{eq:totalflux}). We then mark those points of the grid
for which the value of the total flux satisfies the following condition, as a point of the system's HZ:

\begin{align}
\label{eq:5}
\frac{L_{\Sun}}{{l_{\rm In-Sun}^2}} \, \leq \,
\sum_{i = 1}^N {W_i}\,({T_{\rm star}})\, \frac{L_i/L_{\sun}}{d_i^2}\,\leq \,
\frac{L_{\Sun}}{{l_{\rm Out-Sun}^2}}\,,
\end{align}

\noindent
We repeat this process for one complete orbital period of the system by increasing the time in small increments and allowing the
stars to move in their orbits. When calculating the HZ, we do not consider whether the orbit of the fictitious Earth-like planet
will stay stable. As explained in the next sections, we determine the stability of the HZ by direct integration of the orbit of an
Earth-mass planet at different locations in the HZ. As a result, in some systems, some or all of the HZ will be dynamically unstable 
(see HK13 for several examples). In the following, we present a sample of our calculations as well as the instructions to our interactive 
HZ-calculator website.

\section{Examples}
\label{sec:examples}

To demonstrate the application of our methodology, we calculate the HZ of two multiple stars systems 
KIC\,4150611 and KID\,5653126 from the {\it Kepler} space telescope catalog. As explained below, these systems have stellar and
orbital characteristics that allow for exploring effects of luminosity and distance on the range of the HZ.
We calculate the narrow and empirical HZs in these systems, and show the results for one complete orbit. Following HK13 and KH13,
and from the model by \citet{Kopparapu13a,Kopparapu13b}, we consider the narrow HZ (without cloud feedback) of a Sun-like star to extend 
from 0.97 AU to 1.67 AU, and its empirical HZ to be from 0.75 AU to 1.77 AU. Movies of the time-variation of the HZs 
of our systems can be found at \url{http://astro.twam.info/hz-multi}.

\subsection{KIC\,4150611}
\label{sec:kic4150611}

The KIC\,4150611 (HD\,181469, HIP\,94924) is a system of five stars consisting of a triplet with an A and two K stars, 
and a binary with two F stars. The A-KK and FF systems form a visual pair with a separation of 1.1".
The A-KK triplet has an orbital period of 94.2 days, and the orbital period of the inner KK binary is 1.522 days. 
The effective temperatures of the A and K stars are 8500 K and 4500 K, respective (A. Pr\v sa, private communication).
We consider here only the A-KK system and use our methodology to calculate the location and time-evolution of its HZ
(given the large separation between this system and the FF binary, the effect of the latter is negligible).

As shown by equation (\ref{eq:5}), the calculation of the HZ requires the knowledge of the luminosity, semimajor axis, and spectral weight 
factor of each star. To calculate the semimajor axis and luminosity of the A and K stars, we consider the A star to be $1.5 M_\sun$ 
(corresponding to the lower limit of the mass of a star of spectral type A) and the mass of each K star to be $0.7 {M_\sun}$. 
Using these values of the mass combined with the orbital periods of the KK binary and A-KK system as given above,
the semimajor axis of the KK binary will be 0.029 AU and that of the A-KK system will be equal to 0.58 AU. 
The luminosities are calculated using the mass-luminosity relation, $L={M^{3.5}}$. For the values of the stellar masses 
considered here, the luminosity of the A star will be equal to $4.134 {L_\sun}$ and that of the K stars will be $0.287 {L_\sun}$.
To calculate the spectral weight factor $W(T)$ for each star, we use equation (\ref{eq:weightfactor}). It is important to note that this equation is 
model-dependent and has been derived based on the model of the Sun's HZ by \citet{Kopparapu13a,Kopparapu13b}, which
is valid for stellar temperatures ranging from 2600 K to 7200 K. The latter means that equation (\ref{eq:weightfactor}) may not be
applicable when calculating the spectral weight factor of the A star (8500 K). However, because the purpose of our calculations is to demonstrate
our methodology and how it is used to determine the HZ of multiple star systems, we assume that the model by  
\citet{Kopparapu13a,Kopparapu13b} will maintain the functional form of its temperature dependence (i.e., equation 3) for higher values of 
stellar temperature, and extrapolate this model to effective stellar temperatures of 8500 K. We then calculate the 
spectral weight factor of each star using equation (\ref{eq:weightfactor}). Table \ref{table2} shows these values for the system's A and K stars.

\begin{figure}
\vskip 3in
\centering
\includegraphics[width=0.45\columnwidth]{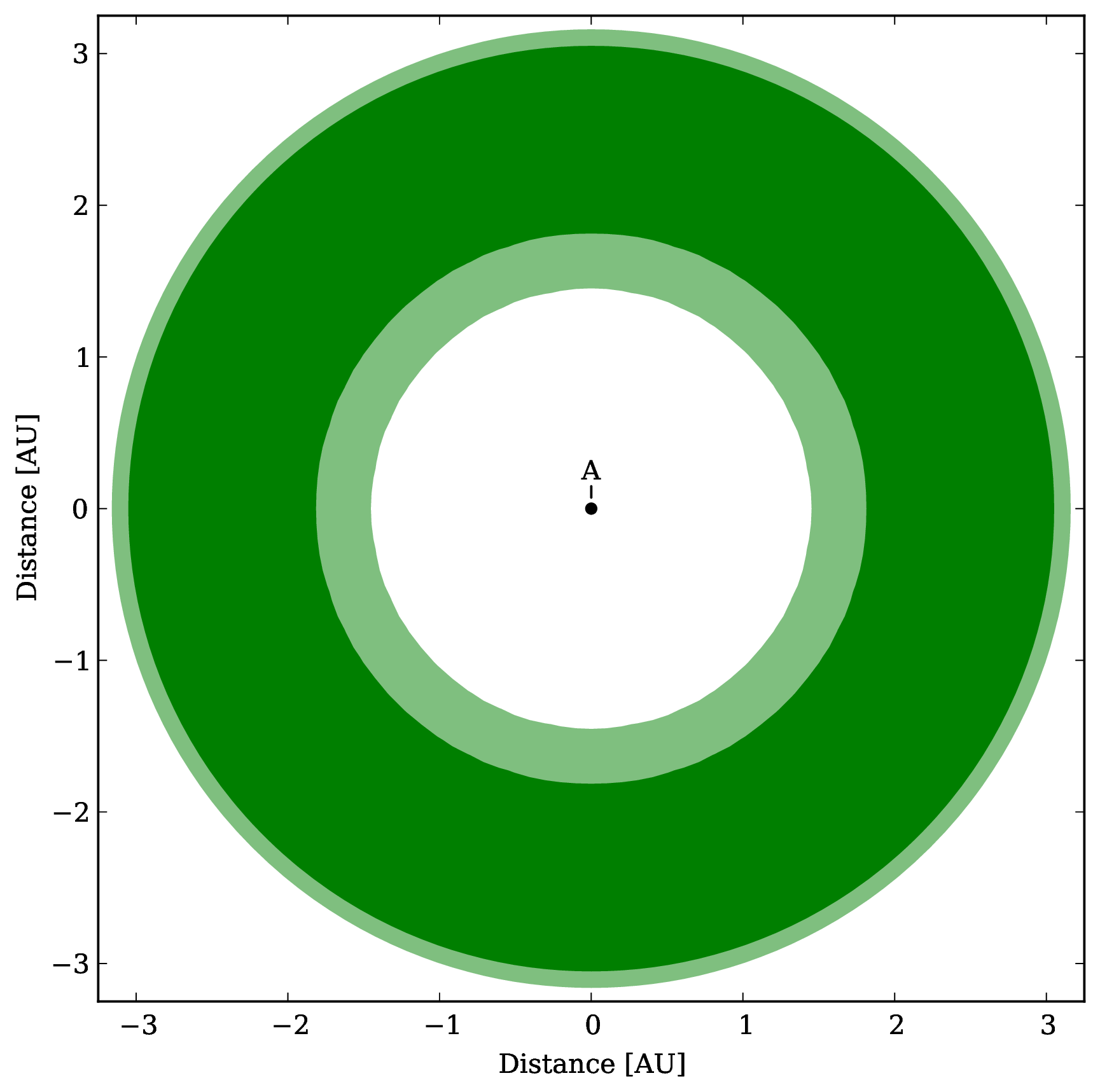}
\includegraphics[width=0.46\columnwidth]{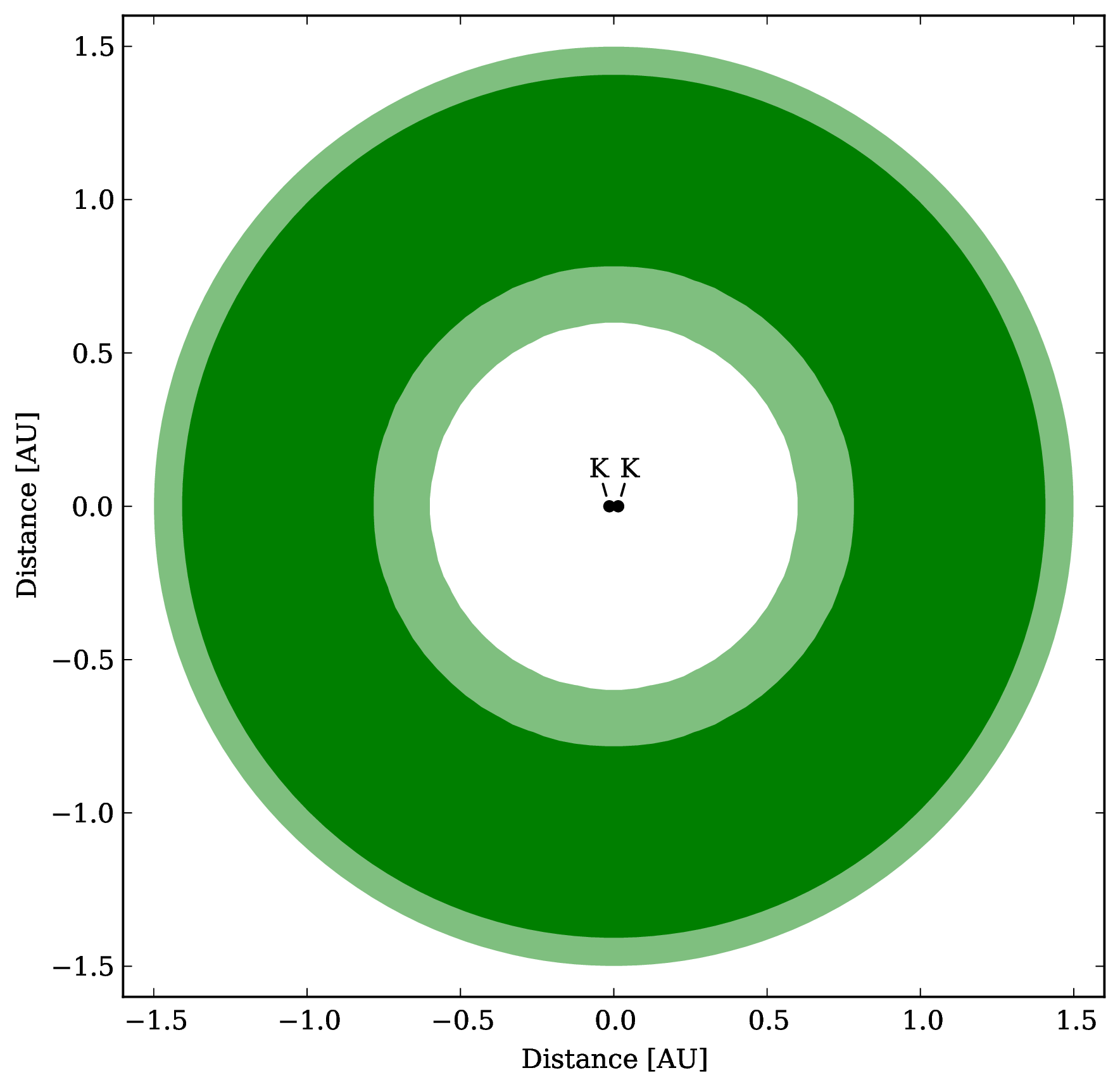}
\vskip 3.8in
\includegraphics[width=0.6\columnwidth]{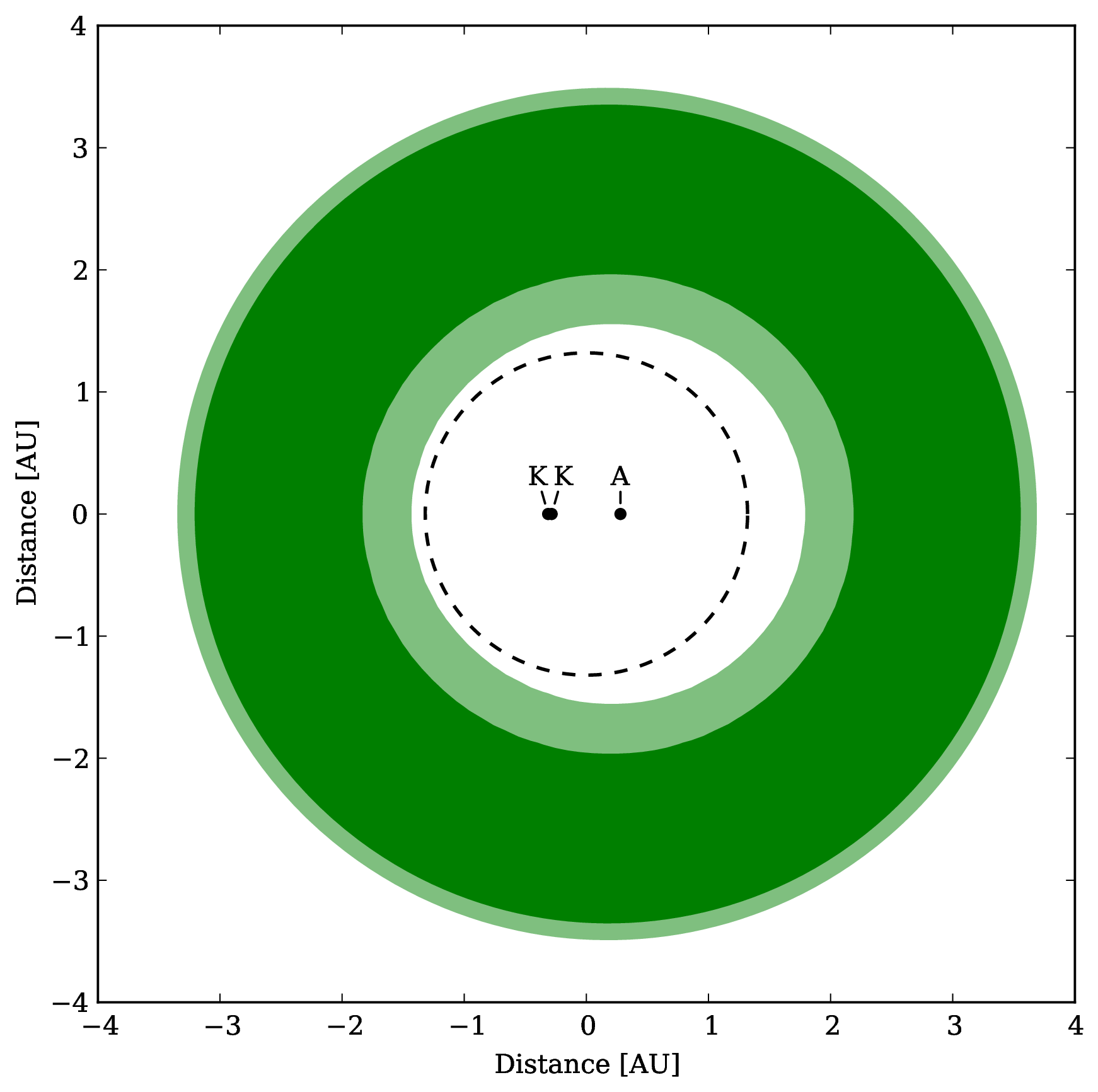}
\caption{Graphs of the HZs of the A-KK triplet in the KIC\,4150611 system and its components.
The top-left panel shows the HZ of the A star, the top-right panel shows that of the KK binary, and the bottom panel shows the
HZ of the A-KK triple star system. Here and in subsequent figures, the dark green corresponds to the narrow and the light green corresponds to the 
empirical HZs. The dashed circle shows the boundary of stability. Interior to this boundary, planetary orbits will become
unstable. A movie of the HZ of this system can be found at http://astro.twam.info/hz-multi.}
\label{fig:4150611}
\end{figure}

Given the short period of the KK binary (1.522 days), the stand-alone HZ of this system (i.e., assuming it is isolated 
and not part of a triple or larger stellar system) will be only at circumbinary distances. Following the methodology by HK13, 
this HZ extends from 0.6 AU to 1.5 AU from the center of mass of the binary. The top-right panel of Figure \ref{fig:4150611}
shows this region. The dark green in this and subsequent figures corresponds to the narrow and the light green corresponds to the 
empirical HZ. Similarly, assuming that the A star is isolated and not part of a multiple star system, its single-star HZ,
shown in the top-left panel of Figure \ref{fig:4150611}, will extend from 1.3 AU to 3.1 AU. As the distance between the A star and the center of mass
of the KK binary in the A-KK system is 20 times larger than the semimajor axis of the binary K stars, 
one would expect that the effect of the A star on the stand-alone
HZ of the KK binary, if not negligible, to be very small. However, as shown in the bottom panel of Figure \ref{fig:4150611}, when these three stars are
considered to be in their current orbital configuration, and assuming that the orbits are co-planar and circular,
the higher luminosity of the A star dominates and causes the entire
HZ around the KK binary to disappear. In this case, the HZ of the A-KK system is primarily due to the A star although the KK
binary does seem to have some small effect. A movie of the HZ of this system can be found at http://astro.twam.info/hz-multi.

\begin{figure}
\centering
\includegraphics[width=1\columnwidth]{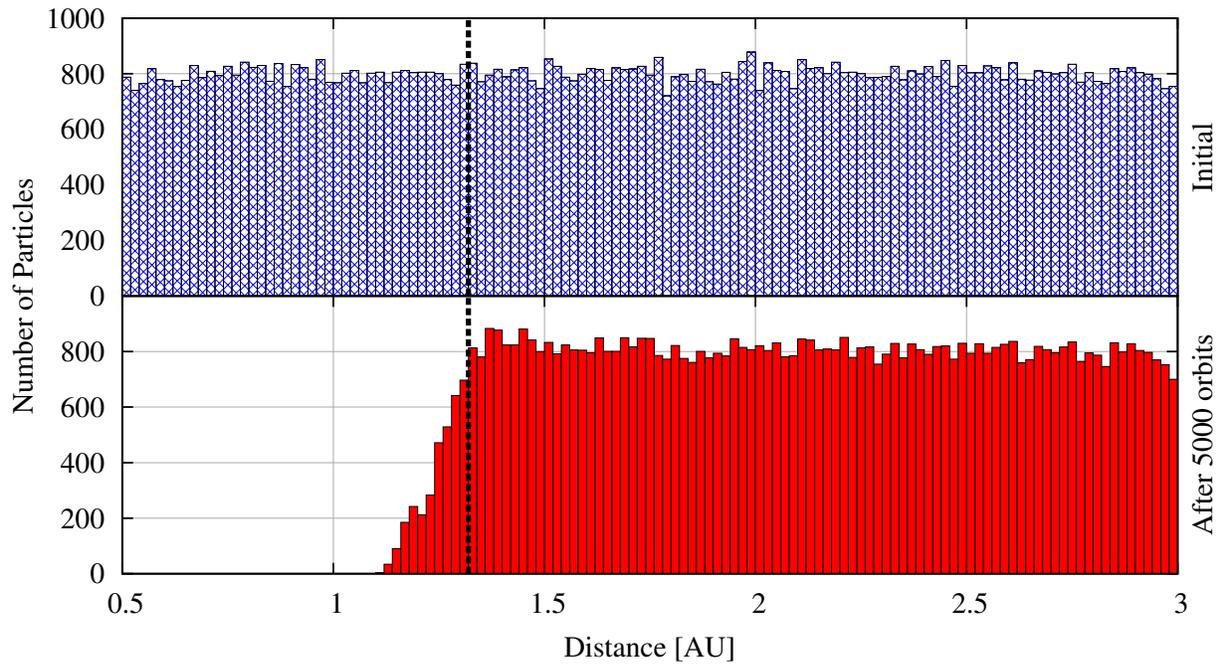}
\caption{Results of the integrations of the orbits of non-interacting Earth-mass planets around the center of mass of the KIC 4150611 system. 
The top panel (blue histogram) shows the initial distribution of planets and the bottom panel (red histogram) corresponds to their 
distribution after integrating their orbits for 5000 orbital periods of star A around the KK binary. The dashed line shows the boundary of orbital 
stability at $ 1.32 $\,AU.}
\label{fig:stability_4150611}
\end{figure}

To determine the orbital stability of an Earth-like planet in the HZ of the A-KK system, we distributed 100,000 non-interacting 
Earth-mass objects between 0.5 AU and 3 AU around the center of mass of the system, and integrated their orbits for 5000 periods of 
the A star using a fifth-order N-Body integrator \citep{Cash90}. The bodies were initially set to be in non-eccentric Keplerian 
orbits. The timesteps of the integrations were taken to 
be 1/100 of the period of the inner KK binary. Figure \ref{fig:stability_4150611} shows the distribution of the Earth-like 
planets at the beginning (top panel, blue)
and end (bottom panel, red) of the simulations. As shown here, at the end of the simulations, a lack of particles appears for distances smaller than 
1.32 AU (shown by the vertical dashed line in Figure \ref{fig:stability_4150611}), 
implying that the orbit of an Earth-mass planet in these regions will be unstable. We, therefore, considered 1.32 AU as 
the stability limit, and have shown this by the dashed circles in Figure \ref{fig:4150611}.

As mentioned above, the stability integrations were carried out for only 5000 orbits of the A star. To determine whether this time of
integration was sufficient, we assumed that the triple star system
of A-KK can be approximated by a binary star system with the A star as its primary and a secondary 
with a mass equal to the sum of the masses of the two K stars $(0.7+0.7=1.4 {M_\sun}$). We recall that the distance between the A star
and the center of mass of the KK binary is 20 times larger than the binary semimajor axis. In this {\it auxiliary} binary system,
an Earth-mass planet in the HZ will have a circumbinary orbit. As shown by \citet{Dvorak86}, \citet{Dvorak89}, and \citet{Holman99}, 
for such planetary orbits, the stability limit is given by

\begin{equation}
\label{eq:ptype-stability}
a_\mathrm{min}= a \left(1.60 + 5.10\,e - 2.22\,e^2 + 4.12\,\mu - 4.27\,e \mu - 5.09\,\mu^2 + 4.61\,e^2 \mu^2  \right),
\end{equation}

\noindent
where $a_{\rm min}$ is the minimum semimajor axis for a planetary orbit to be stable, $a$ is the binary semimajor axis,
$e$ is the binary eccentricity and $\mu={M_{\rm Pr}}/({M_{\rm Pr}}+{M_{\rm Sec}})$ with $M_{\rm Pr}$ and $M_{\rm Sec}$
being the masses of the primary and secondary stars, respectively. Considering that in the actual A-KK system, the orbits are circular
and the A star is at 0.58 AU from the KK binary ($a=0.58$ AU), the stability limit obtained from equation (\ref{eq:ptype-stability}) is at $a_{\rm min}$=1.39 AU.
This value is in good agreement with the 1.32 AU stability limit obtained from our numerical simulations, indicating that stability 
integrations have been carried out for a sufficient amount of time.

\subsubsection{Effect of the eccentricity of the A star}

\begin{figure}
\centering
\includegraphics[width=0.45\columnwidth]{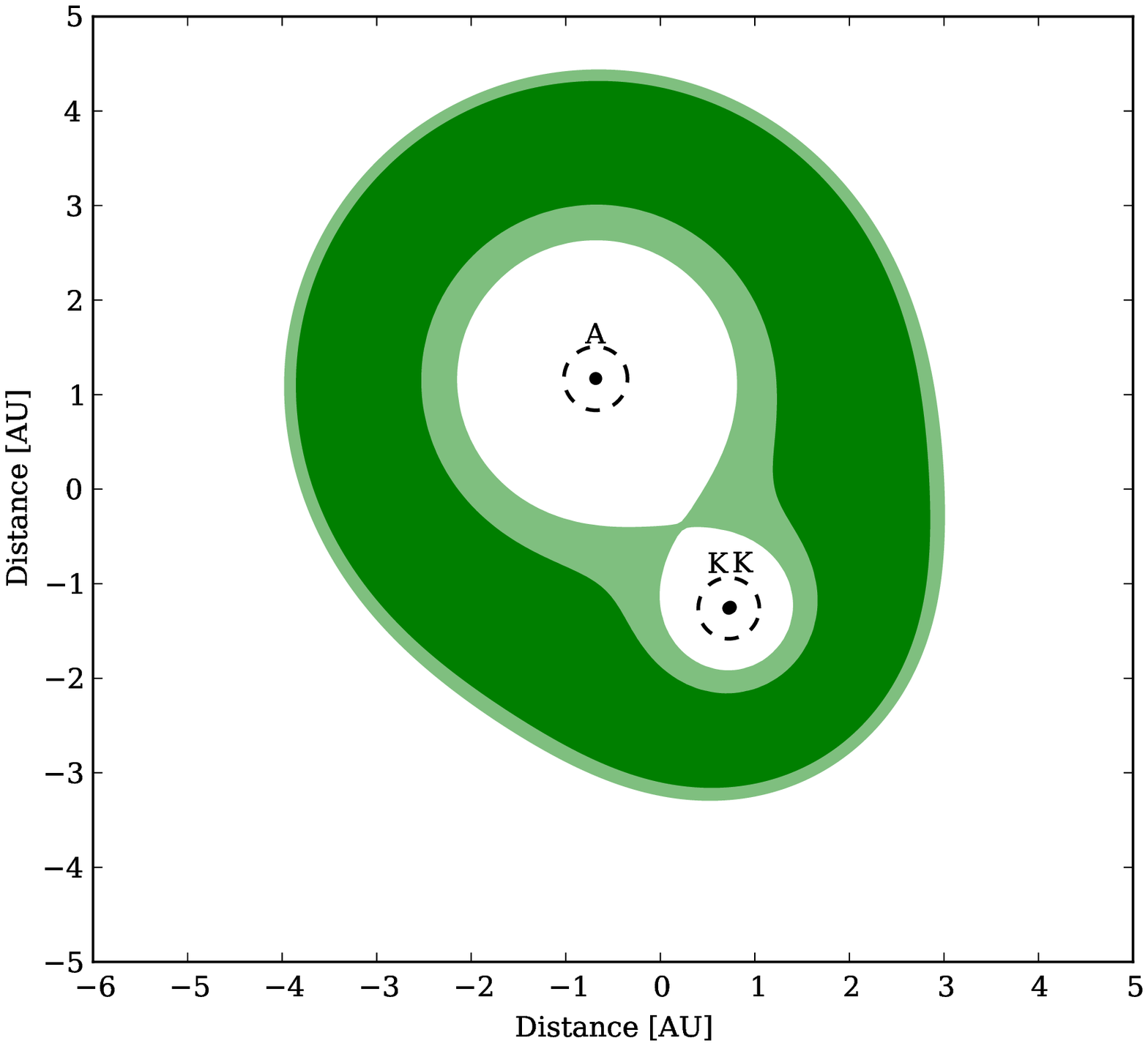}
\includegraphics[width=0.45\columnwidth]{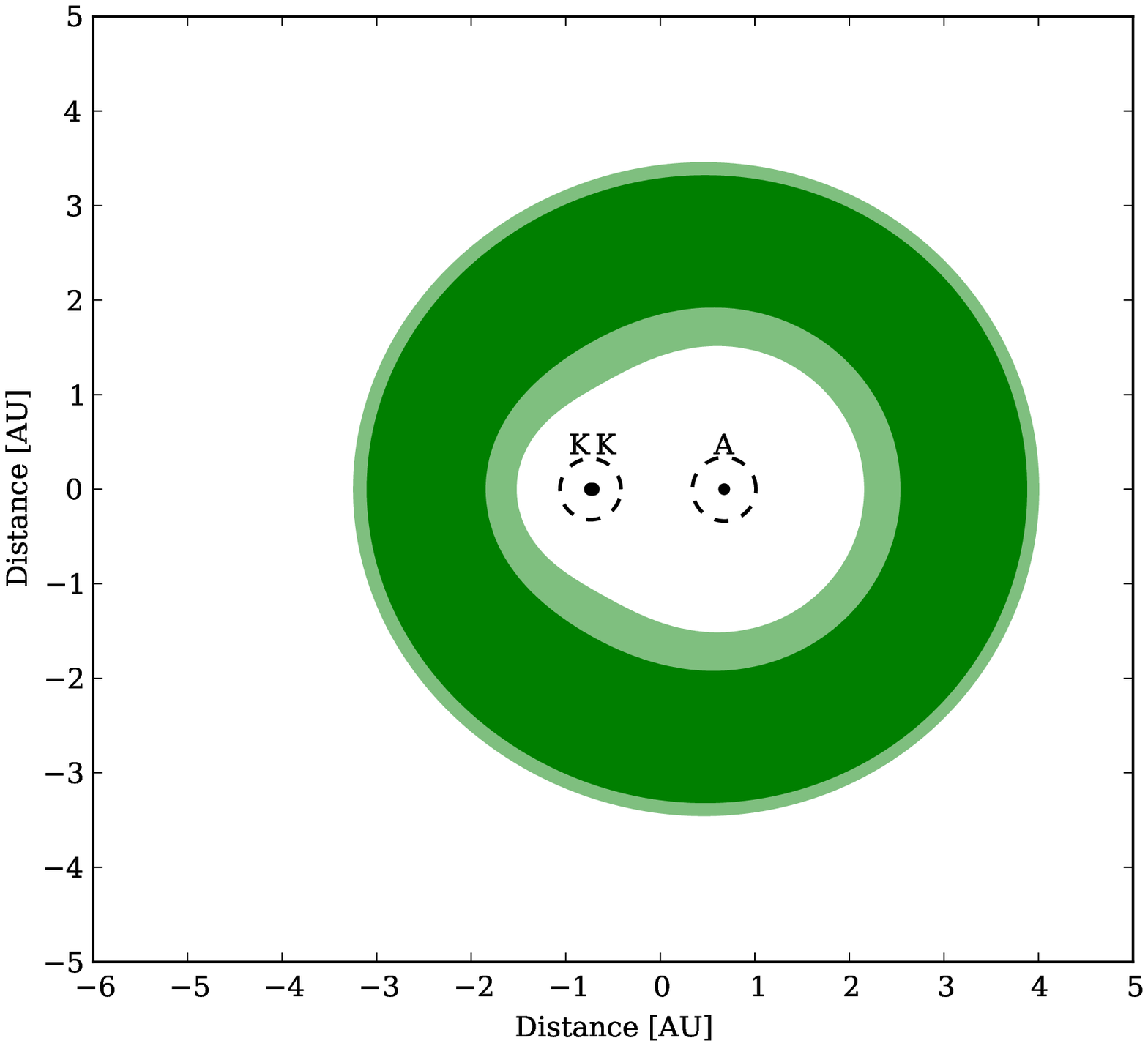}
\vskip 10pt
\includegraphics[width=0.45\columnwidth]{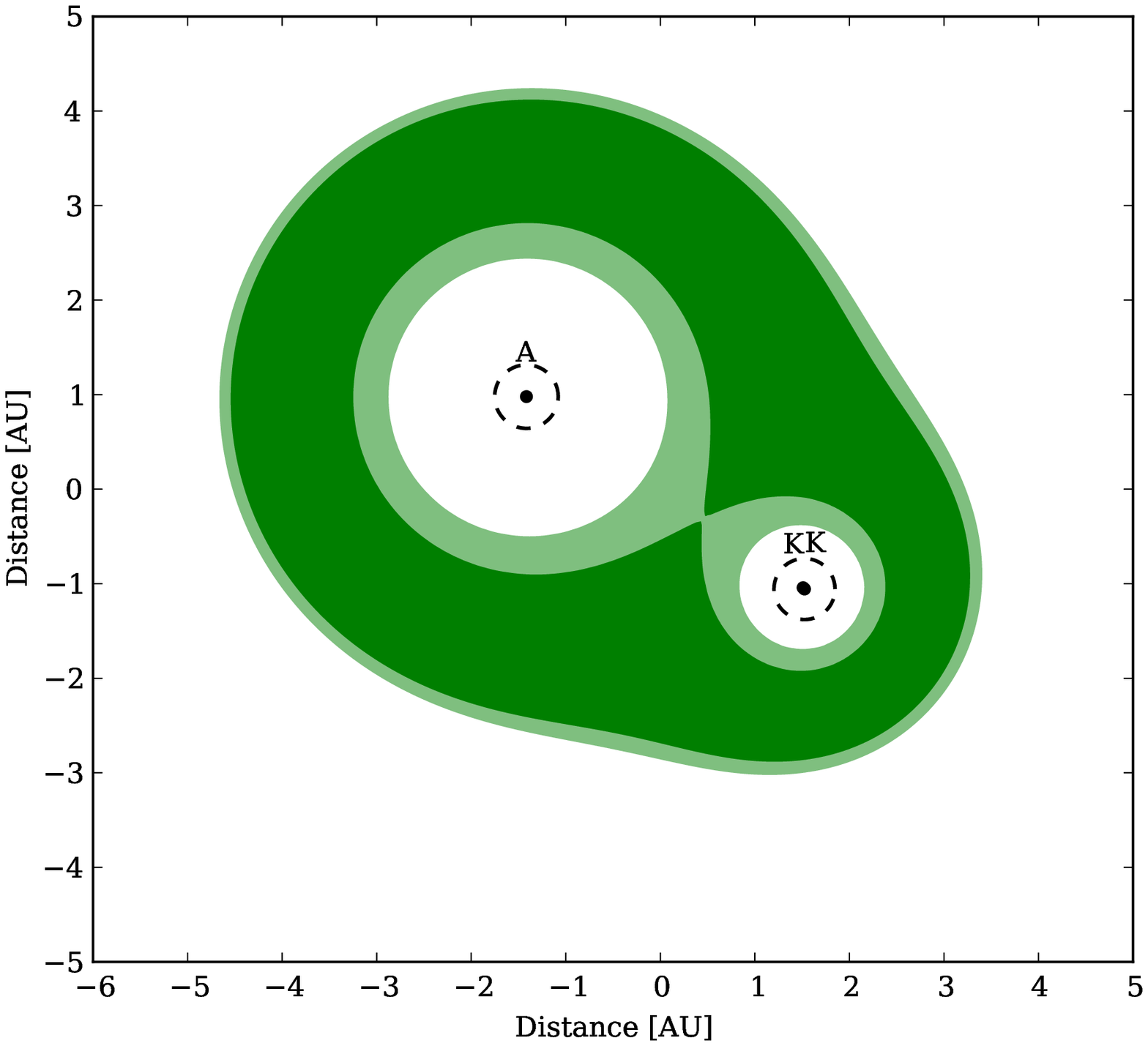}
\includegraphics[width=0.45\columnwidth]{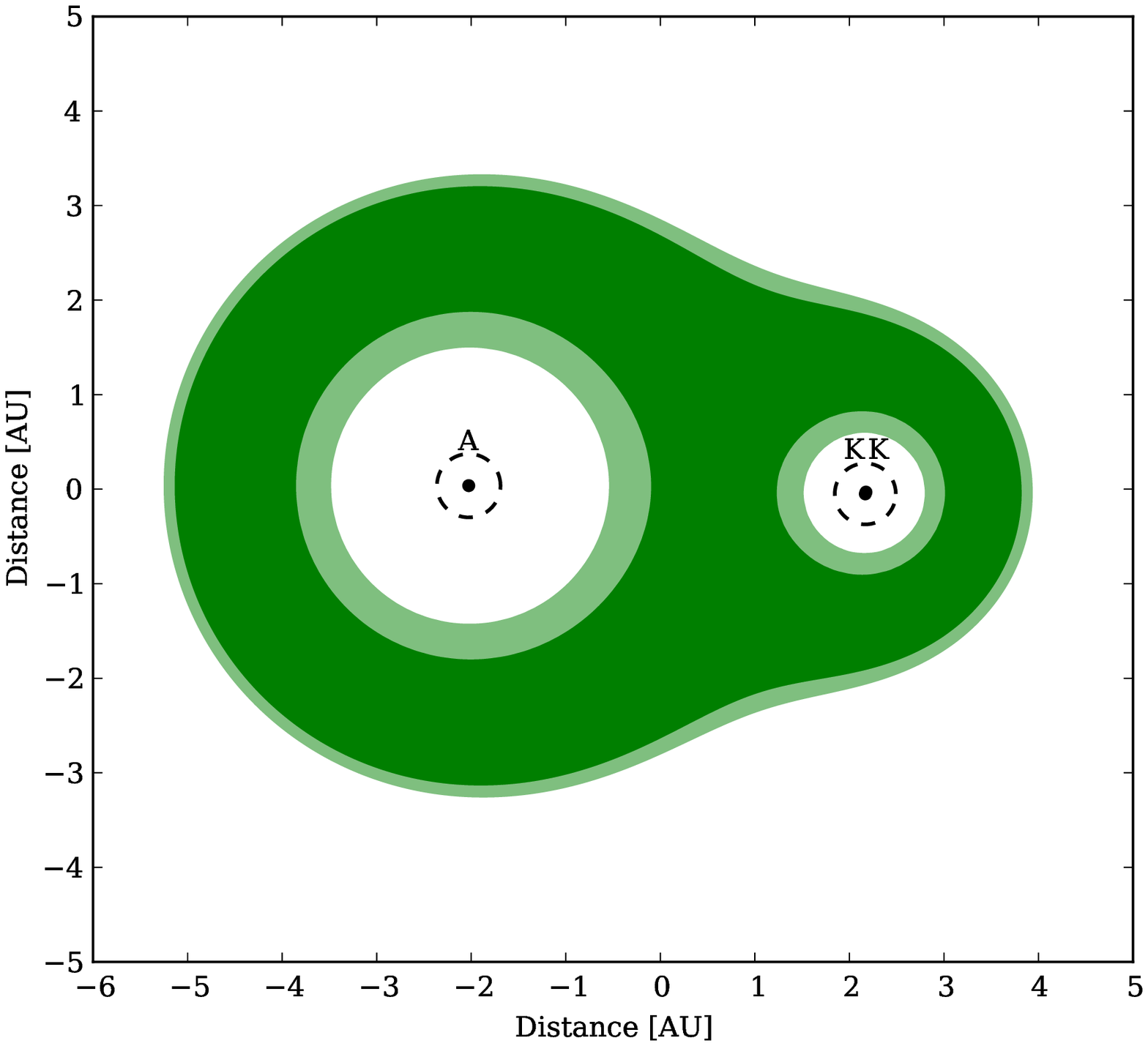}
\caption{The time-evolution of the HZ of the A-KK triplet in the KIC 4150611 system when the orbit of the A star has a semimajor axis of
2.8 AU and an excentricity of 0.5. From top-right panel and in a counter-clockwise rotation, the panels correspond to the A star being at 
${0^\circ},\,{63^\circ},\,{110^\circ},$ and ${180^\circ}$ with respect to the horizontal line passing through zero on the vertical axis. 
The dashed circles correspond to the outer boundary of the stability of planetary orbits.
A movie of the HZ of this system can be found at http://astro.twam.info/hz-multi. }
\label{fig:4150611-model1}
\end{figure}

The fact that the high luminosity of the A star stripped the KK binary from its HZ motivated us to examine how the HZ of the entire
triple system would change, and at what stage the HZ around the KK binary would re-appear, had the A star been at farther distances. 
We, therefore, increased the radial distance of the A star from the center of mass of the system, and determined the smallest distance 
(${r_1}=2.8$ AU) for which the inner edge of the empirical HZ re-appeared around the KK binary. We continued increasing the radial 
distance of the A star until the smallest value (${r_2}=6.45$ AU) for which the outer boundary of the empirical HZ of the KK binary 
re-appeared. Beyond this distance, the two HZs of the A star and KK binary will separate. To study the HZ of the triple system in the 
intermediate distances, we considered two cases. In the first case, we assumed the closest and farthest distances of the A star to be 
$0.5 {r_1}$ and $1.5 {r_1}$, respectively, and calculated the semimajor axis and eccentricity of an elliptical orbit for the A star for which
these distances could be the periastron and apastron $({a_1}=2.8\, {\rm AU}, {e_1}=0.5)$. Figure \ref{fig:4150611-model1} shows 
four snapshots of the HZ of this system
during the orbital motion of the A star. From top-right panel and in a counter-clockwise rotation, the panels correspond to the A star to be at angles
${0^\circ},\,{63^\circ},\,{110^\circ},$ and ${180^\circ}$ with respect to the horizontal line passing through zero on the vertical axis.
The dashed circles correspond to the outer boundaries of planetary stability (planets interior to the dashed circles
will be stable). Around the A star, this boundary is at 0.34 AU,
and around the KK binary is at 0.32 AU. The inner boundary of stability around the entire system is at approximately 10.18 AU (outside this
limit, planets will have stable orbits). 
As shown here, while the A star is still the dominating factor in establishing the boundaries of the narrow HZ, there are distances
where the inner region of the empirical HZ of the KK binary is solely due to the luminosities of its two stars. For a movie of the HZ
of this system see http://astro.twam.info/hz-multi.

\begin{figure}
\centering
\includegraphics[width=0.45\columnwidth]{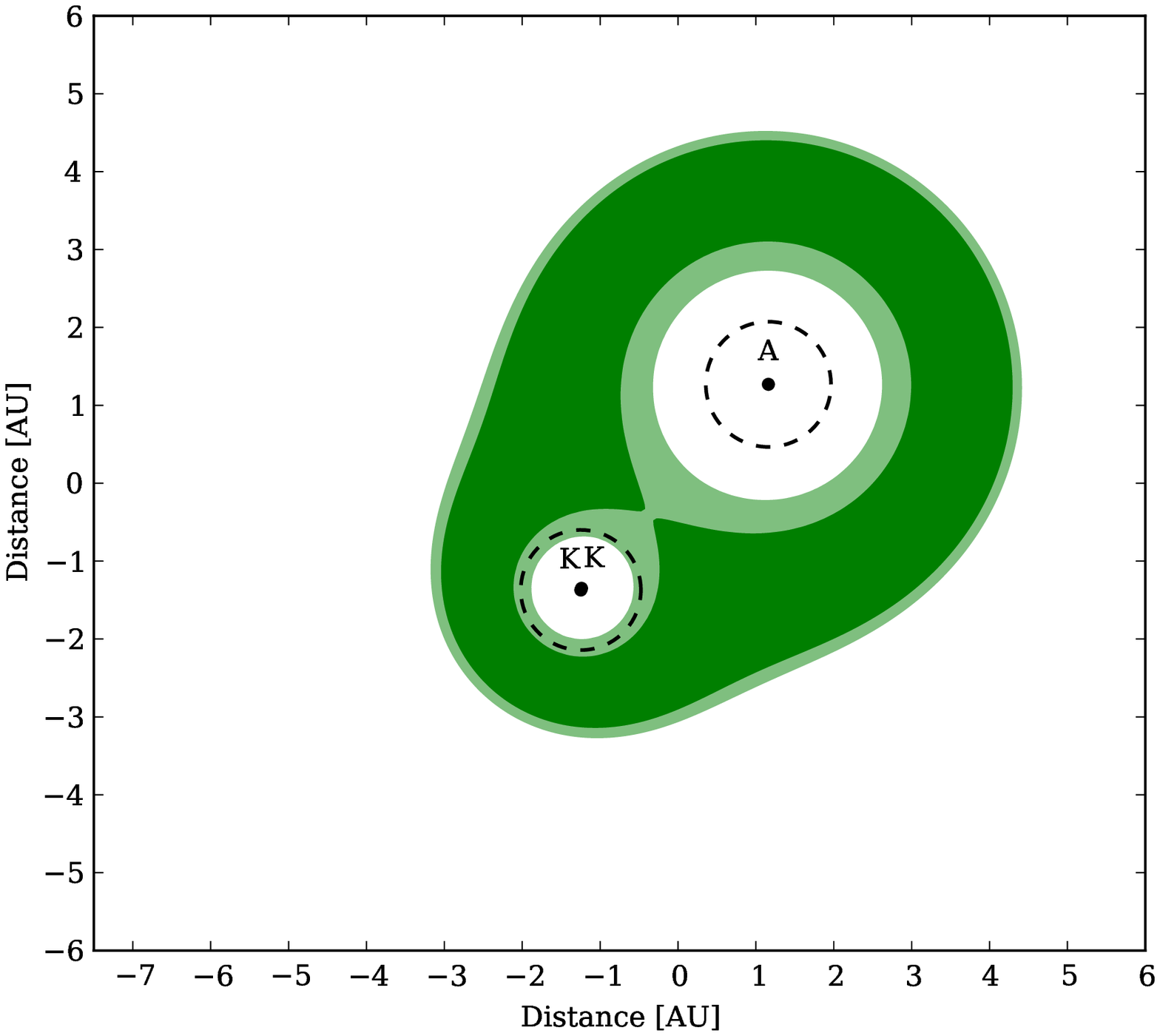}
\includegraphics[width=0.45\columnwidth]{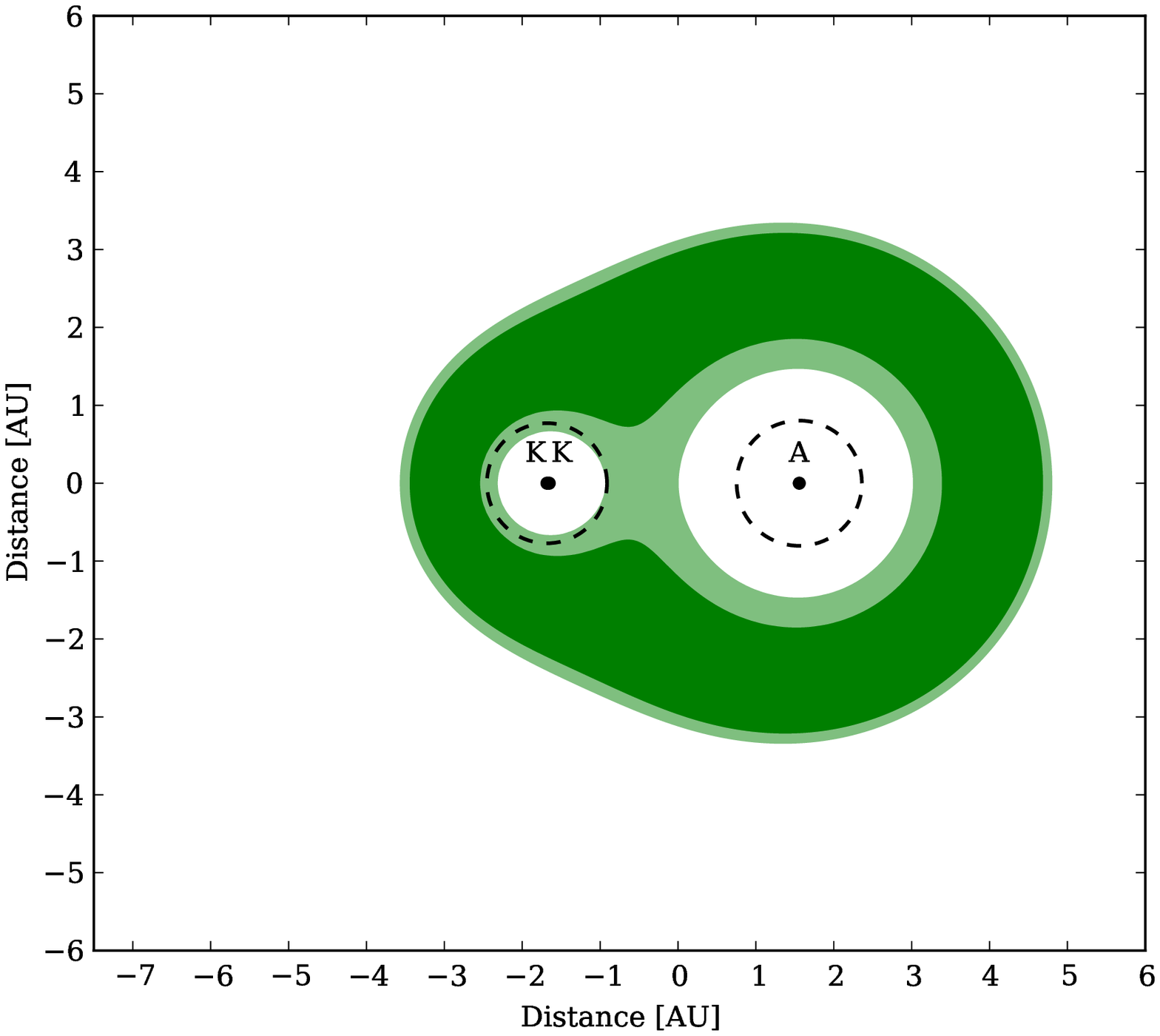}
\vskip 10pt
\includegraphics[width=0.45\columnwidth]{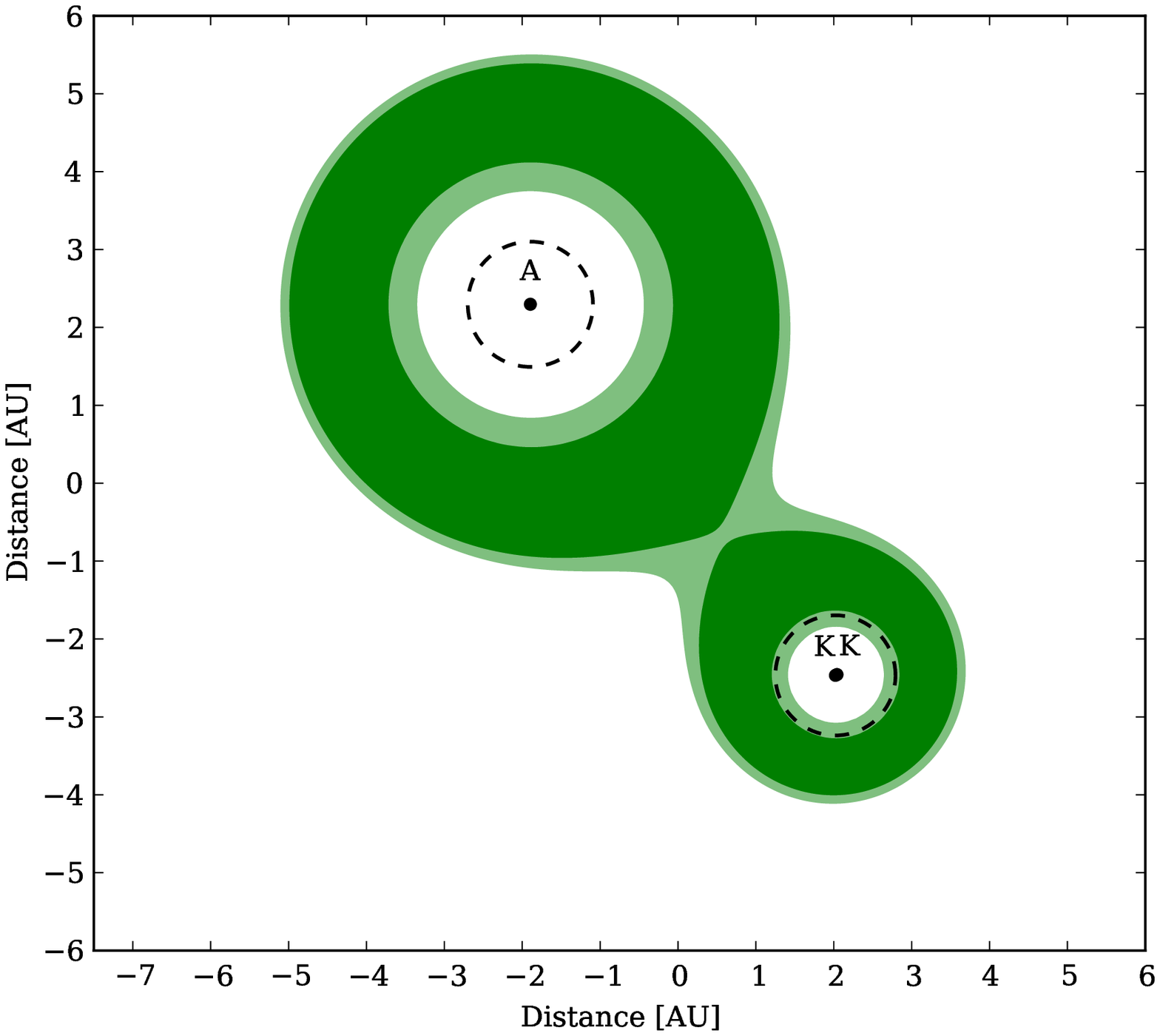}
\includegraphics[width=0.45\columnwidth]{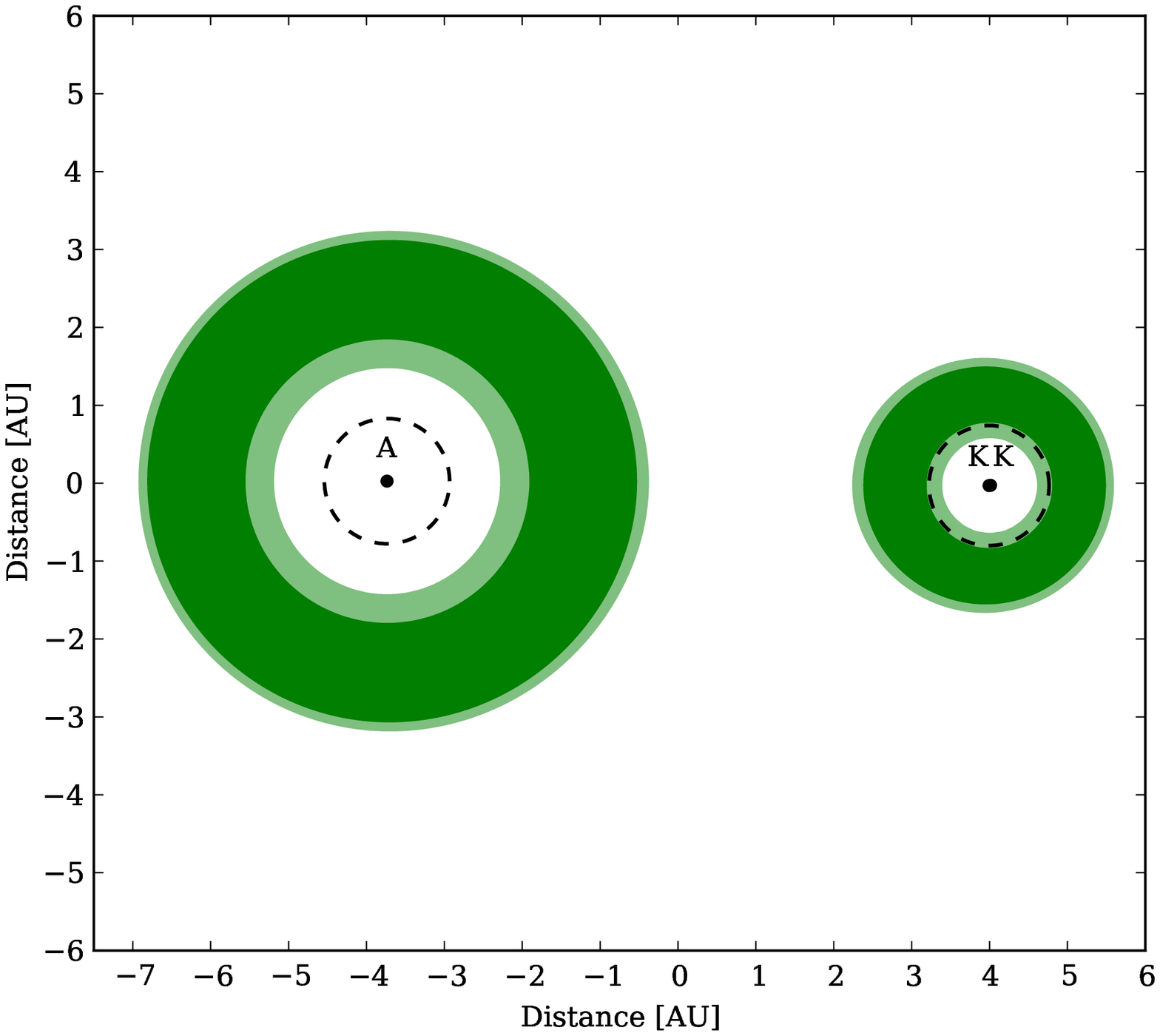}
\caption{The time-evolution of the HZ of the A-KK triplet in the KIC 4150611 system when the orbit of the A star has a semimajor axis of
5.48 AU and an eccentricity of 0.412. From top-right panel in a counter-clockwise rotation, the panels correspond to the A star being at 
${0^\circ},\,{19^\circ},\,{86^\circ},$ and ${180^\circ}$ with respect to the horizontal line passing through zero on the vertical axis.
The dashed circles correspond to the outer boundary of the stability of planetary orbits.
A movie of the HZ of this system can be found at http://astro.twam.info/hz-multi. }
\label{fig:4150611-model2}
\end{figure}

In the second case, we considered the closest and farthest distances of the A star to be 
$0.5 {r_2}$ and $1.5 {r_2}$, respectively. The corresponding elliptical orbit of the A star in this case has a semimajor axis
of 5.48  AU and an eccentricity of 0.412. Figure \ref{fig:4150611-model2} shows the evolution of the HZ during the motion of the A star.
From the top-right panel and in a counter-clockwise rotation, the panels correspond to ${0^\circ},\,{19^\circ},\,{86^\circ},$ and ${180^\circ}$,
respectively. The outer boundaries of planetary orbit stability are at 0.8 AU around the A star and 0.77 AU around the KK binary.
The inner stability limit around the entire system is at 18.95 AU. As shown here, beyond certain distances, the HZ of the KK binary 
reappears, and the HZ of the triple system consists of 
two separate HZs corresponding to those of the A star and KK binary, respectively.
A movie of the HZ of this system can be found at http://astro.twam.info/hz-multi.

\subsection{KID\,5653126}

The A-KK triplet in the multiple star system KIC 4150611 provided an interesting case for studying how the HZ of a binary
can be affected by a highly luminous farther companion. To study the reverse situation, that is, the possible effect of a less luminous
farther star on the HZ of a binary with more luminous components, we considered the hierarchical triple star system KID 5653126.
This system consists of a close binary (hereafter labeled as A and B) with a period of 38.5 days and a farther companion (hereafter labeled as C)
orbiting the AB binary in an 800-day orbit. The stellar properties of this system are unknown. To ensure that the farther
companion would be a less luminous star, we consider the star A to be similar to the primary of the Kepler 47 system and 
stars B and C to be similar to the primary and secondary stars of Kepler 64, respectively. Table \ref{table3} shows the values of the mass, 
luminosity, and effective temperatures of these stars. From the values of the mass and orbital periods of the system, the AB binary 
will have a semimajor axis of ${a_{\rm AB}}=0.31$ AU and that of star C, with respect to the center of mass of the binary system,
will be ${a_{\rm C}}=2.42$ AU. In contrary to the A-KK triplet in the KIC\,4150611 system, the separation 
of star C is only 7.8 times larger than the semimajor axis of the inner binary AB, implying that this star may have a noticeable
effect on the HZ of the binary system. 

\begin{figure}
\vskip -0.2in
\centering
\includegraphics[width=0.42\columnwidth]{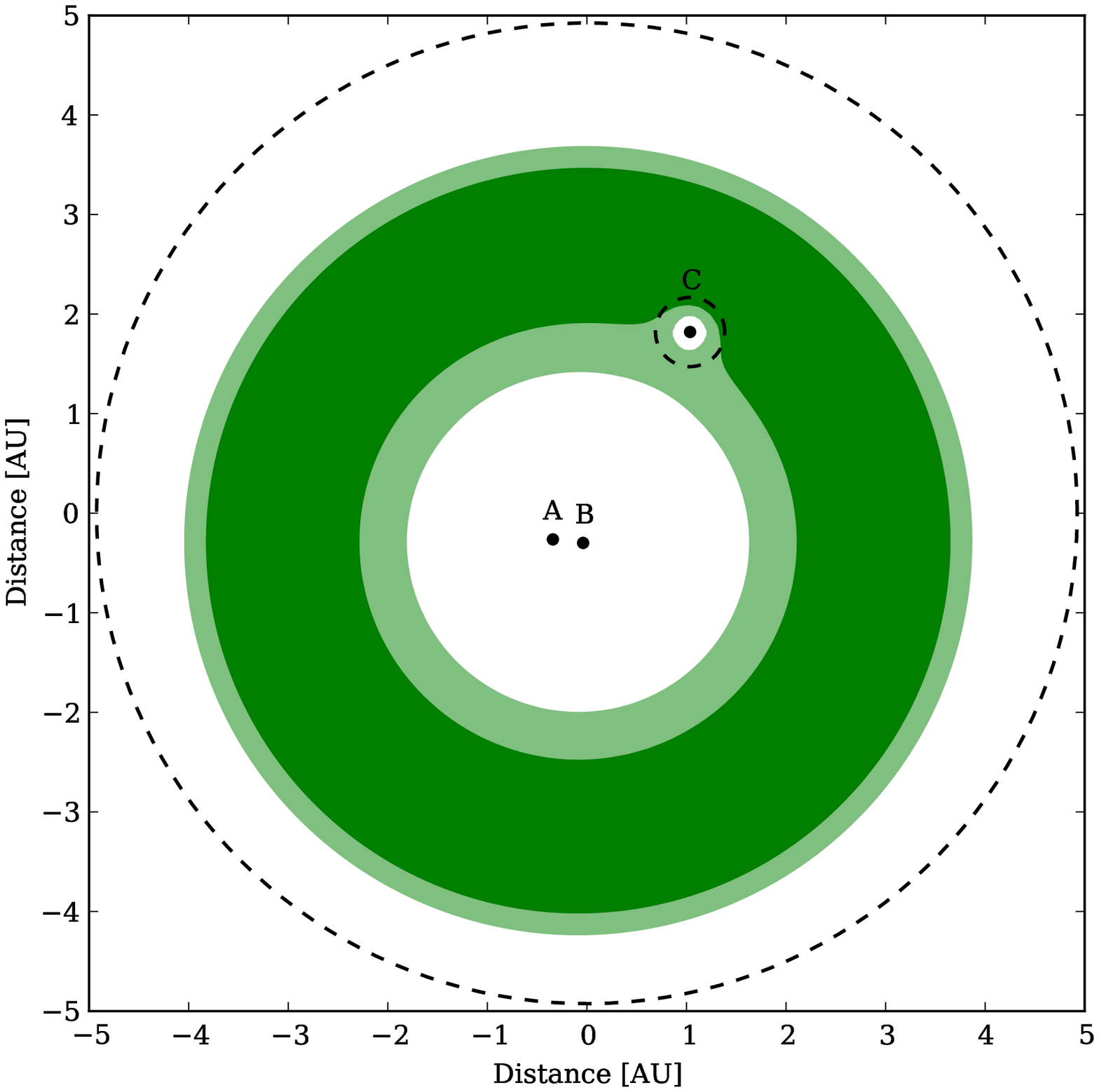}
\includegraphics[width=0.42\columnwidth]{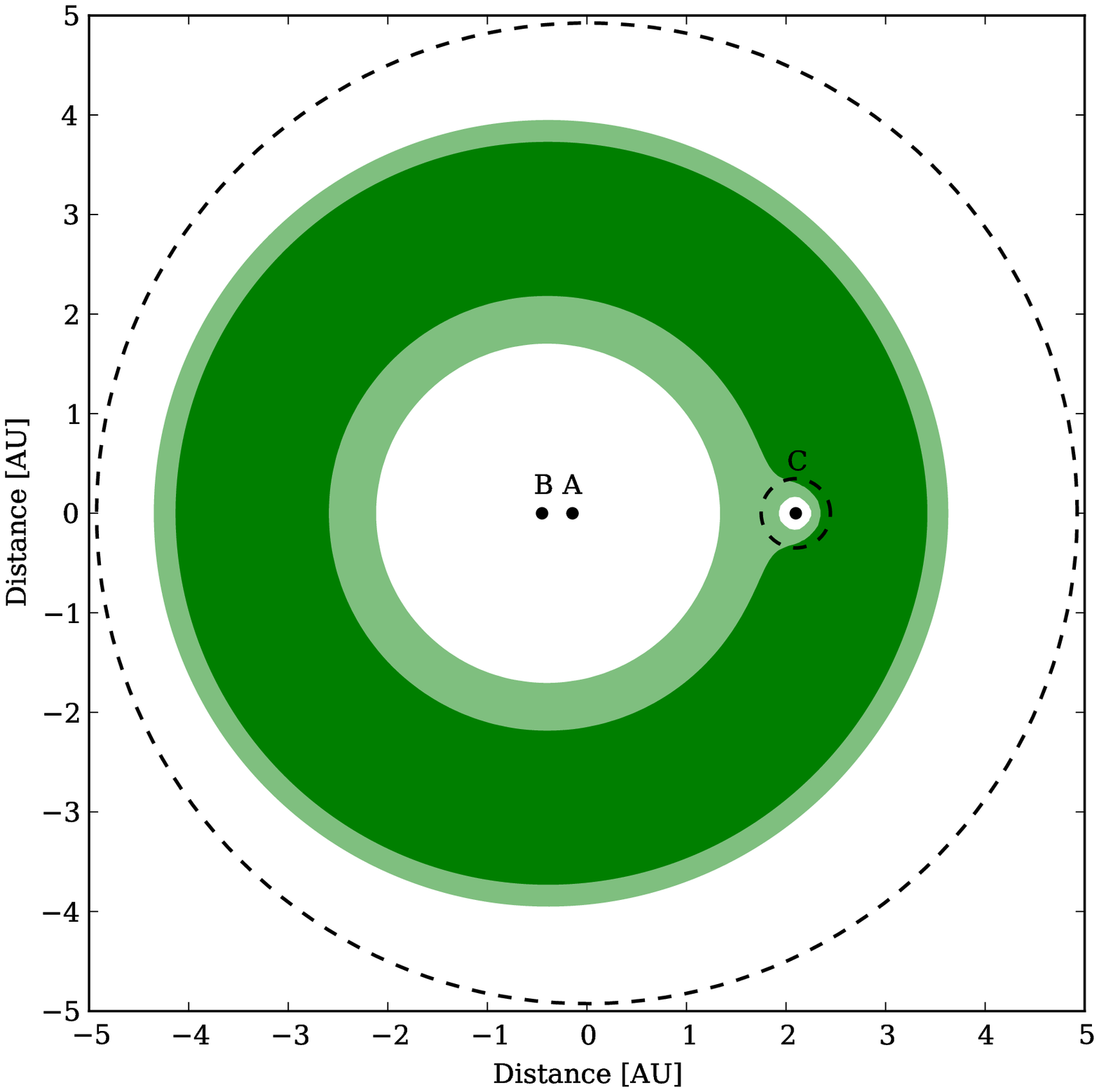}
\includegraphics[width=0.42\columnwidth]{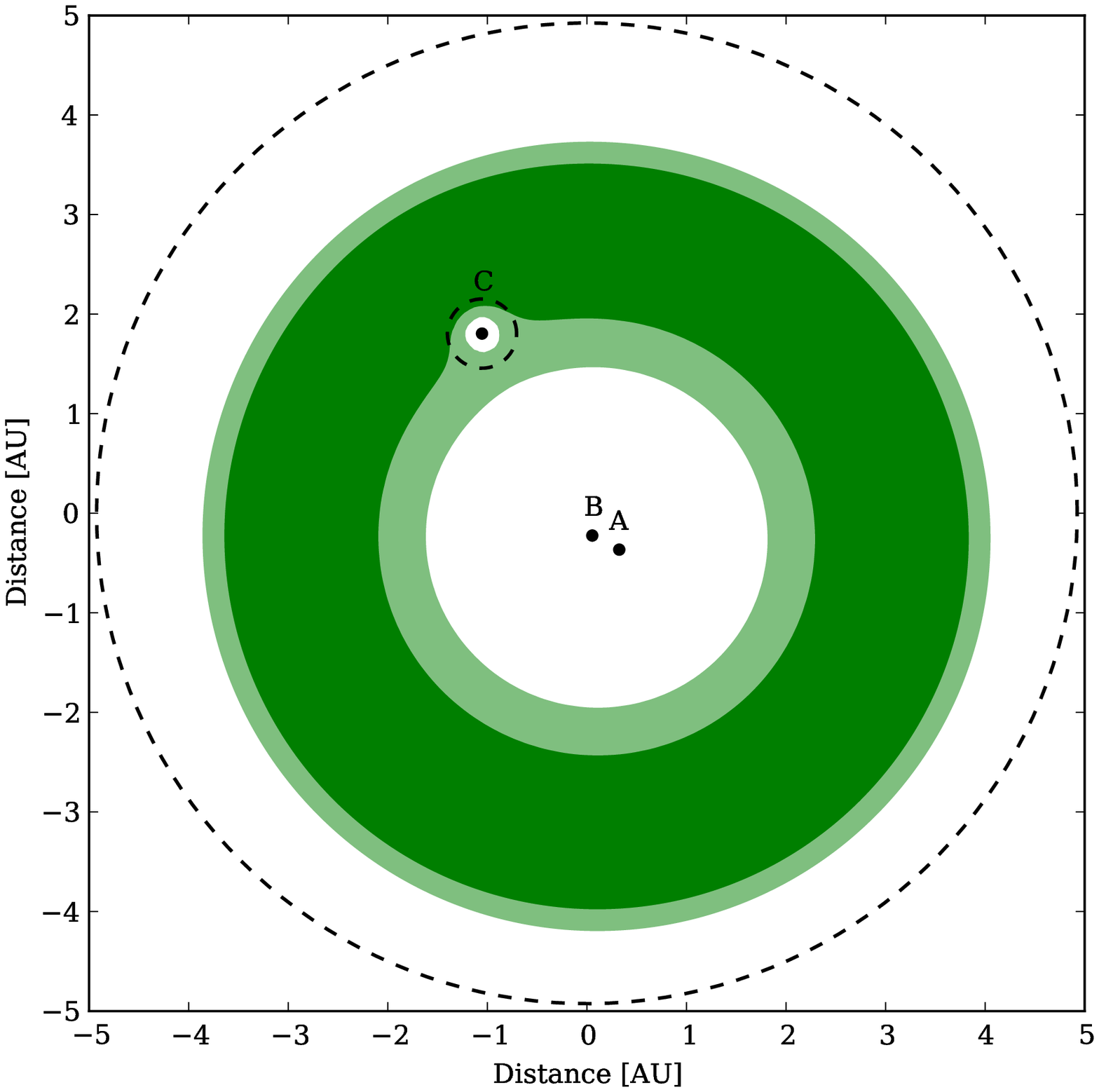}
\includegraphics[width=0.42\columnwidth]{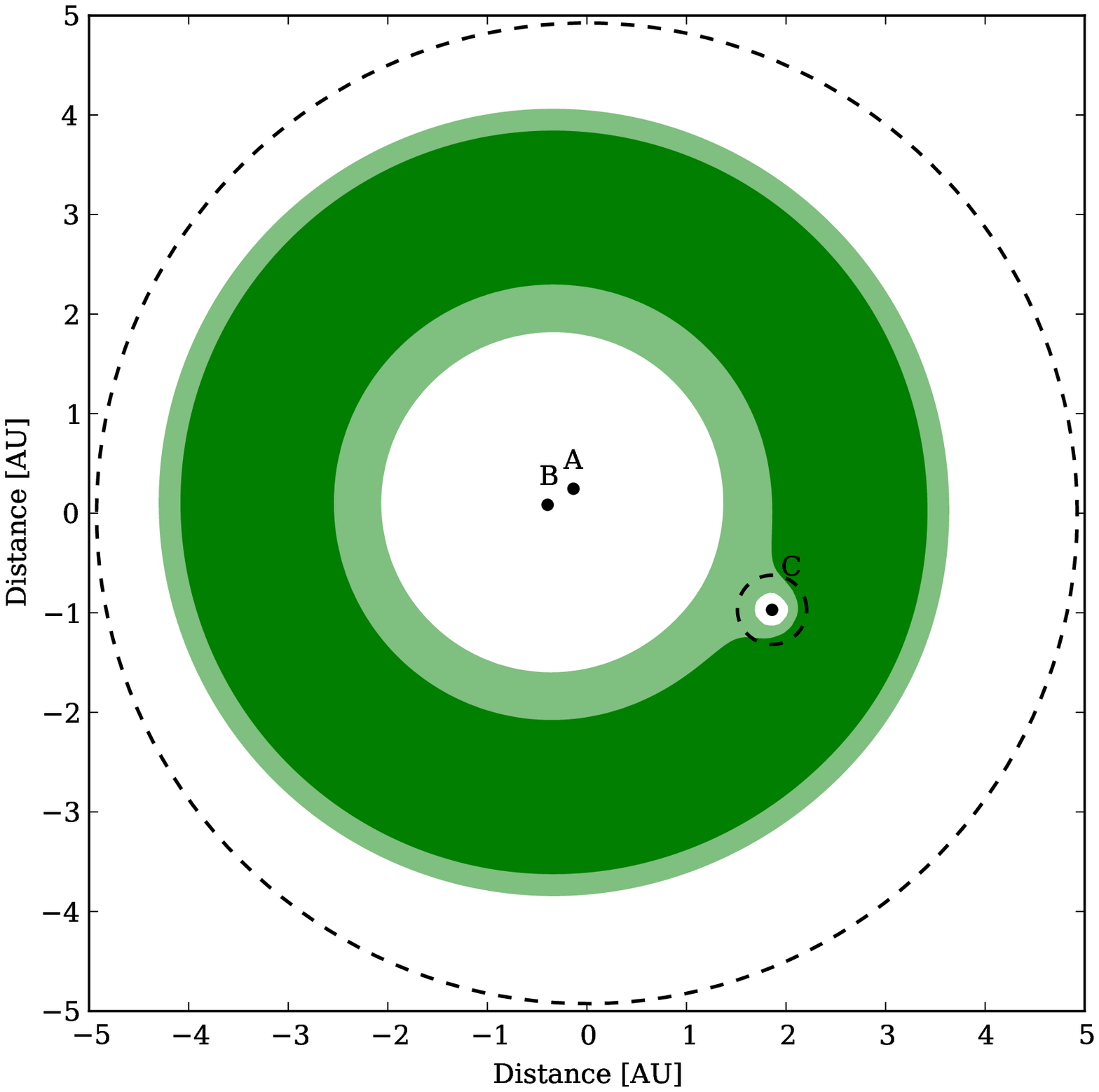}
\includegraphics[width=0.42\columnwidth]{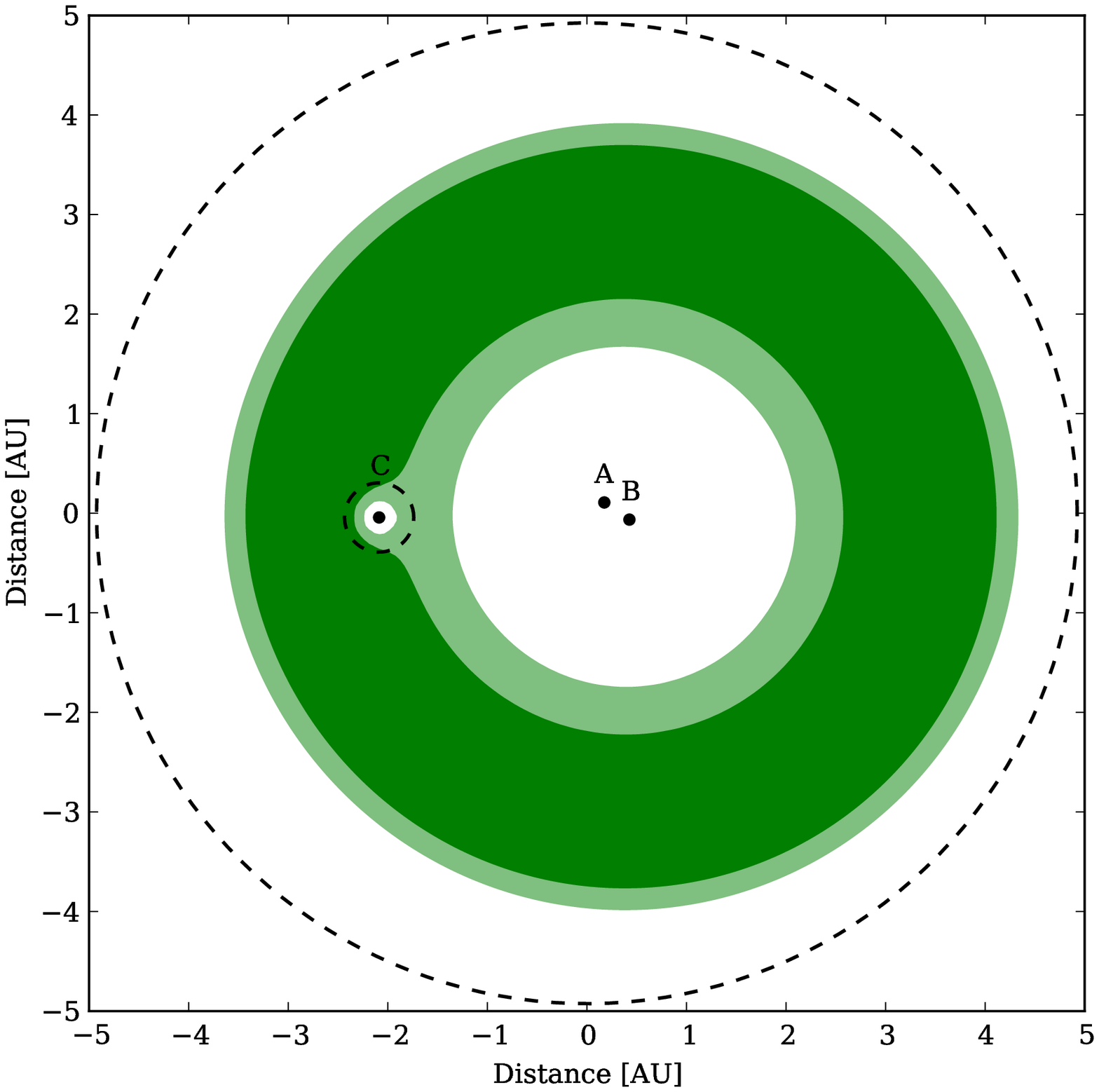}
\includegraphics[width=0.42\columnwidth]{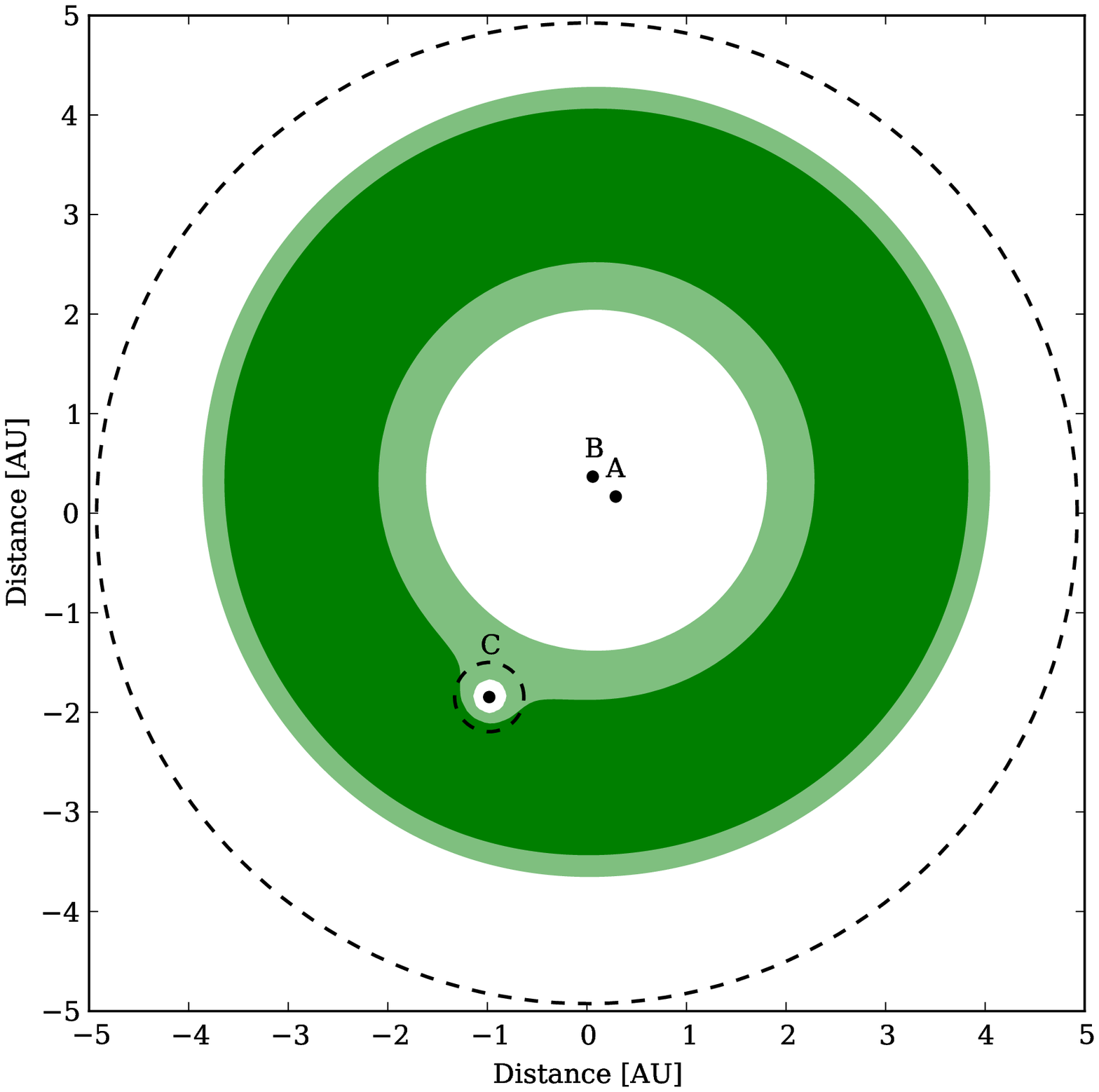}
\caption{Graphs of the HZ of the KID 5653126 system when the three stars are in circular orbits. From top-right panel and counter-clockwise, 
the figures show the changes in the boundaries of the HZ during one revolution of the star C when this star is at 
${0^\circ},\,{60^\circ},\,{120^\circ},\,{180^\circ},\,{240^\circ}$ and ${330^\circ}$, respectively. The dashed circles corresponds to the 
limits of planetary orbit stability (see figure \ref{fig:stability_5653126}).
A movie of the HZ of this system can be found at http://astro.twam.info/hz-multi.}
\label{fig:5653126}
\end{figure}

Figure \ref{fig:5653126} shows the HZ of the system when the binary AB and star C are in circular orbits.
Starting from the upper-right panel and counter-clockwise, the figures show the evolution of the HZ during one revolution of 
the star C around the binary when this star is at ${0^\circ},\,{60^\circ},\,{120^\circ},\,{180^\circ},\,{240^\circ}$ and ${330^\circ}$,
respectively. As shown here, while the HZ of the system is primarily due to the AB binary, the radiation from star C excludes the 
region around this star from the system's narrow HZ and slightly extends the empirical HZ of the AB binary to farther distances
(movies of the HZ at http://astro.twam.info/hz-multi).

\begin{figure}
\centering
\includegraphics[width=1\columnwidth]{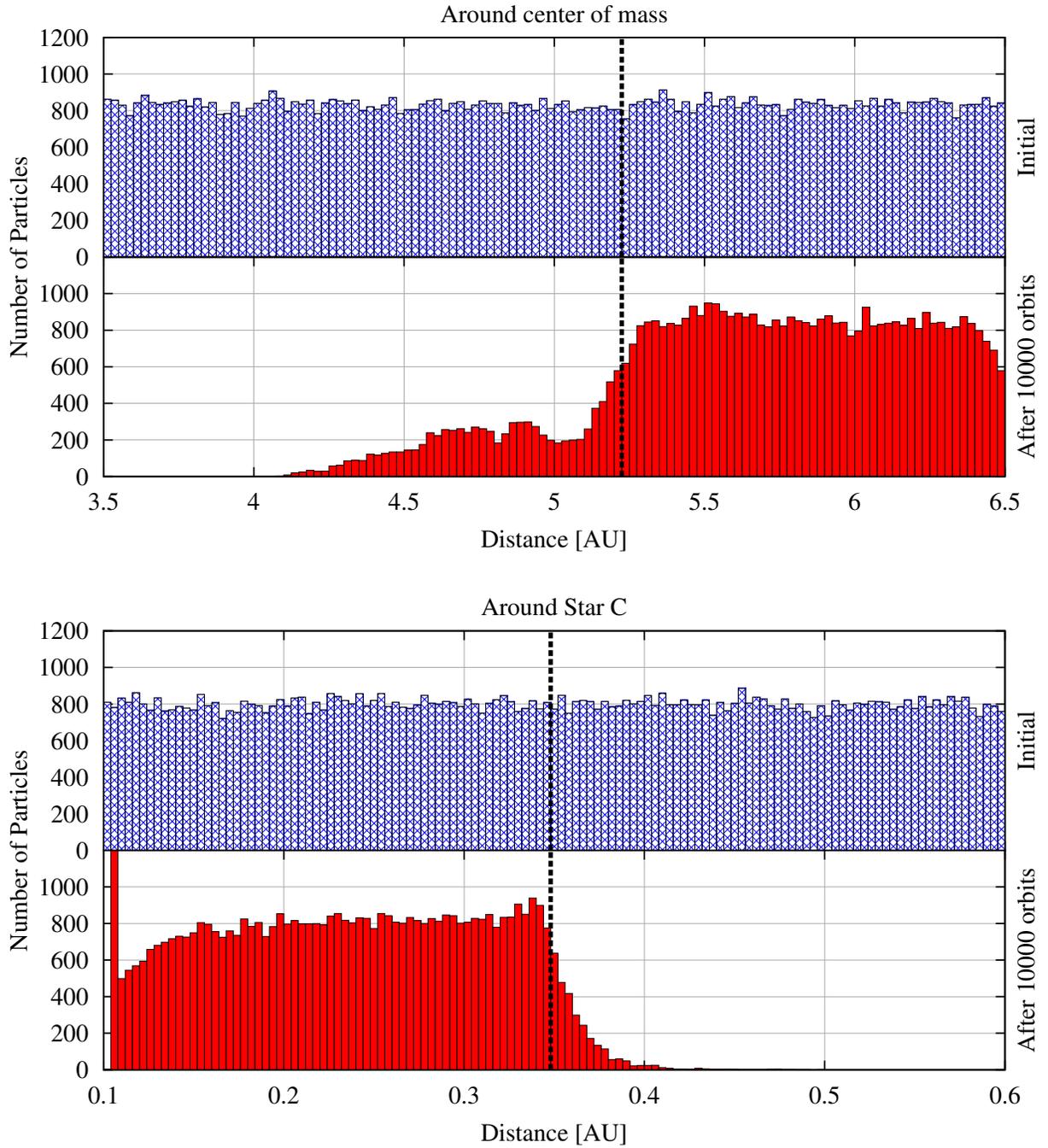}
\caption{
Results of the integrations of non-interacting Earth-mass planets around the center of mass of the KID 5653126 system (top) 
and around the star C (bottom). In each panel, the upper graph (blue histogram) shows the initial distribution of the planets 
and the lower graph (red histogram) corresponds to their distribution after integrating their orbits for 10000 orbital periods 
of the star C around the binary AB. The dashed line shows the boundary of orbital stability at 5.225 AU around the center of mass, and at 
$ 0.348$ AU around star C.}
\label{fig:stability_5653126}
\end{figure}

To determine the stability of an Earth-like planet in the HZ of the KID 5653126 system, similar to the case of KIC 4150611,
we distributed a large number of non-interacting, Earth-mass planets around the center of mass of the AB-C system,
and integrated their orbits for 10000 orbital periods of the star C using the N-body integrator explained in section \ref{sec:kic4150611}.
Figure \ref{fig:stability_5653126} shows the distributions of the Earth-mass planets at the beginning and end of the integrations.
The top panel of this figure corresponds to the stability around the entire system and the bottom panel is for the stability around 
the star C. As shown here, the stability limit around the system's center of mass is at approximately 5.225 AU (shown by a dashed line
in the top panel of Figure \ref{fig:stability_5653126} and by a dashed circle in Figure \ref{fig:5653126}) indicating that the majority of the HZ of the system is unstable. 
As indicated by the bottom panel of Figure \ref{fig:stability_5653126}, however, a small region of stability exists interior to 0.348 AU around the star C 
(shown by a dashed line in the bottom panel of Figure \ref{fig:stability_5653126} and a dashed circle around this star in Figure \ref{fig:5653126})
where an Earth-mass planet can have a stable orbit in a small portion of the empirical HZ around this star.

To determine whether the above-mentioned integration time was sufficient for calculating the stability limit around the entire three stars, 
we approximated the AB-C system with a circular P-type binary.
In this binary, the mass of the primary star is equal to the sum of the masses of stars A and B, and the binary semimajor
axis is 2.42 AU. Using equation (\ref{eq:ptype-stability}), the stability limit of this P-type 
system will be at $\sim$5 AU, which is in a very good agreement with the 5.225 AU that was obtained from direct integrations.

To examine whether the time of the integration was sufficient for calculating the stability limit around the star C, we approximate
the AB-C system with an S-type binary, considering star C to be the primary star. As shown by 
\citet{Holman99} and \citet{Rabl88}, the critical stability limit around each star of an S-type binary (interior to which
a planet will have a long-term stable orbit) can be calculated using

\begin{align}
\label{eq:stype-stability} a_\mathrm{max} = a \left(0.464 - 0.38\,\mu - 0.631\,e + 0.586\,\mu e  + 0.15\,e^2 - 0.198\,\mu e^2 \right)\,.
\end{align}

\noindent
Considering a circular binary with a semimajor axis of 2.42 AU and a mass-ratio of 
$\mu=({M_{\rm A}}+{M_{\rm B}})/({M_{\rm A}}+{M_{\rm B}}+{M_{\rm C}})$, 
the outer boundary of the stable region around star C will be at 0.328 AU. This is in a very good agreement with the 0.348 AU limit obtained 
from direct integrations confirming that the time of integrations for our stability analysis was sufficient.

\subsubsection{Effect of the eccentricity of the star C}

\begin{figure}
\centering
\includegraphics[width=0.48\columnwidth]{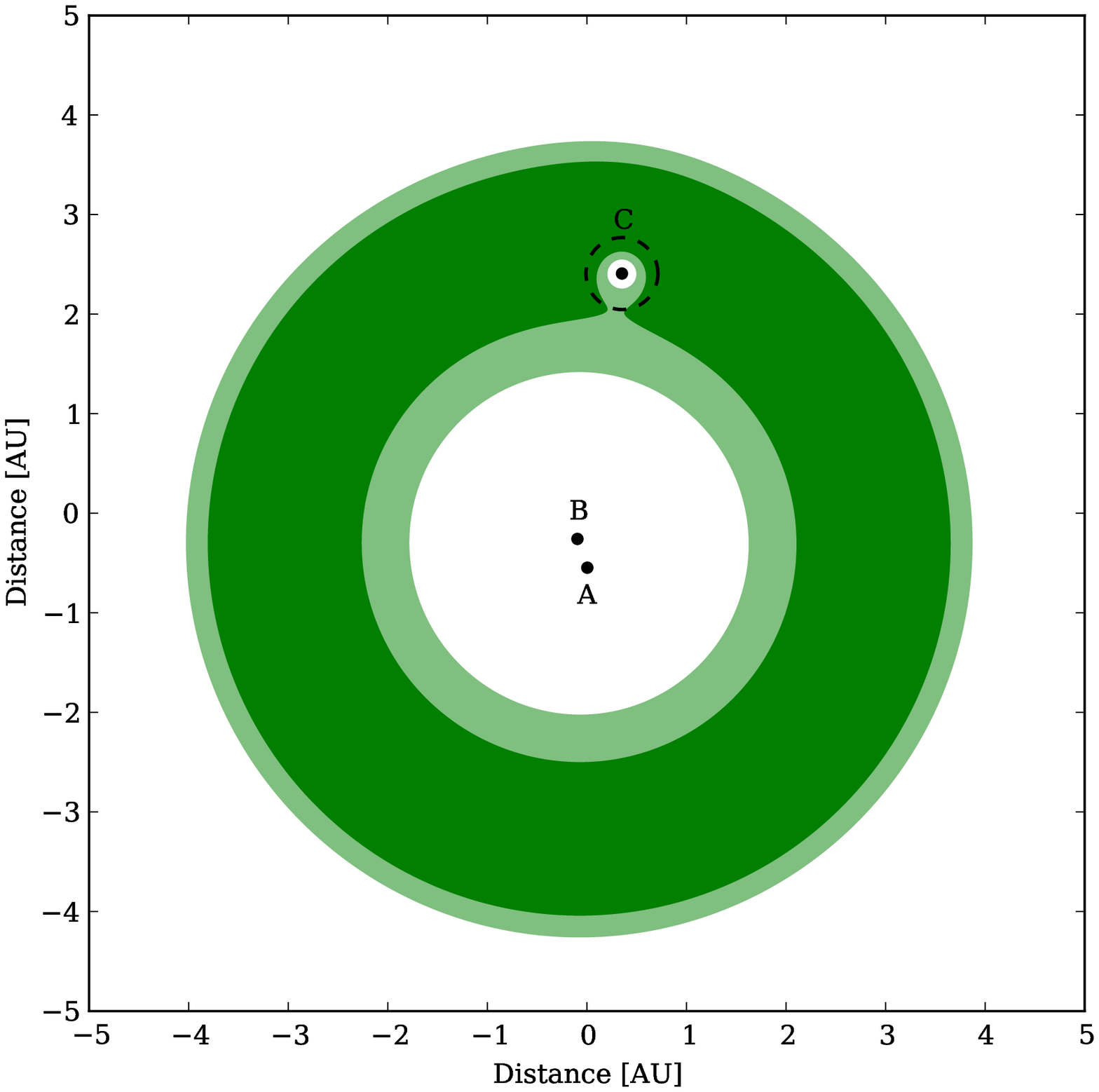}
\includegraphics[width=0.48\columnwidth]{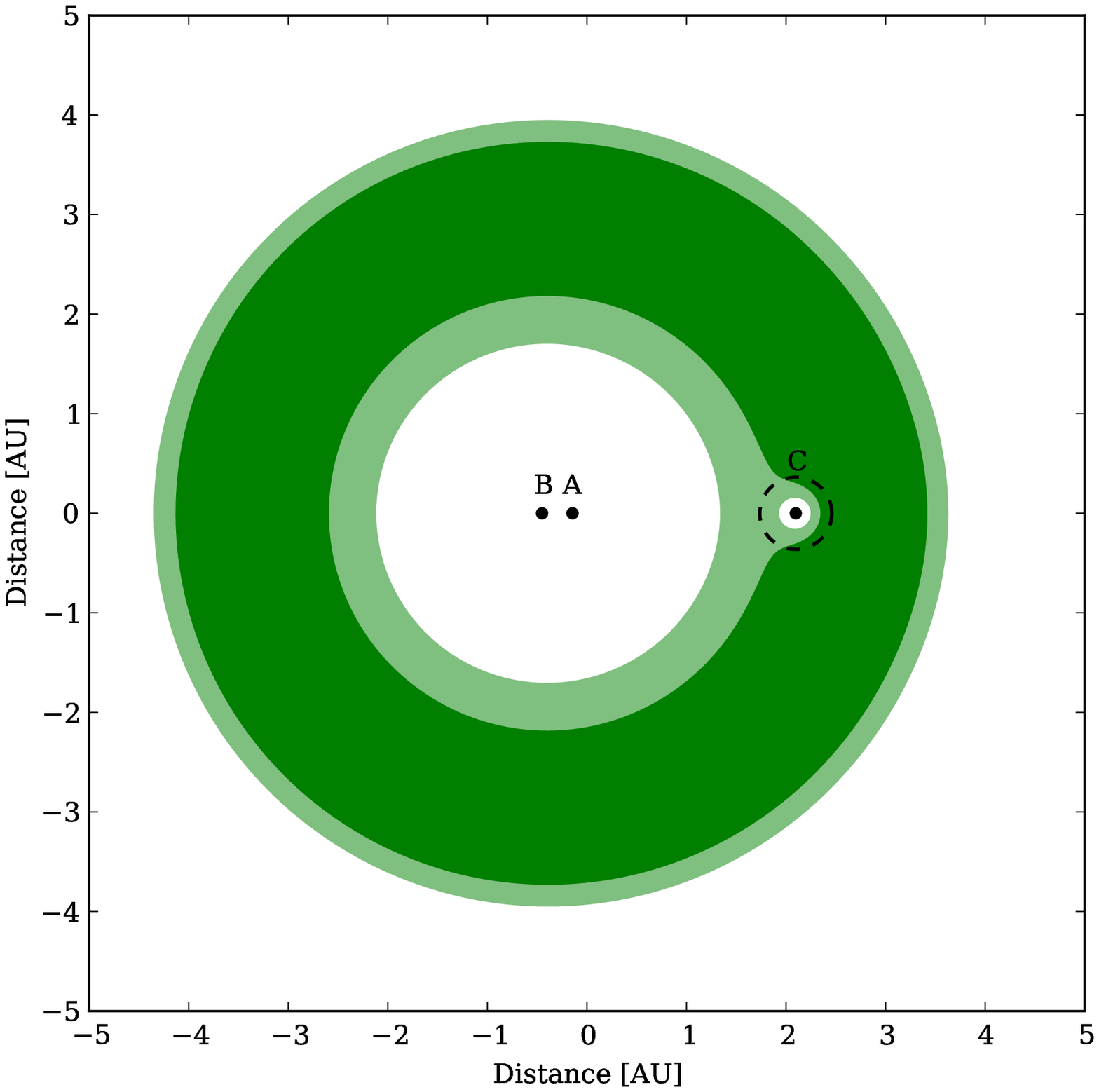}
\vskip 5pt
\includegraphics[width=0.48\columnwidth]{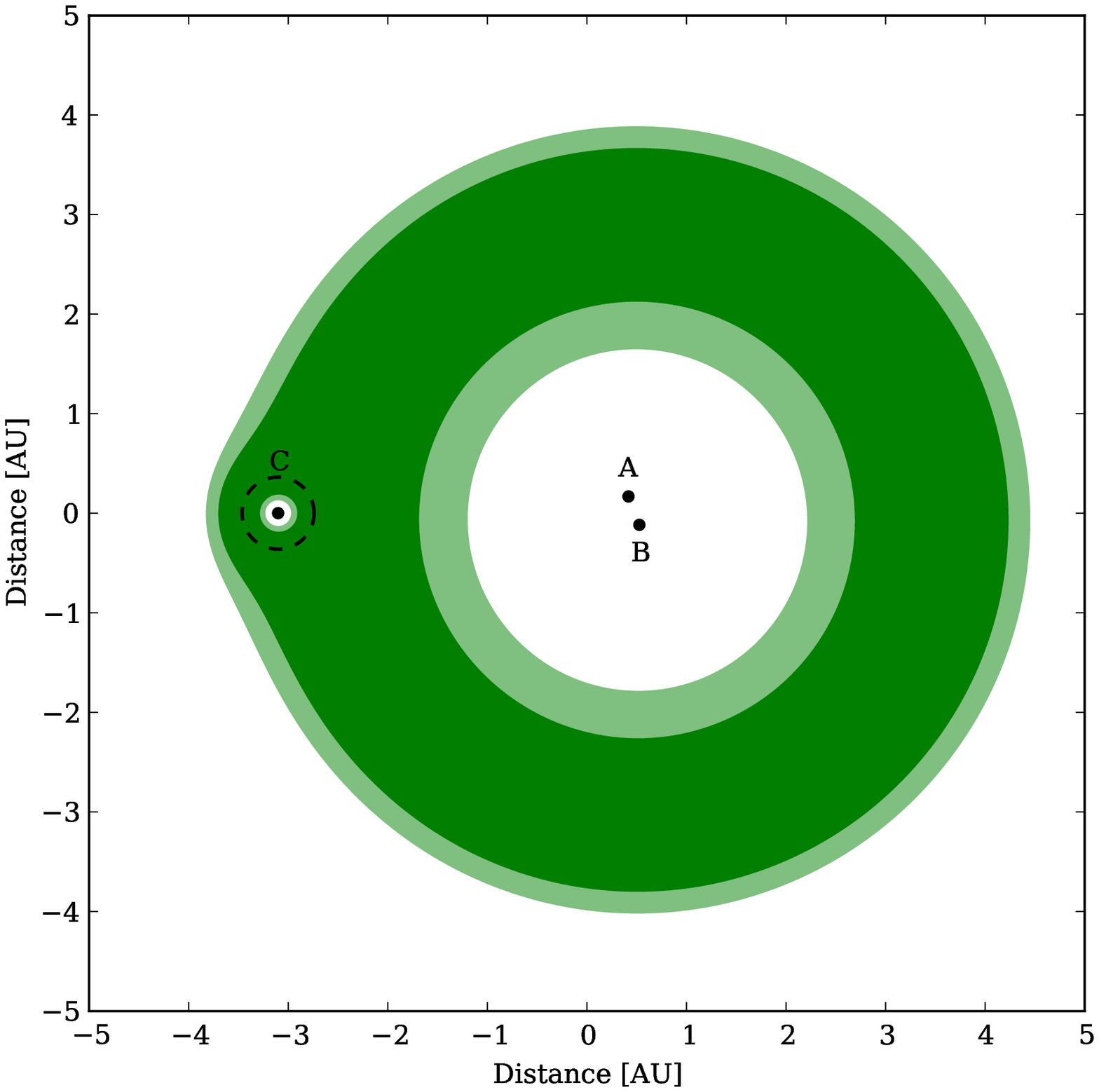}
\includegraphics[width=0.48\columnwidth]{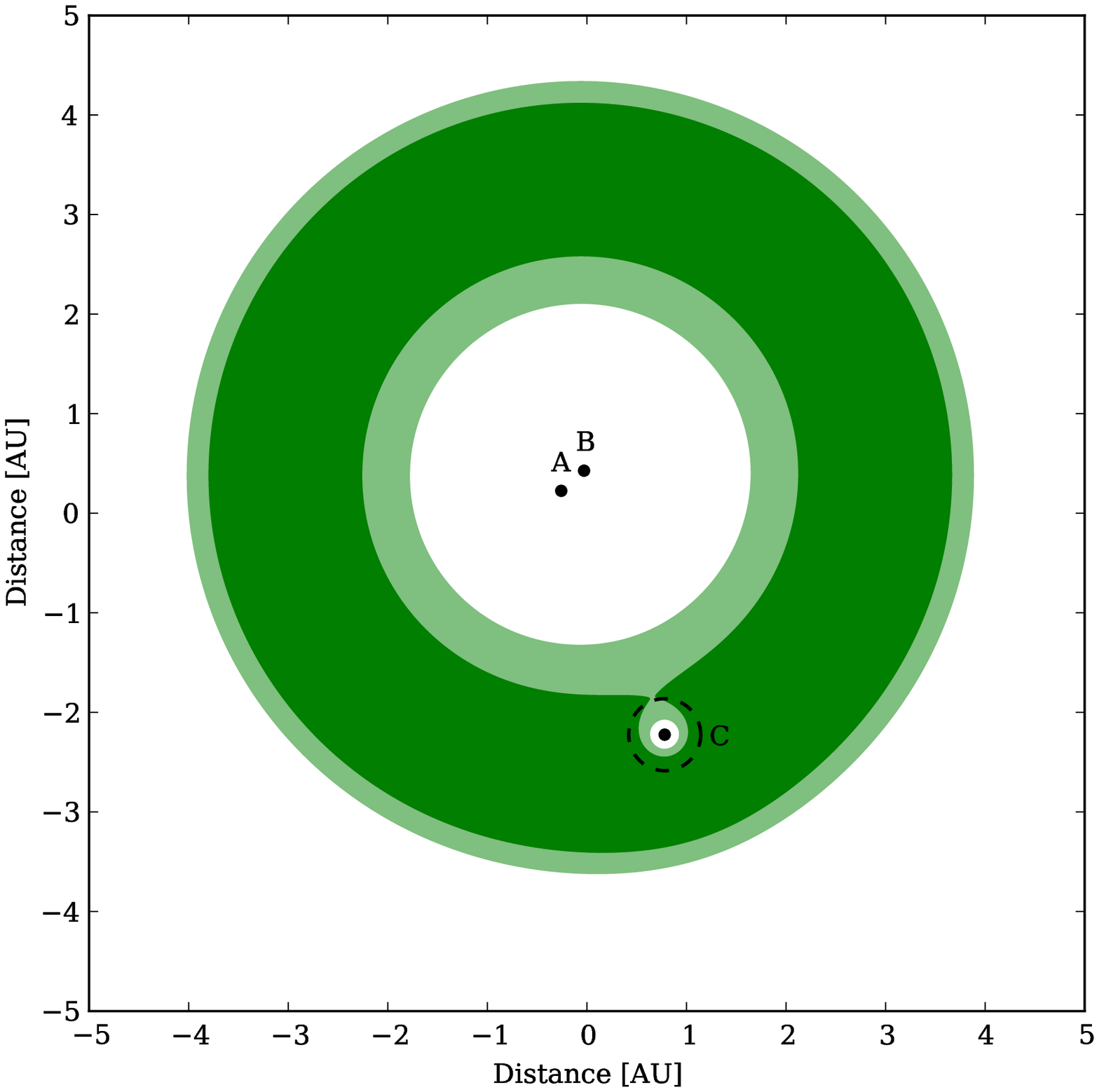}
\caption{Graphs of the HZ of the KID 5653126 system when the star C has a semimajor axis of 3.01 AU and an eccentricity of 0.195.
From top-right panel and in a counter-clockwise rotation, the panels correspond to the star C being at
$,{0^\circ}, {60^\circ}, {180^\circ}$ and $309^\circ$, respectively. 
A movie of the HZ of this system can be found at http://astro.twam.info/hz-multi.}
\label{HZ-5653126C-circular}
\end{figure}

\begin{figure}
\centering
\includegraphics[width=1\columnwidth]{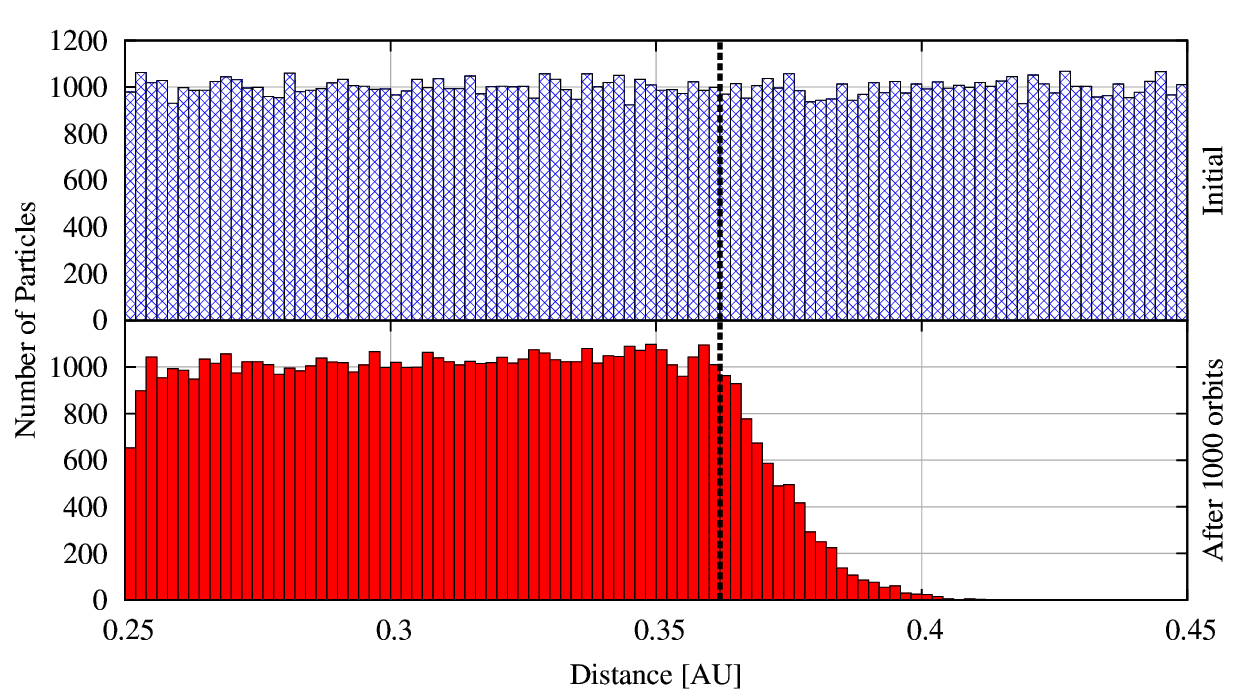}
\caption{Results of the integrations of non-interacting Earth-mass planets around star C in the system of 
Figure \ref{HZ-5653126C-circular}. 
The upper panel (blue histogram) shows the initial distribution of the planets and the lower panel (red histogram) 
corresponds to their distribution after integrating their orbits for 10000 orbital periods of star C. The dashed line 
shows the outer boundary of orbital stability at 0.362 AU.}
\label{Stability-5653126C-circular}
\end{figure}

The fact that during its motion around the AB binary, the star C removes part of the HZ of the system that is in its vicinity motived us to
examine how the HZ would change if the orbit of this star were eccentric. Similar to the case of KIC 4150611, we considered two cases.
In the first case, we determined the values of the semimajor axis and eccentricity of star C for which this star would break away
from the inner part of the empirical HZ of the system while having a periastron distance equal to 2.42 AU and an apastron interior
to the outer edge of the system's empirical HZ. In this case ${a_{\rm C}}=3.01$ AU, and $e_{\rm C}=0.195$. Figure \ref{HZ-5653126C-circular} 
shows the HZ of the system in one orbit of the star C around the binary. From the upper-right panel and in a counter-clockwise rotation,
the panels correspond to star C at ${0^\circ}, {60^\circ}, {180^\circ}$ and $309^\circ$. As shown here, the inner part of 
the empirical HZ of the system expands temporarily as the star C moves away from this region. The outer edge of the HZ is also extended 
to larger distances as this star approaches its apastron position. Figure \ref{Stability-5653126C-circular} shows the outer boundary of
orbital stability for Earth-mass planets around star C. Integrations were carried out for 1000 orbital periods of planet C.
As indicated by the vertical dashed line, 
planets with semimajor axes smaller than 0.362 AU will have stable
orbits around this star. Figure \ref{HZ-5653126C-circular} shows this region with a dashed circle around star C. As shown in this figure, the
empirical HZ around this star maintains stability for planetary orbits as this star rotates around the AB binary.
For a movie of the HZ of this system, we refer the reader to the website http://astro.twam.info/hz-multi.

\begin{figure}
\vskip -.2in
\centering
\includegraphics[width=0.33\columnwidth]{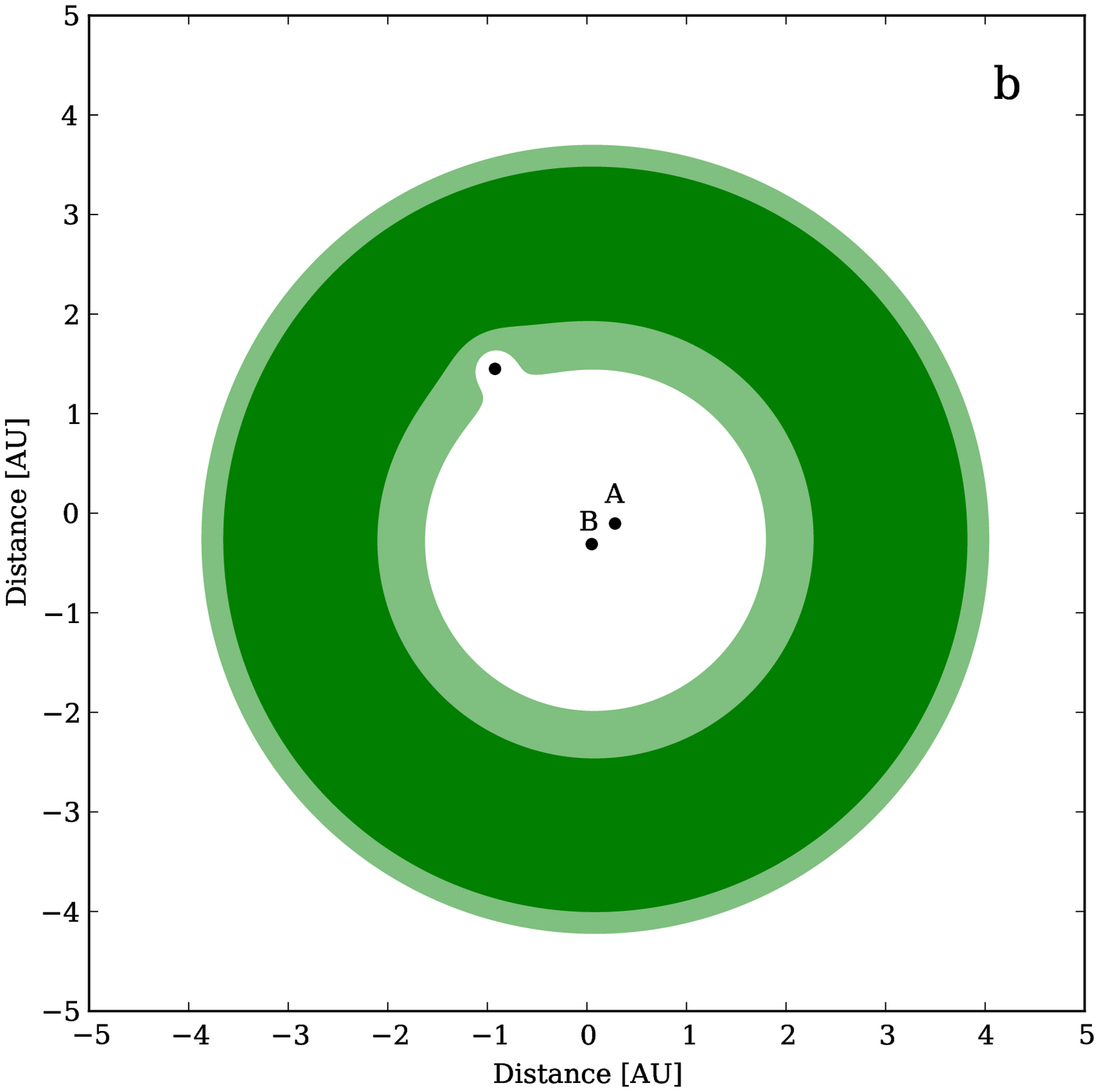}
\includegraphics[width=0.33\columnwidth]{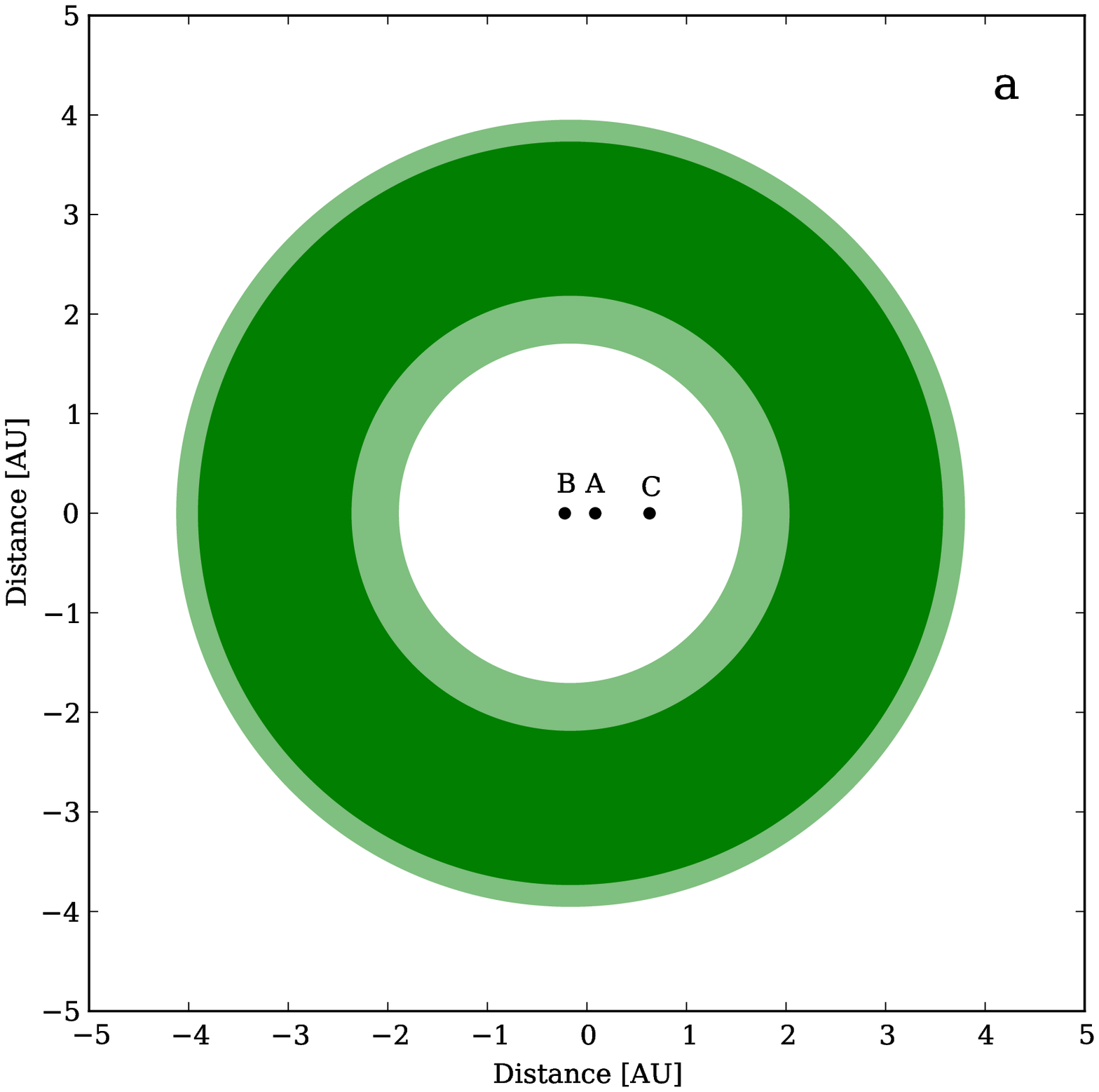}
\includegraphics[width=0.33\columnwidth]{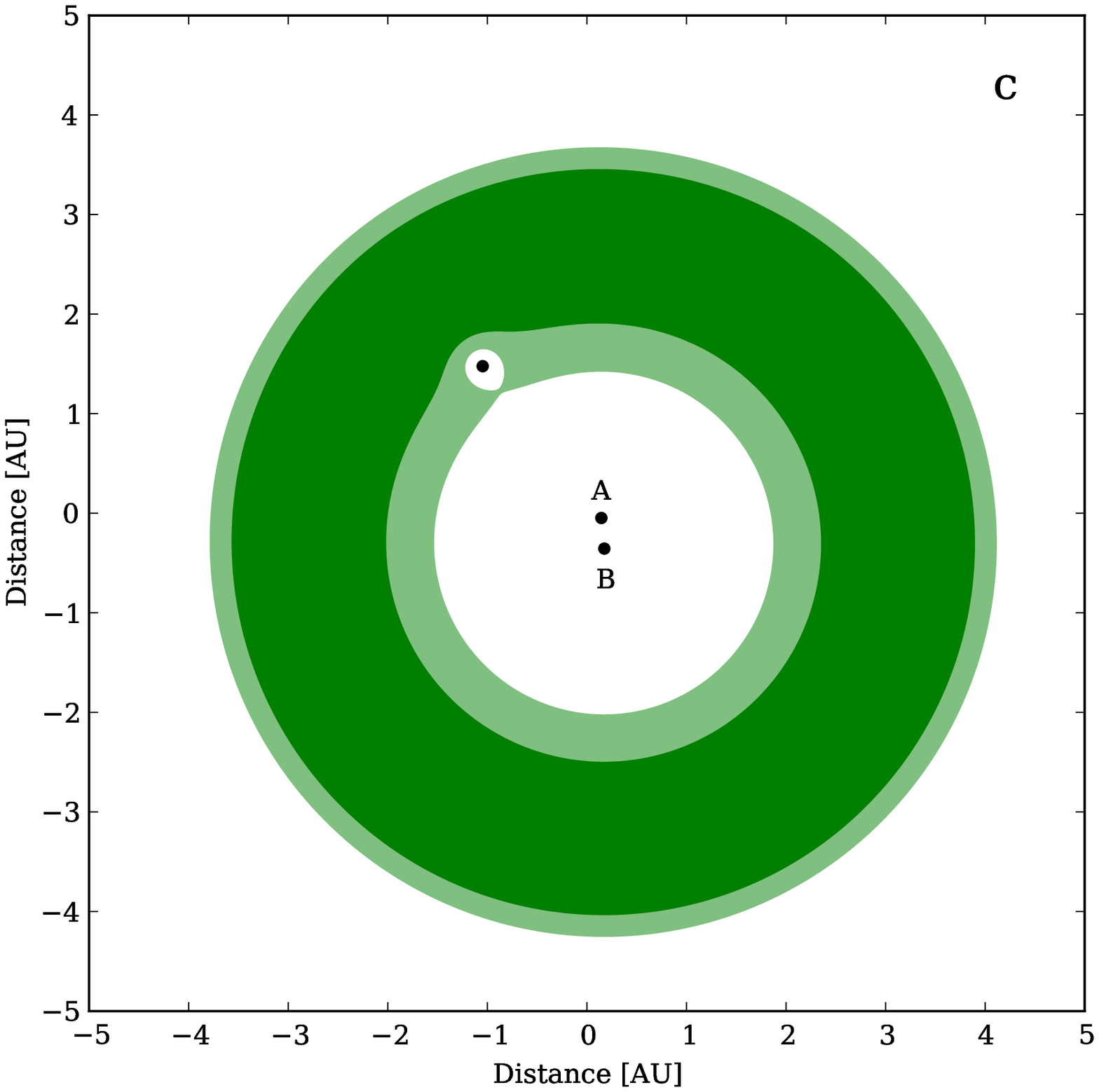}
\includegraphics[width=0.33\columnwidth]{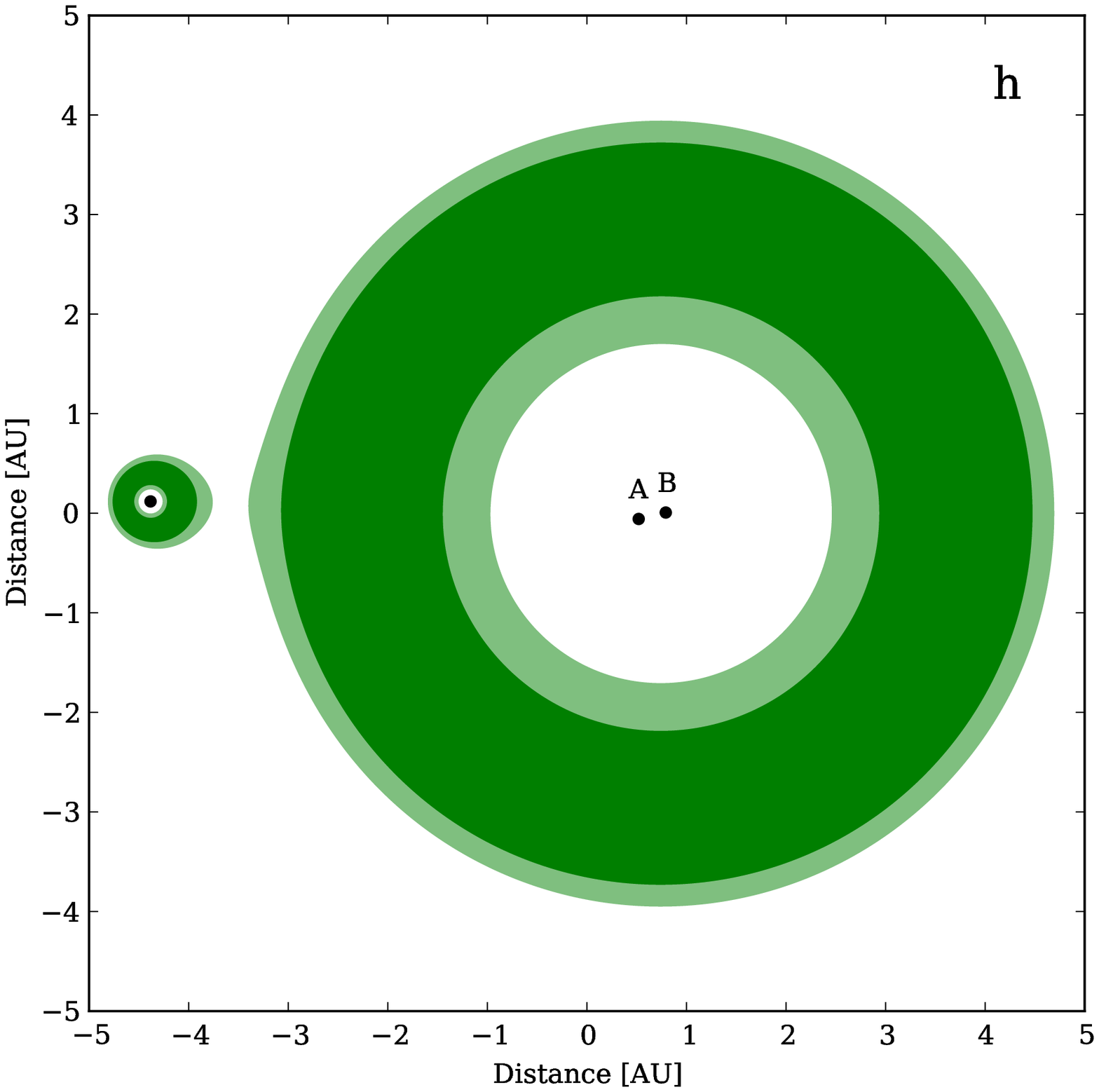}
\includegraphics[width=0.33\columnwidth]{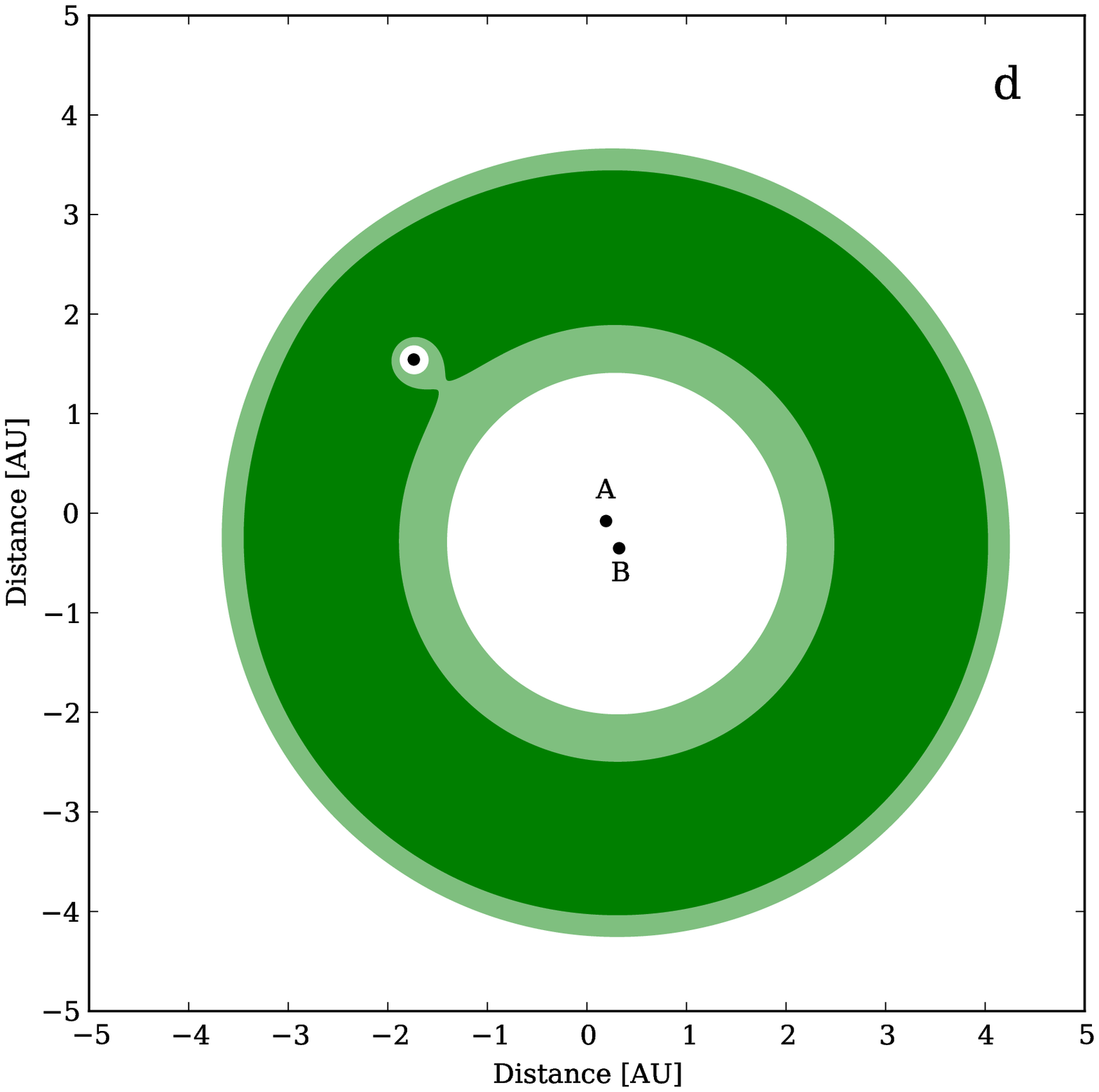}
\includegraphics[width=0.33\columnwidth]{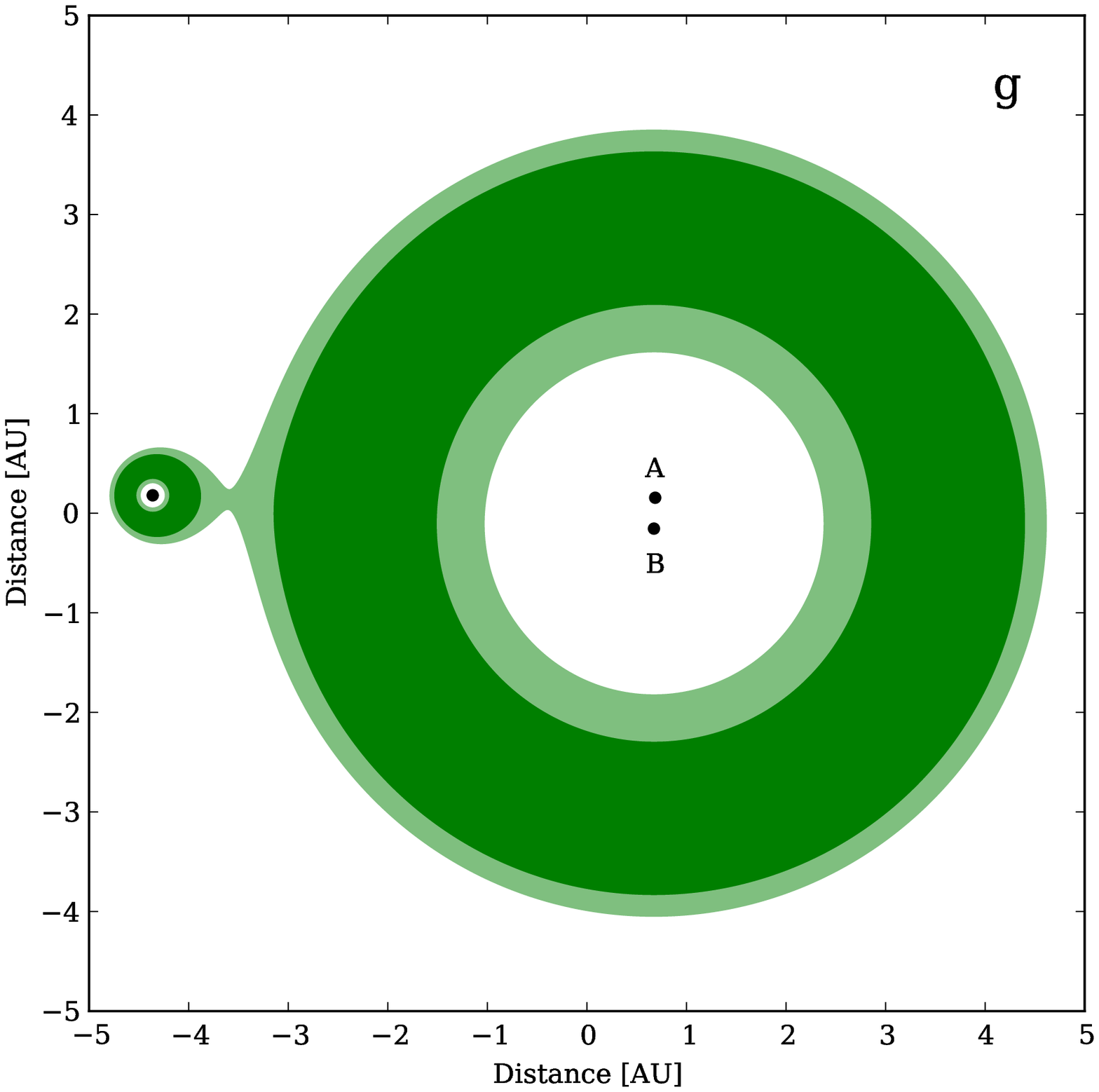}
\includegraphics[width=0.33\columnwidth]{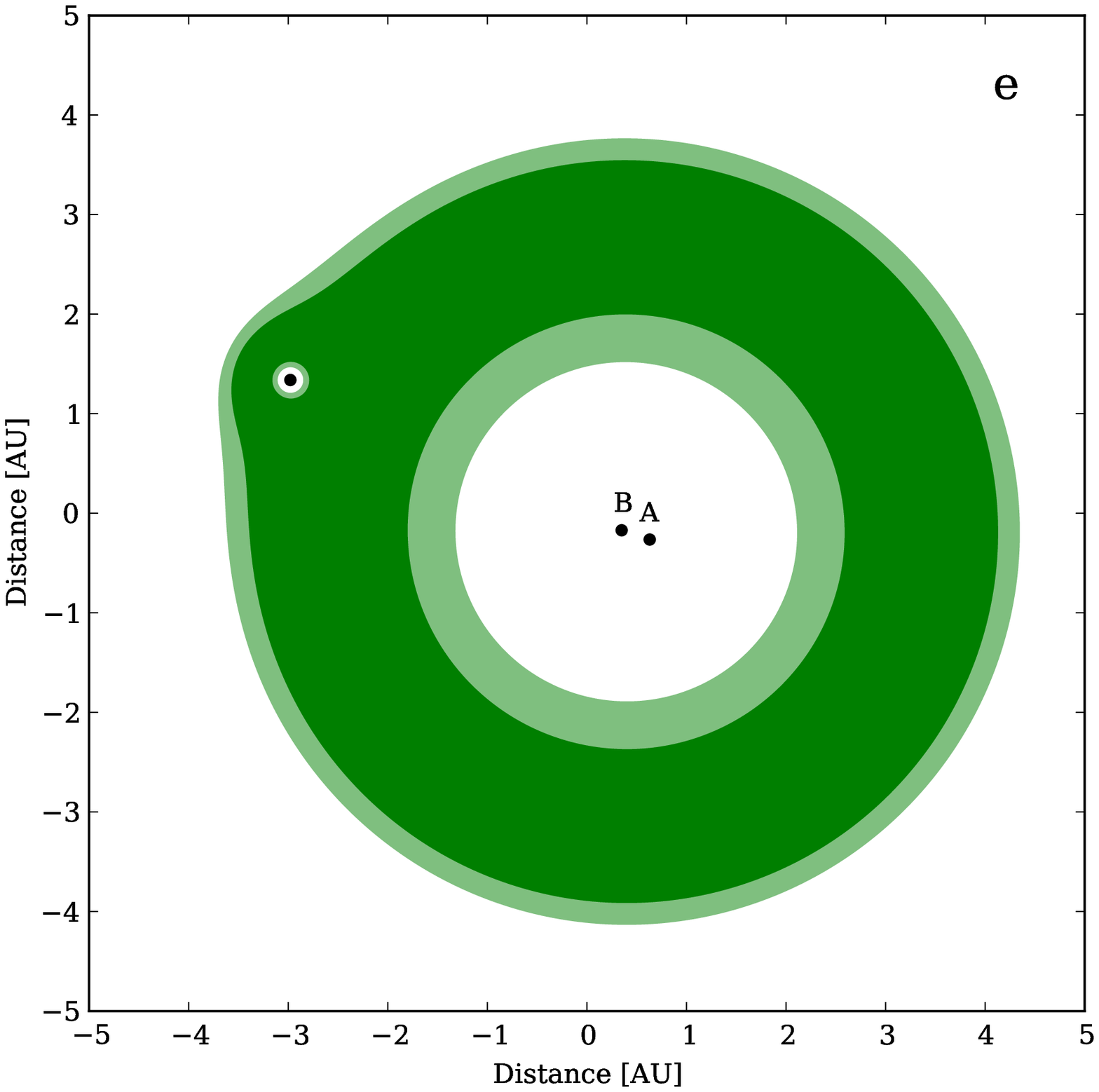}
\includegraphics[width=0.33\columnwidth]{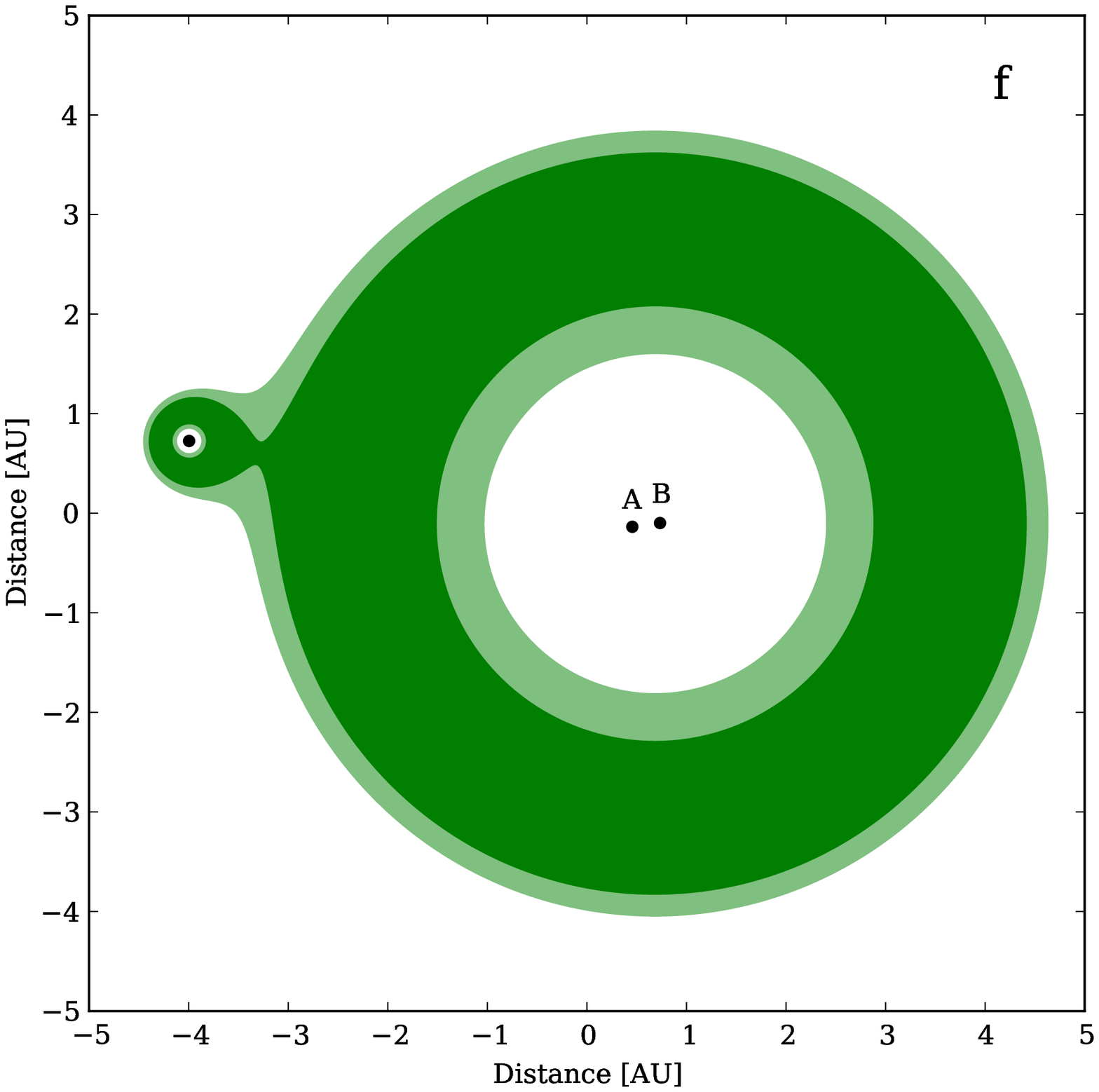}
\vskip -0.15in
\caption{Graphs of the HZ of the KID 5653126 system when the star C has a semimajor axis of 2.42 AU and an eccentricity of 0.7.
From top-right panel and in a counter-clockwise rotation (from a to h), the panels correspond to the star C being at
${0^\circ}, {26^\circ}, {28^\circ}, {41^\circ}, {52^\circ}, {73^\circ}, {117^\circ}, {150^\circ}$ and ${154^\circ}$, respetively.
A movie of the HZ of this system can be found at http://astro.twam.info/hz-multi.}
\label{HZ-5653126C-elliptical}
\end{figure}

In the second case, we considered the star C to have a semimajor axis equal to 2.42 AU and changed the value of its orbital eccentricity
until this star left the HZ of the binary $({e_{\rm C}}=0.7)$. Figure \ref{HZ-5653126C-elliptical}  
shows this for half of the orbital period of star C. From top-right
and in a counter-clockwise rotation, the panels show this star at 
${0^\circ}, {26^\circ}, {28^\circ}, {41^\circ}, {52^\circ}, {73^\circ}, {117^\circ}, {150^\circ}$ and ${154^\circ}$. 
As shown here, the motion of star C disturbs the HZ of the AB binary
by either excluding regions of it that are around this star, or extending its inner and outer boundaries. The capability of the HZ of the
AB binary to accommodate stable planetary orbits is also affected by the motion of the star C. Except for when this star is either entirely
inside the inner boundary of the system's empirical HZ or is outside the HZ of the AB binary, the stability of the system's HZ is limited 
to only a small region around star C. When star C is separated from the AB binary, planetary stability is maintained in both HZs.
However, as this star enters the inner boundary of the empirical HZ of the system, the stability region around it becomes unstable due
to the perturbation of the AB binary. This all means that during the motion of the stars around their center of mass,
an Earth-like planet may not maintain a long-term stable orbit in the HZ of the system. For more details, we refer the reader to the 
website \url{http://astro.twam.info/hz-multi} where movies of the time-variations of the HZs shown in figures \ref{HZ-5653126C-circular}  
and \ref{HZ-5653126C-elliptical} can be found.

\section{Interactive Website}
\label{sec:website}

To streamline the calculations of the HZ in binary and multiple star systems, we have developed an interactive website where
by inputting the orbital and physical properties of the stars, the HZ of the system is calculated for the time at which the 
stars are in the configuration that corresponds to their prescribed orbital parameters. The website can be found at

\vskip 3pt
\begin{center}
\url{http://astro.twam.info/hz}
\end{center}

\begin{figure*}
\centering
\includegraphics[width=\textwidth]{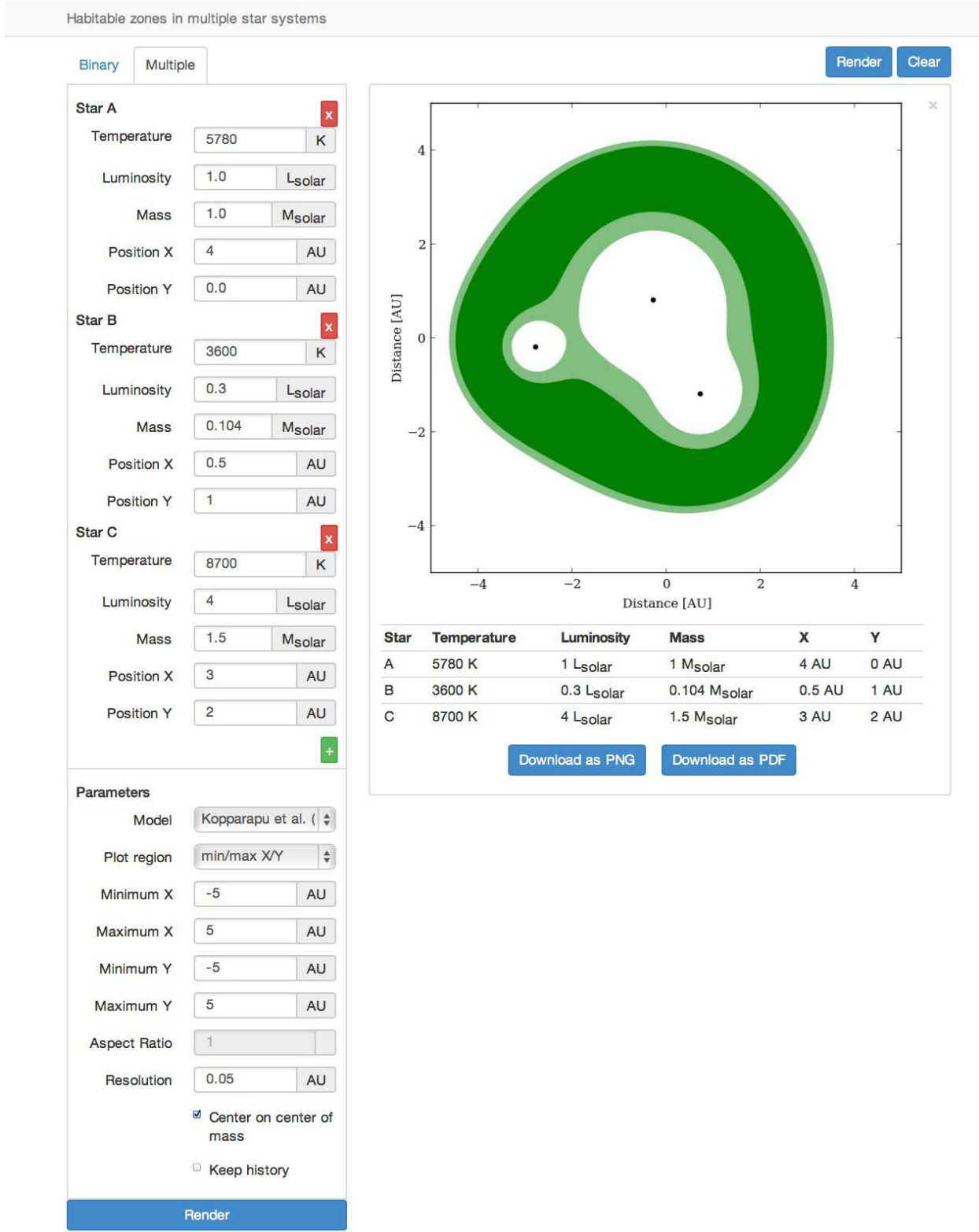}
\caption{A screenshot of the interactive website http://astro.twam.info/hz for calculating the HZ of binary and multiple star systems.}
\label{fig:website}
\end{figure*}

\noindent
Figure \ref{fig:website} shows a screenshot of the website.  
As shown on the top of the figure, the website gives the option for the calculations to be performed for a binary or 
a multiple star system. We chose to separate these two cases because in calculating the HZ of binary star systems, the website
has the capability of using equations (\ref{eq:ptype-stability}) and (\ref{eq:stype-stability}) to determine the boundaries of planetary stability, as well. However,
for a stellar system with more than 2 stars, the stability limits can be obtained only by direct integration of the
equation of motion of planetary bodies at different locations in the system. 
Also, for binary stars, the website offers the choice of setting the value of the mean-anomaly, which
can be used to obtain snapshots of the HZ during the rotation of the two stars around their common center of mass.

The next set of input parameters are the effective temperature, luminosity, and mass of each star as well as the star's
position. The temperature and luminosity of a star are used to calculate its 
spectral weight factor and the boundary of the system's HZ. The mass of each star is used to calculate boundaries of planetary stability
in binary star systems, and for the calculation of the center of mass in multiple star systems.
Since the value of the spectral weight factor and the locations of the boundaries of the HZ are model-dependent, the website offers
the option to choose between three models of the Sun's HZ: The model by \citet{Kasting93}\footnote{The calculations of the spectral weight factor
for this model is carried out using the formulas given by \citet{Underwood03}.}, \citet{Selsis07}, and 
\citet{Kopparapu13b}. The remaining input parameters are for determining the range of plotting the HZ and its resolution. 

After entering all parameters and choosing
the options, the HZ of the system can be displayed by clicking on the "Render" button. The website has been programmed to stop the calculations
and ask for reducing the resolution if the calculations take longer then 10 seconds. Once the HZ of the system is calculated, the 
displayed image can be saved as a vectorized PDF file, or a rasterized PNG image. Figure \ref{fig:website} 
shows as an example the input parameters and the calculated HZ of a GKM 
triple star system. In the next section, we show the application of the website by calculating the HZ of some interesting 
analytical solutions to the three-body problem. We would like to emphasize that the systems studied in the next section have been chosen 
for the mere purpose of demonstrating the use of our interactive website. We, therefore, will not be concerned about the possibility of the
formation of terrestrial-class planets in these systems, and the orbital stability of Earth-like bodies in their HZs (although when possible,
we will address the latter).

\subsection{Interesting Examples: Equilateral Three-Star System}

Among the currently known solutions of the general three-body problem, only a few are stable.
The most well-known stable solutions are the equilateral configurations were three stars (with different masses) rotate around their
common center of mass in an equilateral triangle at all times. The orbits of the stars can be circular or elliptical. In the latter
configuration, the distances between the stars vary with time whereas in the circular case, the distances stay constant. 

\begin{figure}
\centering
\includegraphics[width=0.48\columnwidth]{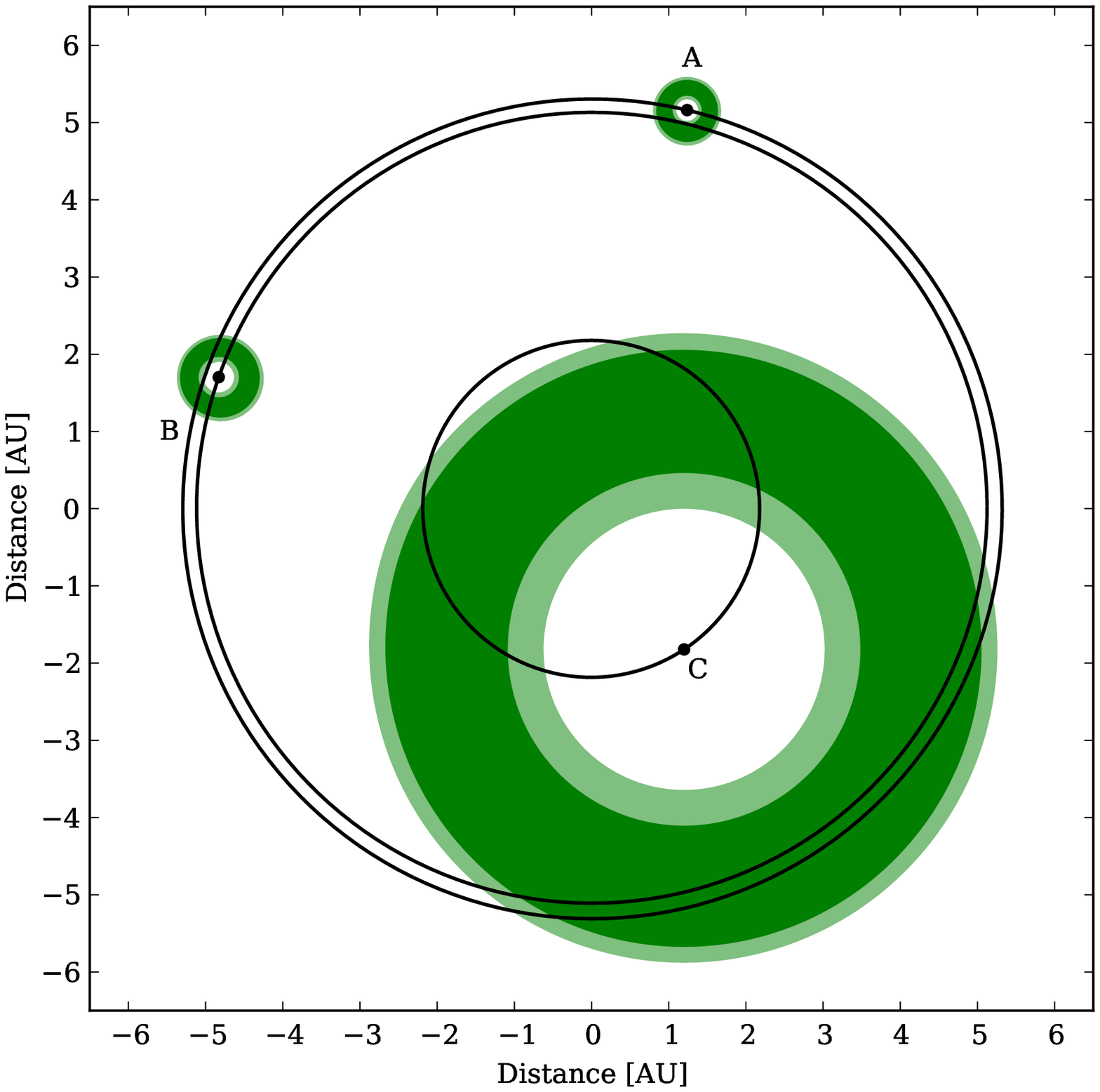}
\includegraphics[width=0.48\columnwidth]{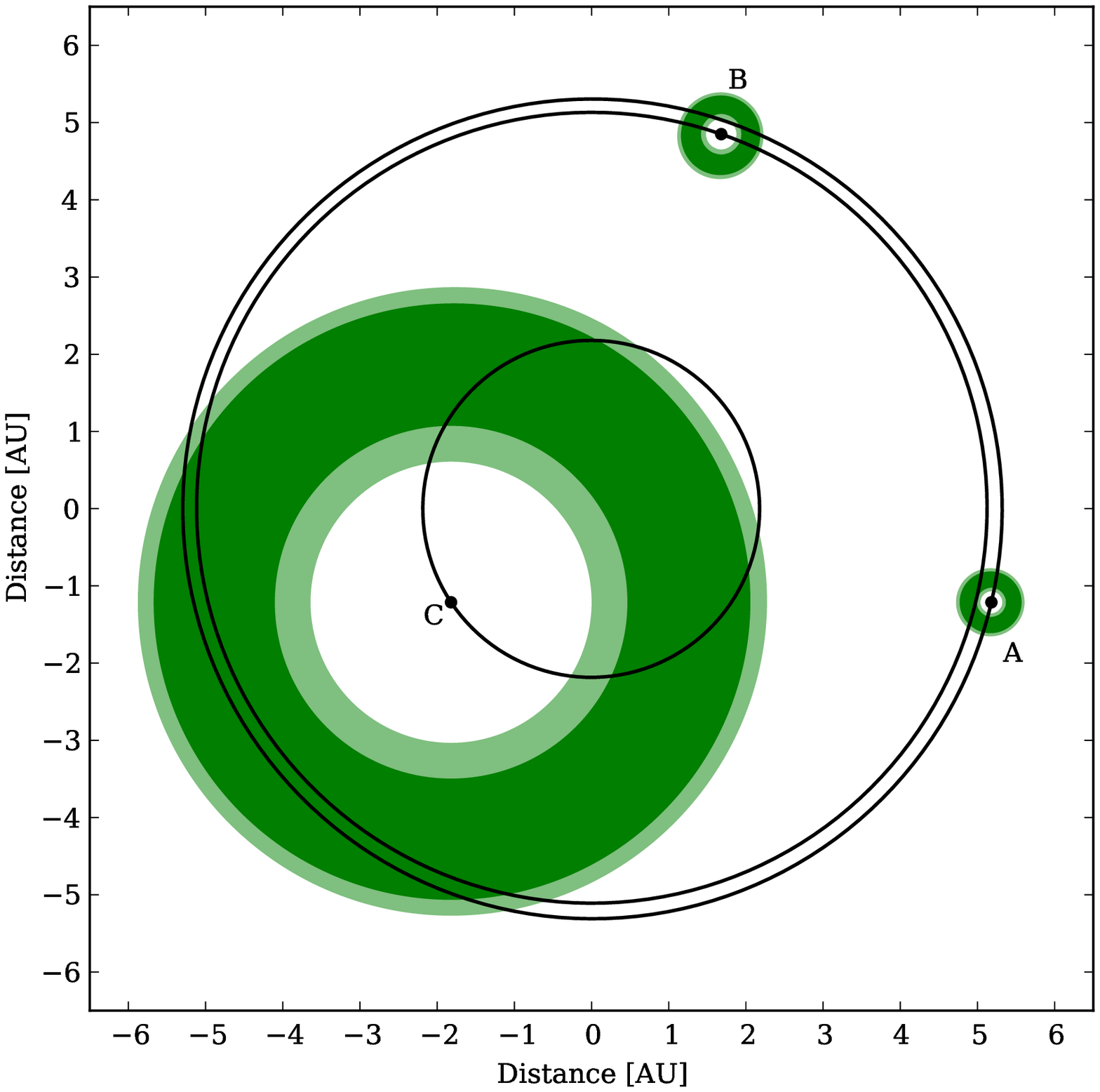}
\includegraphics[width=0.48\columnwidth]{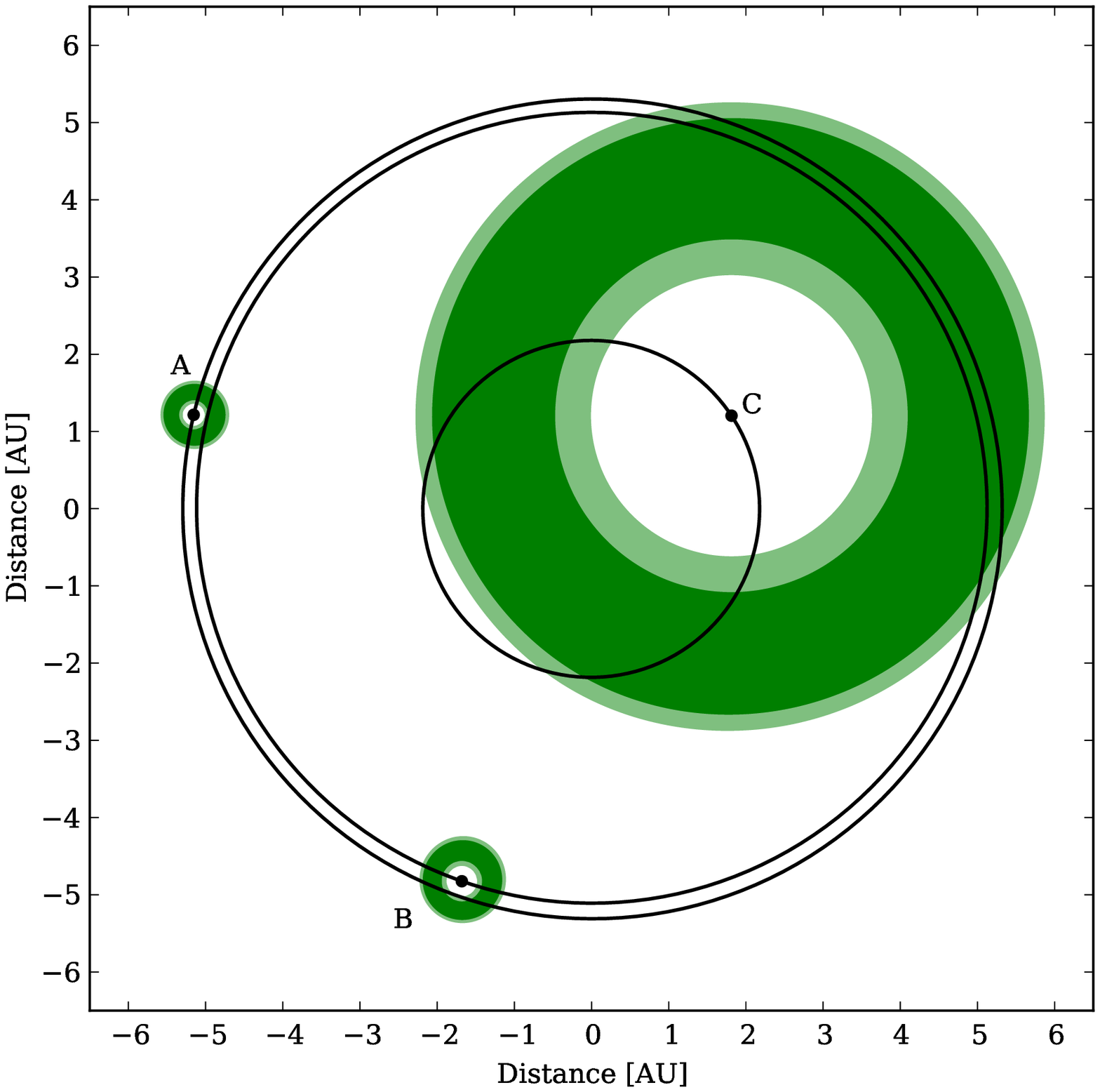}
\includegraphics[width=0.48\columnwidth]{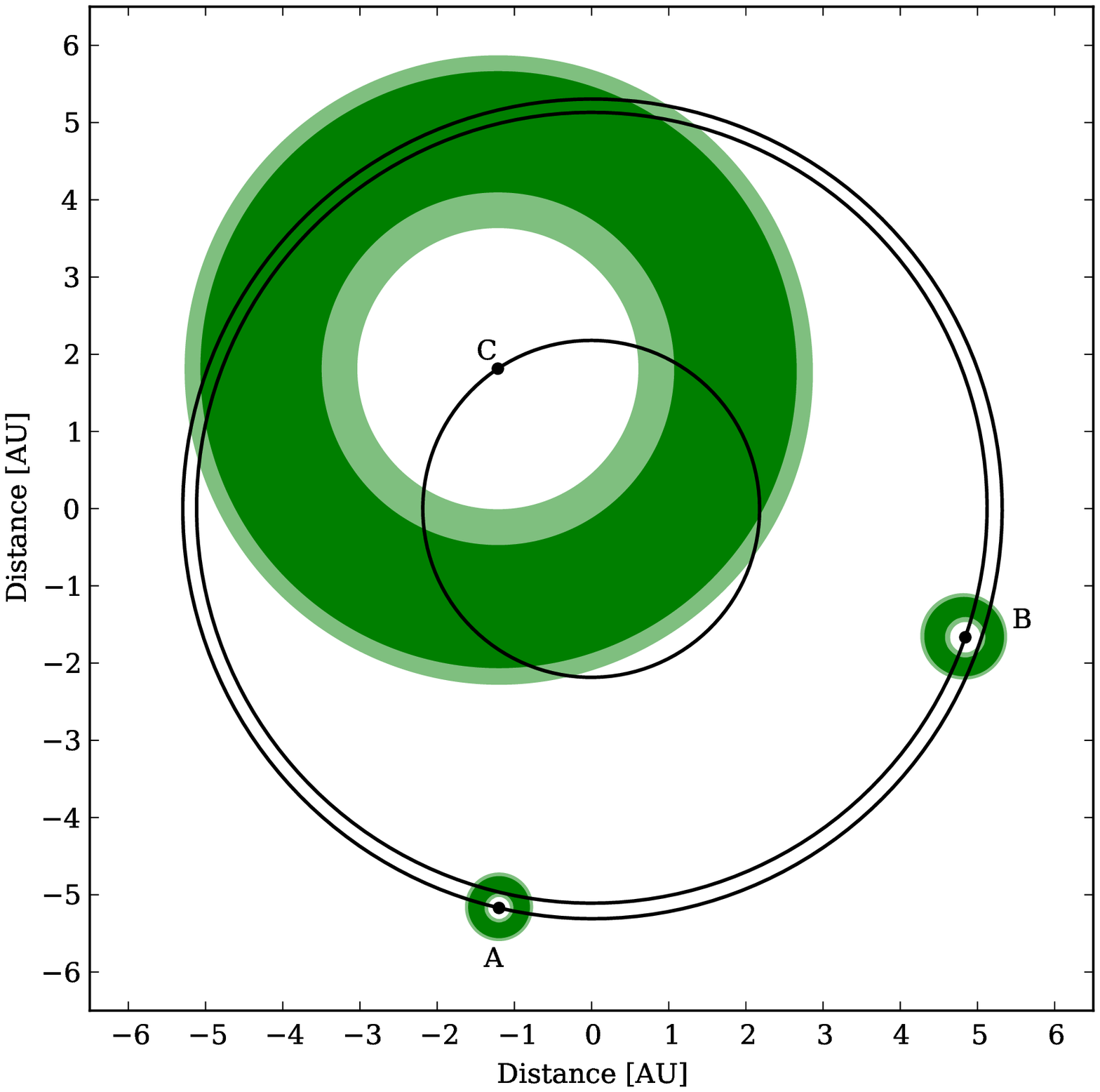}
\caption{Habitable zone of a triple star system with the stars on an equilateral triangle as an analytical solution to the 
general three-body problem. From top-right panel and counter-clockwise, the figures show the evolution of the HZ of the system for one
complete revolution around its center of mass. The stars masses, luminosities, and temperatures as well as their initial positions and velocities
are given in Table \ref{table4}. The orbits of the stars around the center of mass ($0,0$) are shown in black.
A movie of the HZ of this system can be found at http://astro.twam.info/hz-multi.}
\label{fig:equilateral-circle}
\end{figure}

\begin{figure}
\vskip -.2in
\centering
\includegraphics[width=0.33\columnwidth]{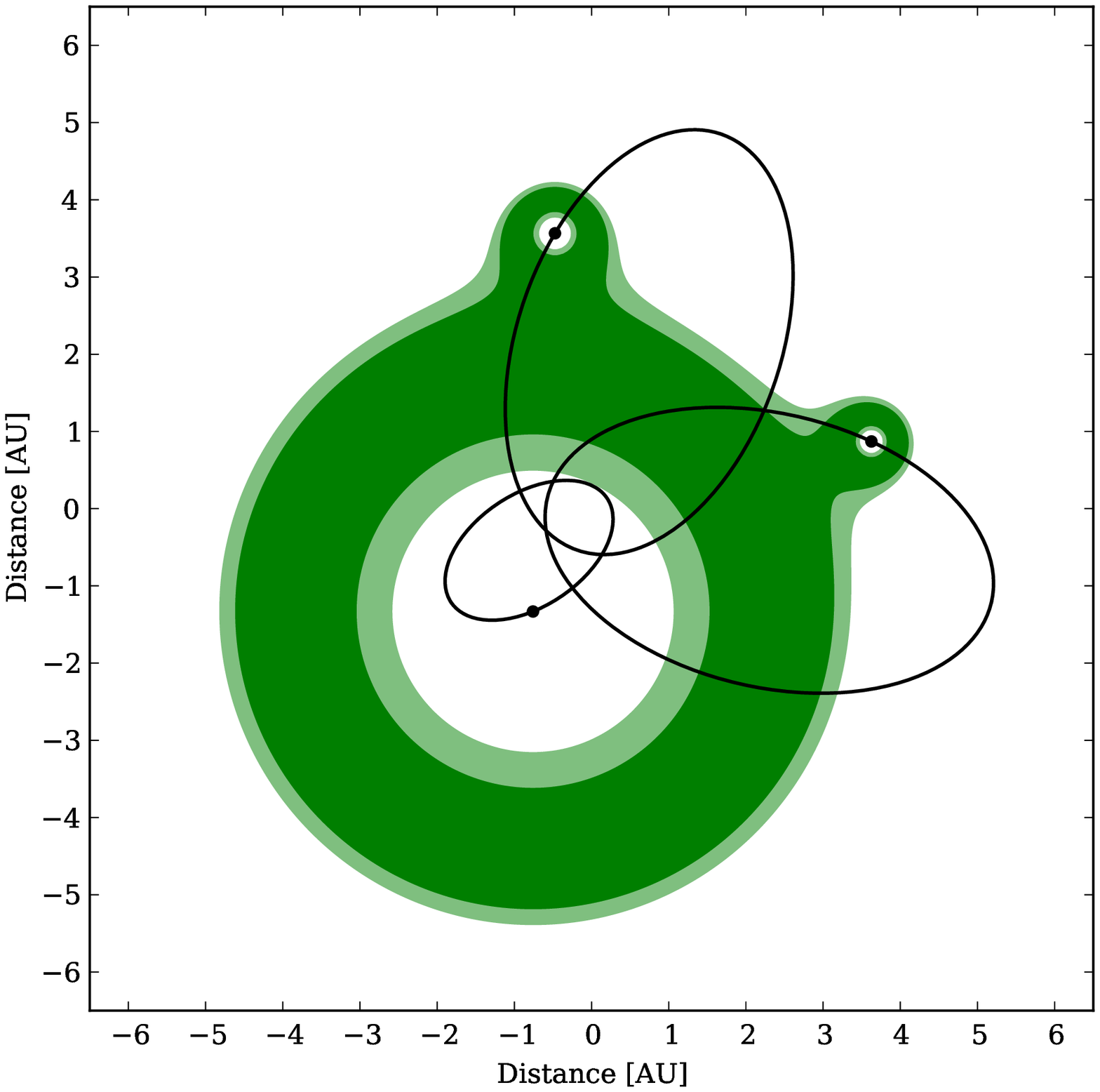}
\includegraphics[width=0.33\columnwidth]{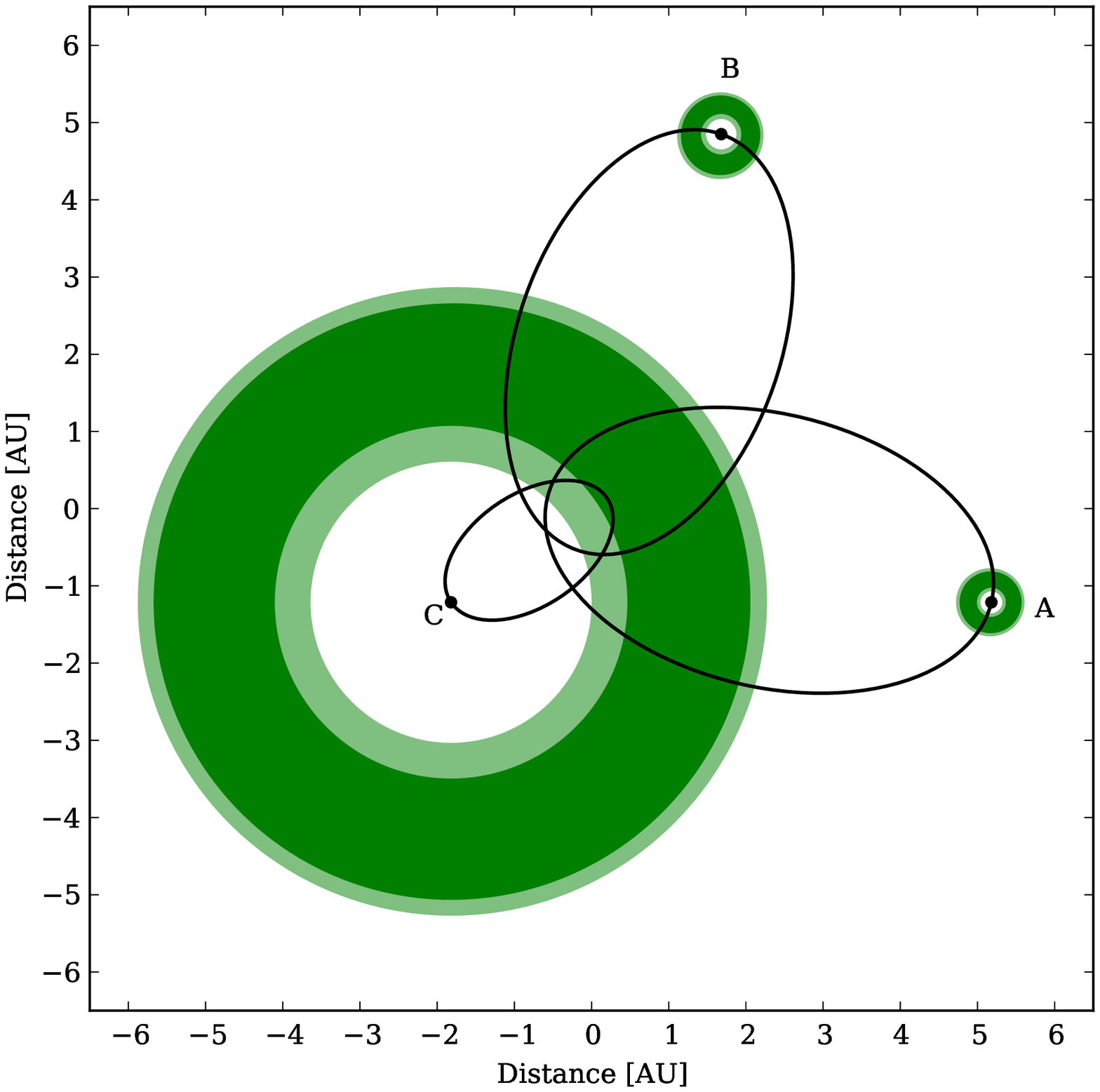}
\includegraphics[width=0.33\columnwidth]{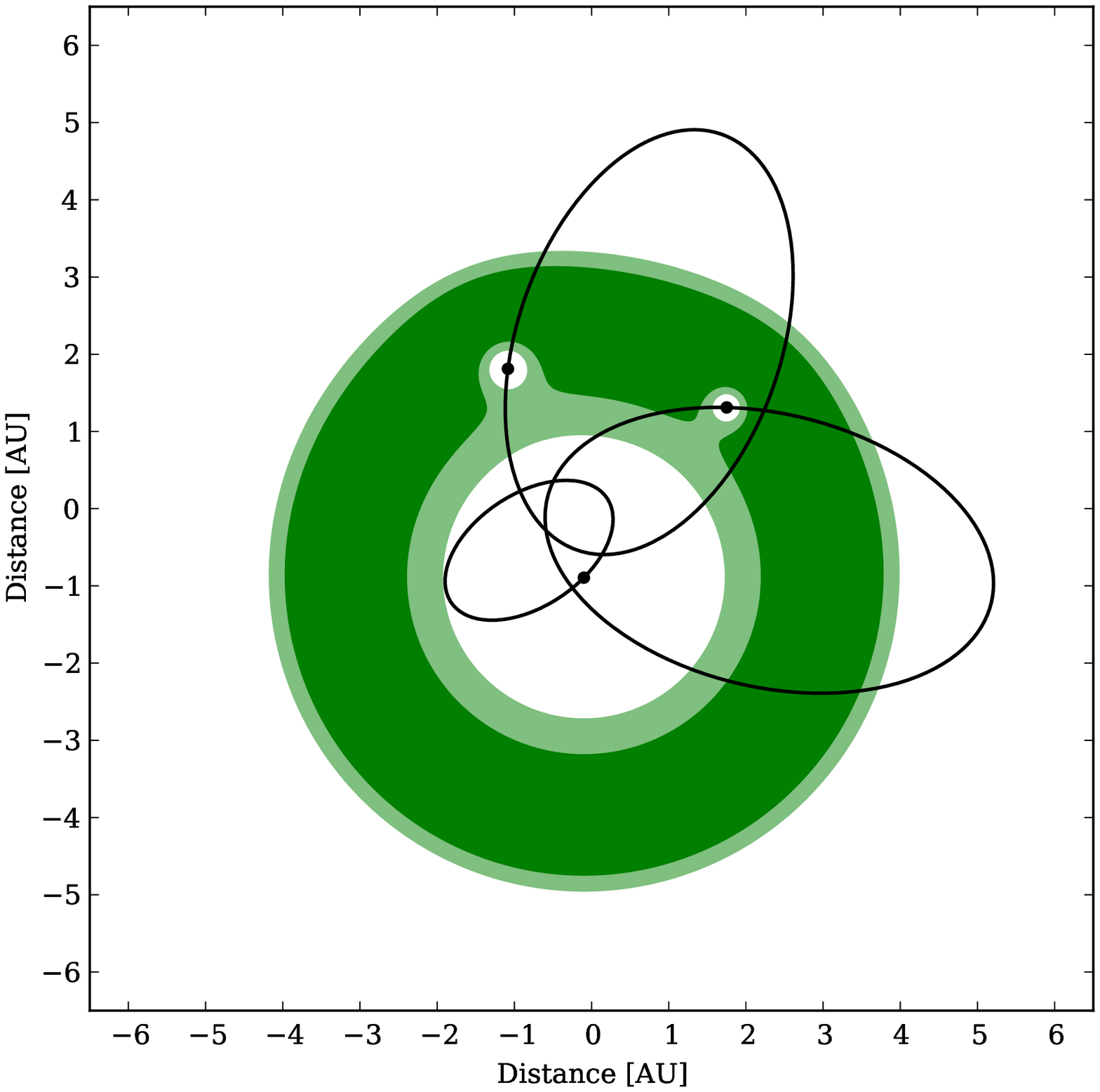}
\includegraphics[width=0.33\columnwidth]{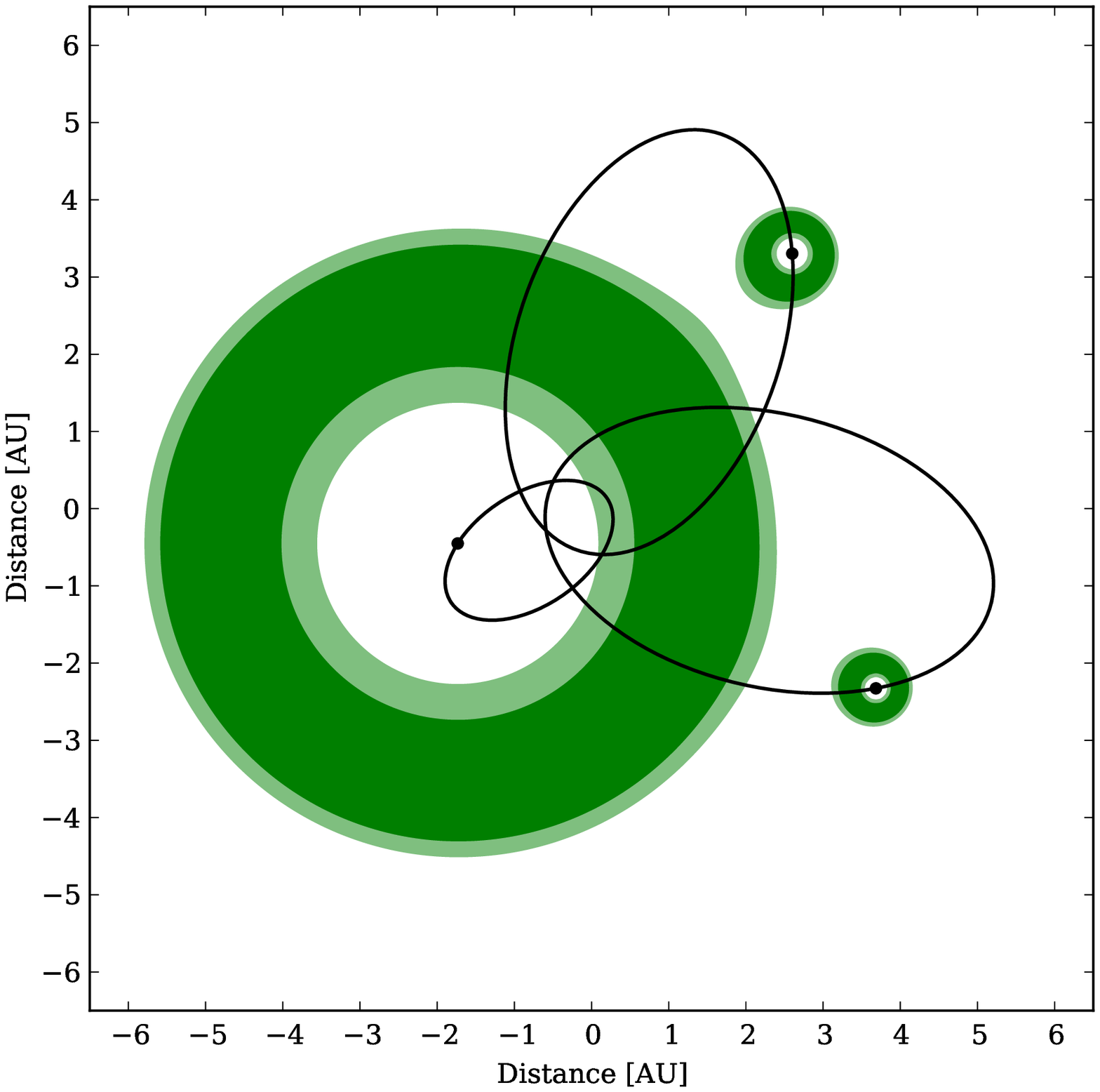}
\includegraphics[width=0.33\columnwidth]{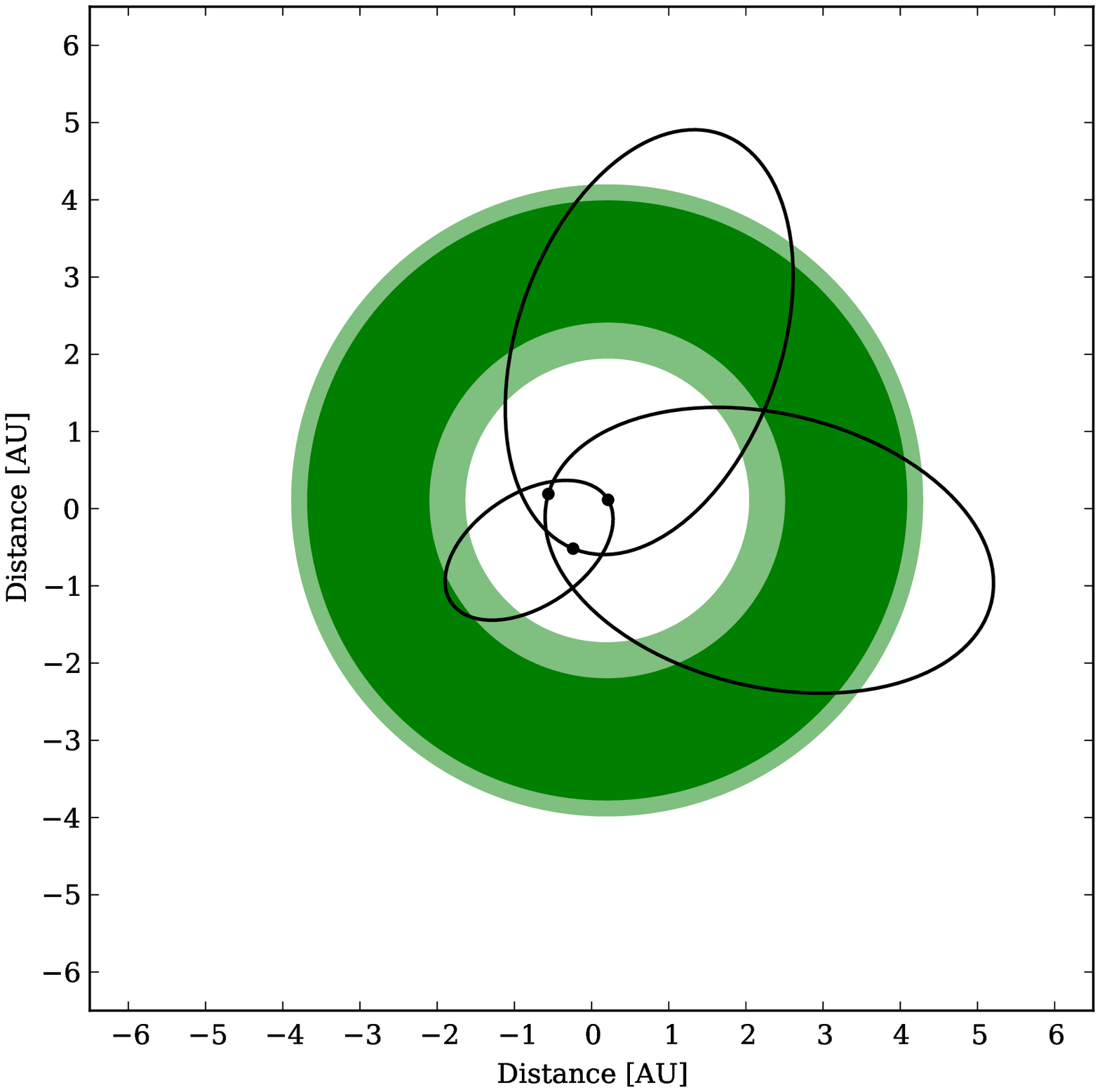}
\includegraphics[width=0.33\columnwidth]{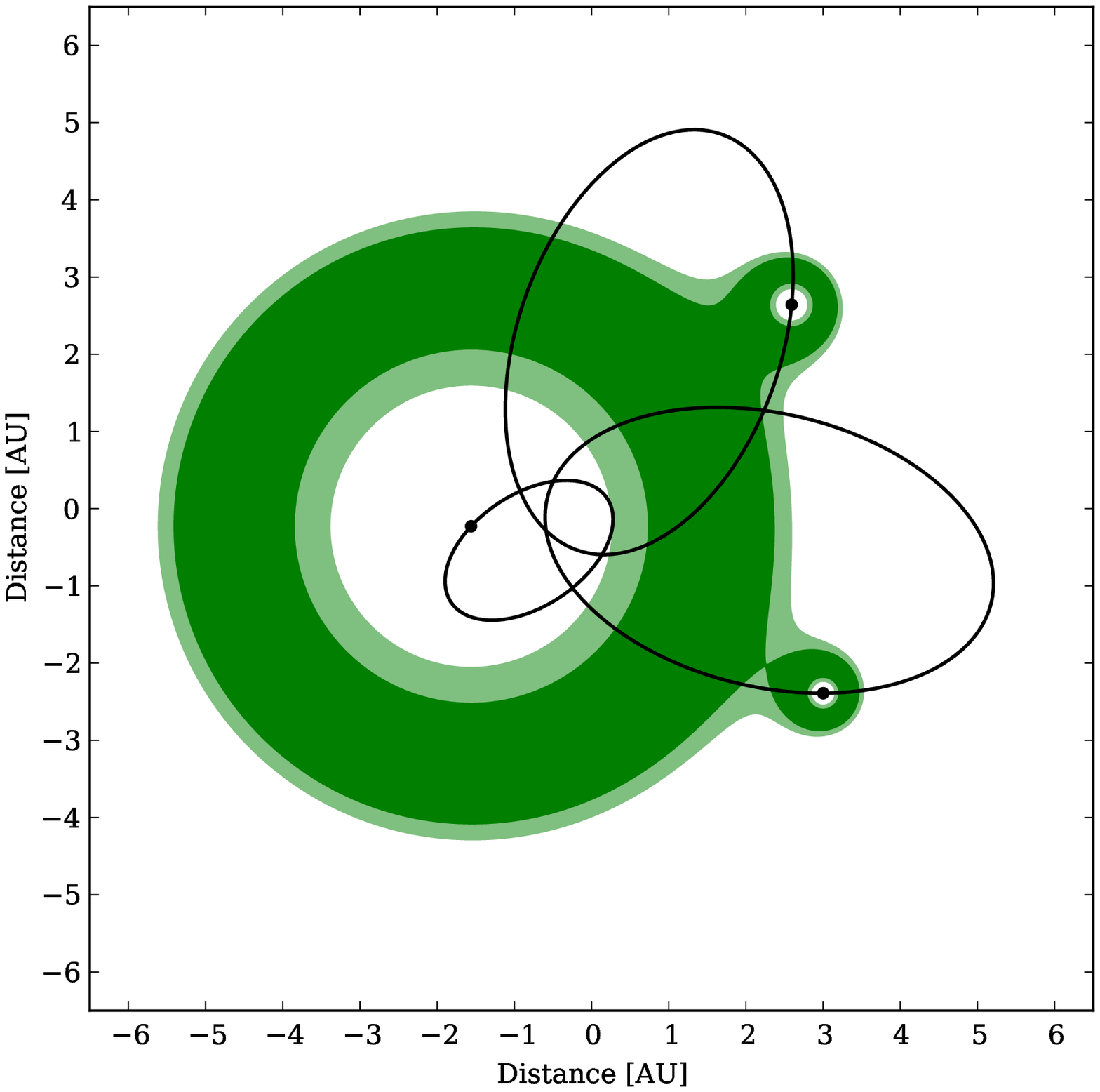}
\includegraphics[width=0.33\columnwidth]{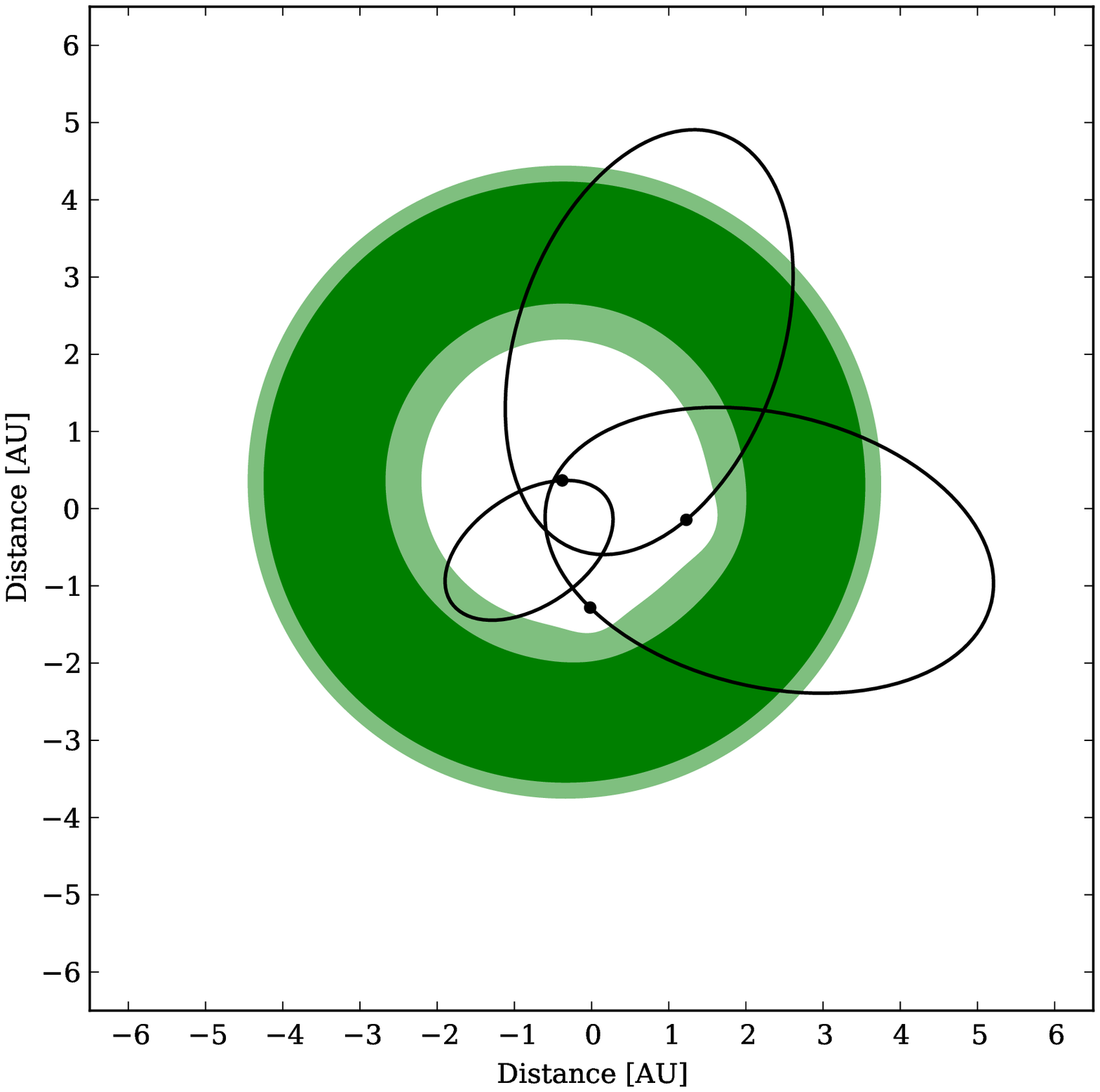}
\includegraphics[width=0.33\columnwidth]{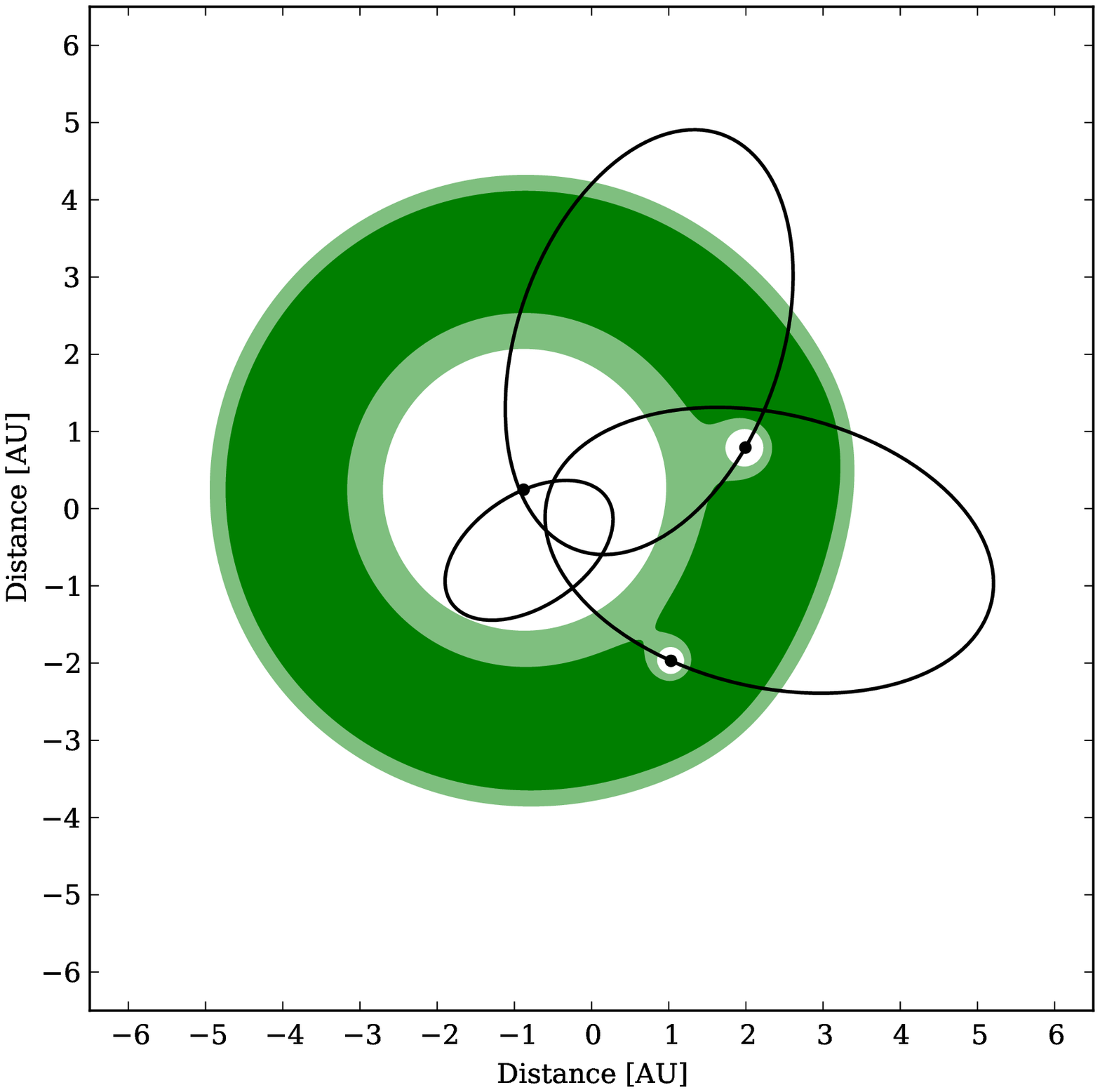}
\vskip -0.18in
\caption{HZ of the same triple star systems as in Figure \ref{fig:equilateral-circle} with stars being in elliptical orbits. 
From top-right panel and counter-clockwise, the figures show the evolution of the HZ of the system for one
complete revolution around its center of mass. The stars masses, luminosities, and temperatures as well as their initial 
positions and velocities are given in Table \ref{table4}.
The orbits of the stars around the center of mass ($0,0$) are shown in black.
A movie of the HZ of this system can be found at http://astro.twam.info/hz-multi.}
\label{fig:equilateral-elliptical}
\end{figure}

We calculated the HZ of the system using our interactive website for different values of the mass, luminosity and temperature of the
three stars, and for the model of the Sun's HZ by \citet{Kopparapu13a,Kopparapu13b}. 
Table \ref{table4} shows a sample of our systems for a circular (Figure \ref{fig:equilateral-circle}) and an elliptical
(Figure \ref{fig:equilateral-elliptical}) configuration. The initial positions $(X,Y)$ and velocities $({V_X},{V_Y})$ 
of the stars are also shown. Figure \ref{fig:equilateral-circle} shows four snapshots of the HZ of the circular case
along with the orbits of the stars (black, solid circles) 
for one complete revolution of the system (from top-right panel and in a counter-clockwise rotation). Since in this configuration, the
distances between the stars do not change, one would expect that for the types of the stars considered here, the mutual stellar 
interactions will be so small that planetary orbits maintain stability in the stars' individual HZs. Figure \ref{fig:equilateral-elliptical}, 
shows similar stars in an elliptical configurations. 
A shown here, although at times during their orbital motions, the stars are so far away from one another that they maintain their
individual HZs, their subsequent close approaches causes the HZ of the system to change and the interactions among them may become so strong
that the orbits of Earth-mass objects in the HZ will become unstable. 
Movies of the time-evolution of the HZ of the system can be found at \url{http://astro.twam.info/hz-multi}.

\subsection{Interesting Examples: Three Stars in a Figure-Eight Orbit}

Another interesting solution to the general three-body problem is when three equal-mass bodies revolve around their center of mass
in a figure-eight orbit \citep{Moore93,Chenciner00}. Although such orbital configuration is unlikely to appear in nature, it would
be interesting to calculate its HZ and determine how it evolves as the stars move in their orbits. 

\begin{figure}
\centering
\includegraphics[width=\columnwidth]{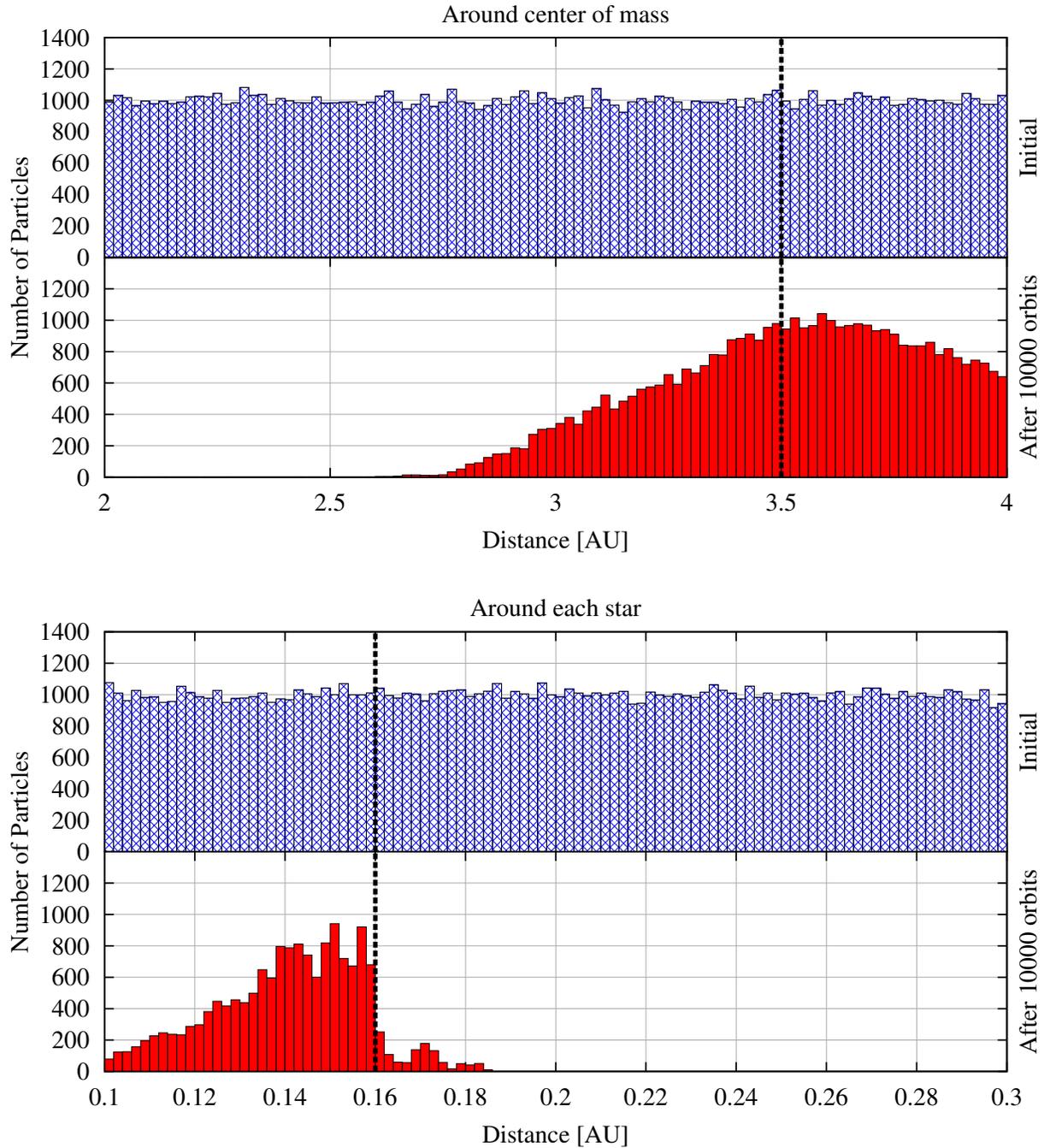}
\caption{Results of the integrations of non-interacting Earth-mass planets around the center of mass of a triple stars system
with Sun-like stars in a figure-eight orbit. The blue histograms show the initial distribution of the planets and the red histograms 
correspond to their distribution after integrating their orbits for 10000 orbital periods of one star around the center of 
mass of the system (the orbital periods of all stars around the center of mass are identical). 
As shown here, Earth-mass planets initially at distances larger than 3.5 AU (top panel) and 
interior to 0.160 AU (bottom panel) around each star, maintained their orbits for the duration of the integration.}
\label{fig:eight-stability}
\end{figure}

\begin{figure}
\centering
\includegraphics[width=0.48\columnwidth]{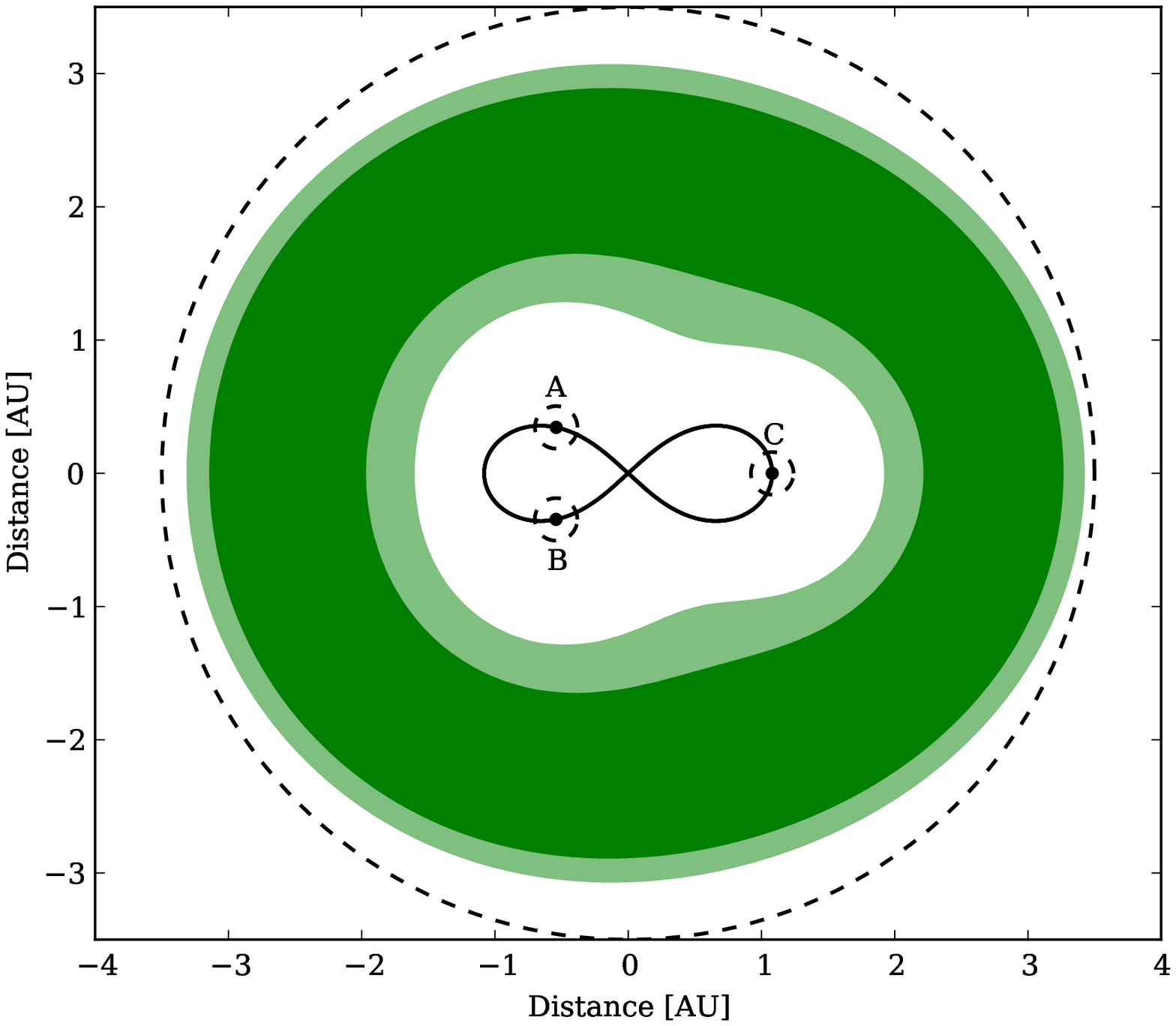}
\includegraphics[width=0.48\columnwidth]{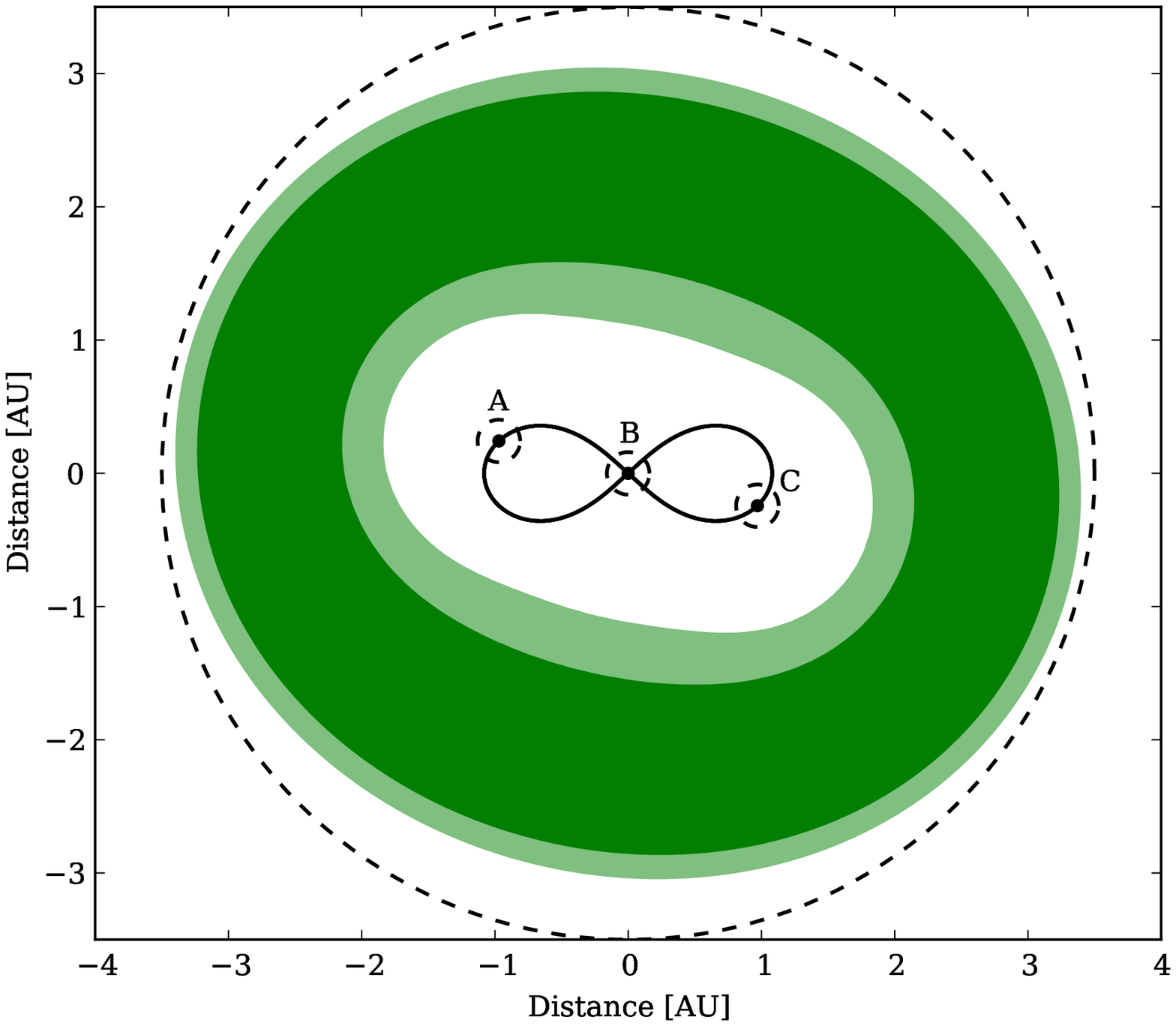}
\vskip 20pt
\includegraphics[width=0.48\columnwidth]{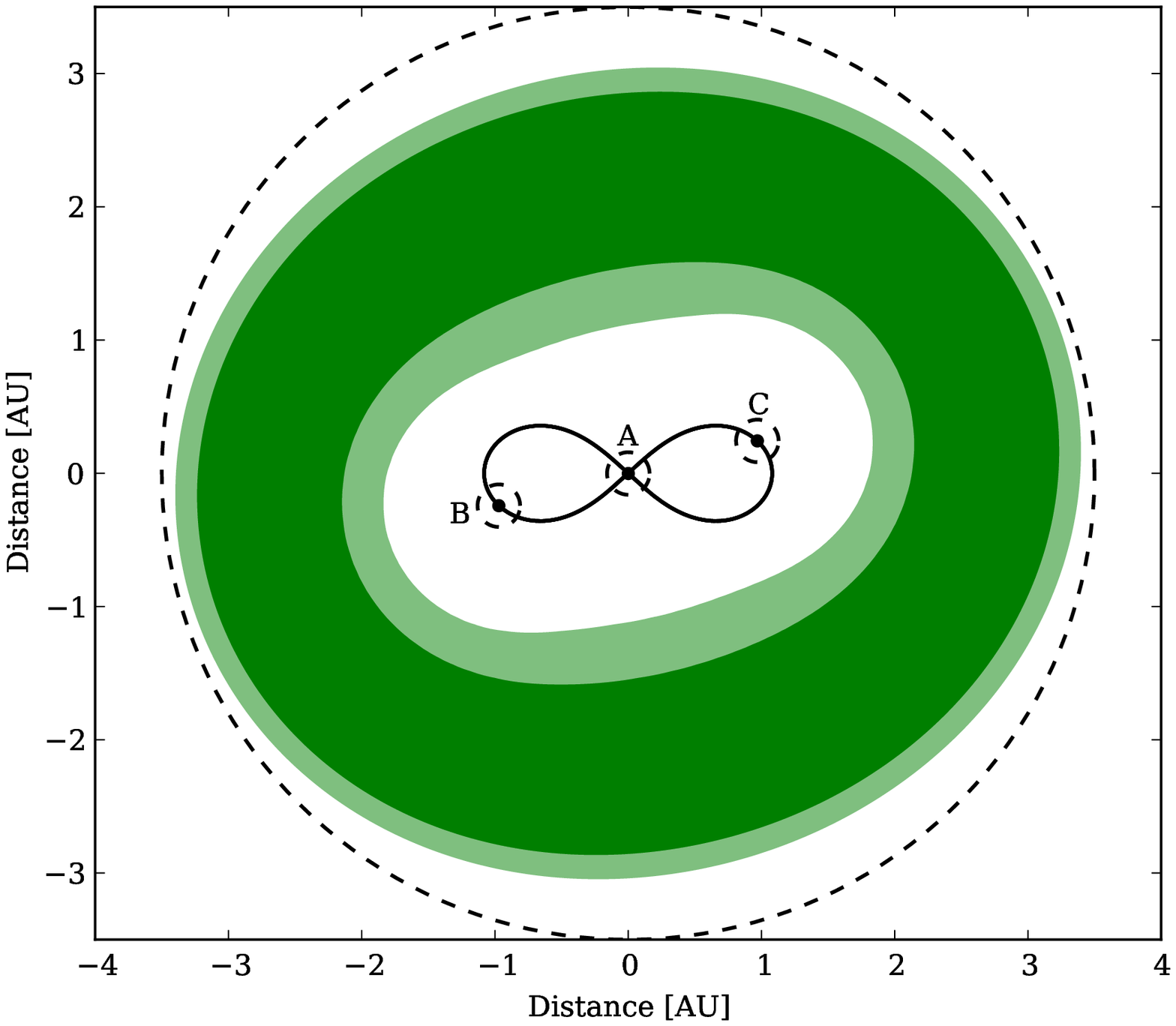}
\includegraphics[width=0.48\columnwidth]{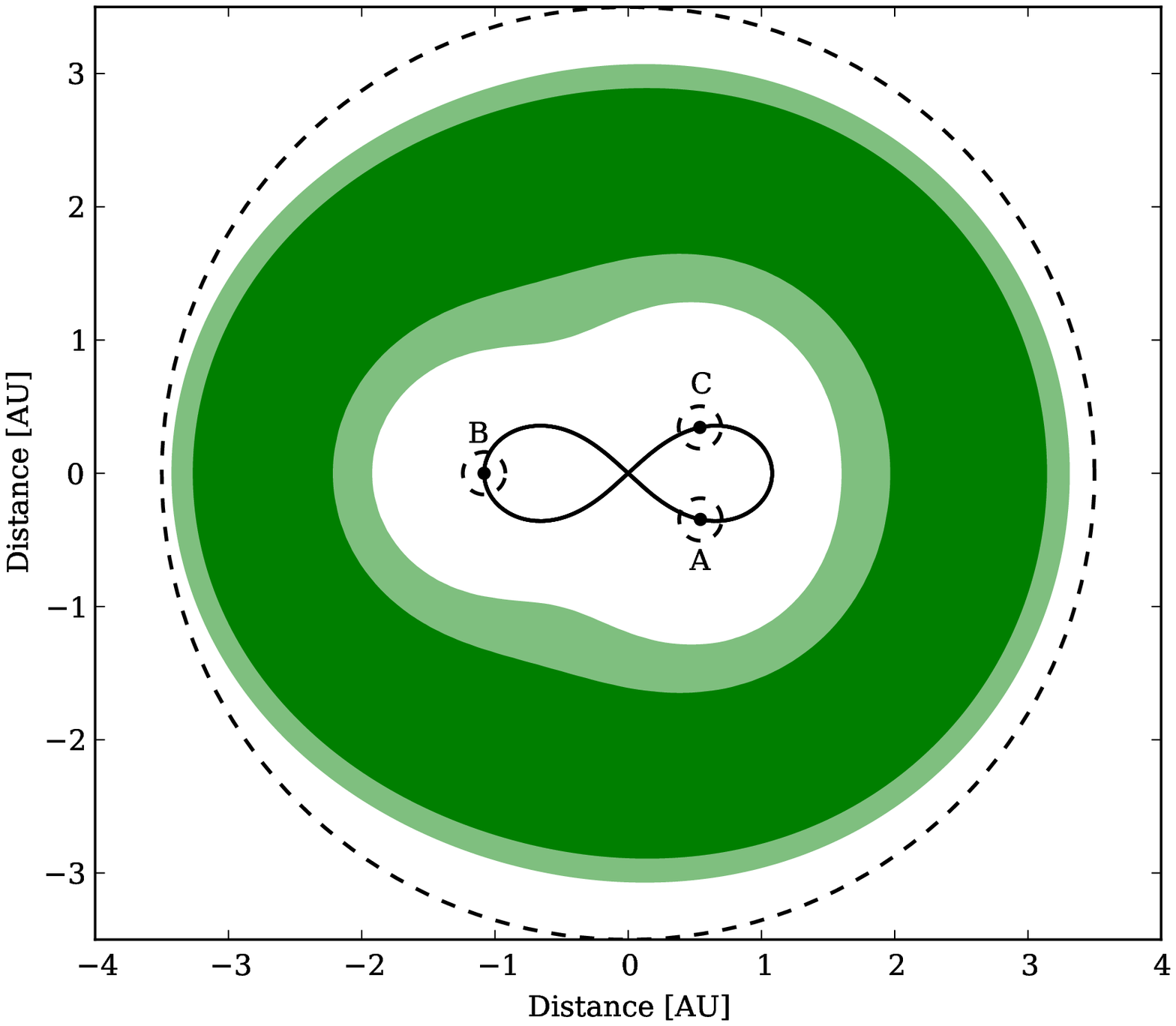}
\vskip 10pt
\caption{The HZ of a triple star system with Sun-like stars in a figure-eight orbit. From top-right panel, and in a counter-clockwise
rotation, the figure show the evolution of the HZ of the system for one complete revolution around its center of mass. The initial
orbital elements and velocities of the stars are given in Table \ref{table5}. 
The dashed circles correspond to the boundaries of planetary orbit stability. As shown here, the HZ encompasses the entire system
and is dynamically unstable. A movie of the HZ of this system can be found at http://astro.twam.info/hz-multi.}
\label{fig:eight-Sunlike}
\end{figure}

Using our interactive website, we calculated the HZ of such systems for different values of the mass and temperature of each star.
Results of stability analysis indicated that in general in a triple star system in a figure-eight orbit, the orbit of an Earth-mass planet
will be stable at large distances around the entire system and in a small region around each star where the perturbations of other stars do 
not affect its motion. Figure \ref{fig:eight-stability} shows the results of these integrations for a system with three
Sun-like stars ($T=5780$ K). Integrations were carried out for 10000 orbital period of one star around the center of mass of the system.
The values of the initial positions $(X,Y)$ and velocities $({V_X},{V_Y})$ 
of these stars are given in Table \ref{table5}.
As indicated by the vertical dashed line in the figure, the regions of planetary stability are at distances larger than 3.5 (AU) around 
the entire system, and smaller than 0.160 (AU), around each star. Figure \ref{fig:eight-Sunlike} shows the HZ of this system for one 
orbital period (starting from upper-right panel and counter-clockwise). The dashed
circles in this figure correspond to the above-mentioned stability limits. Given the luminosities of the stars, the HZ of the system
encompasses the entire triple stars system and its outer boundary is inside the stability limit. The latter implies that an Earth-like planet will
not be able to maintain a stable orbit in this HZ.

\begin{figure}
\centering
\includegraphics[width=0.48\columnwidth]{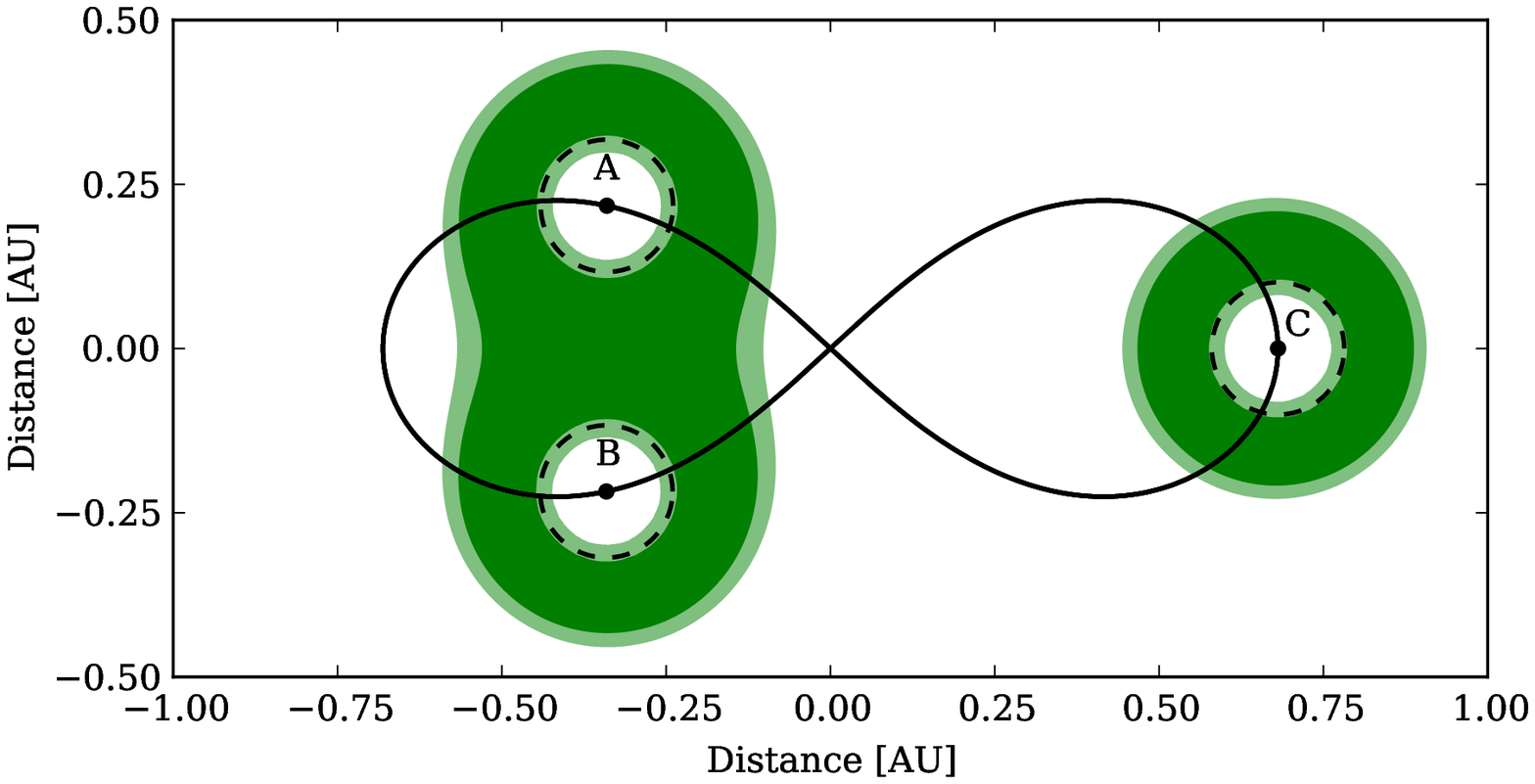}
\includegraphics[width=0.48\columnwidth]{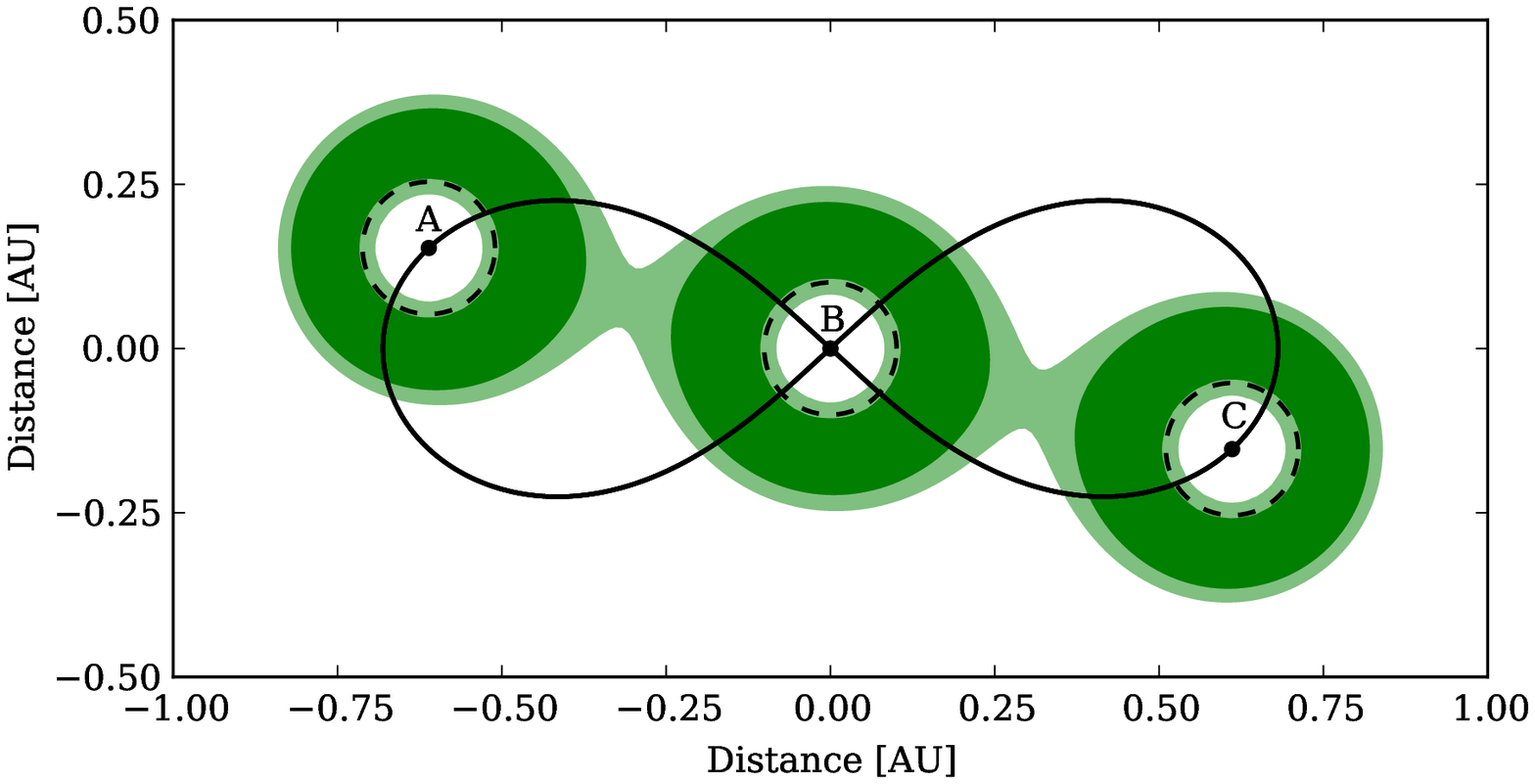}
\includegraphics[width=0.48\columnwidth]{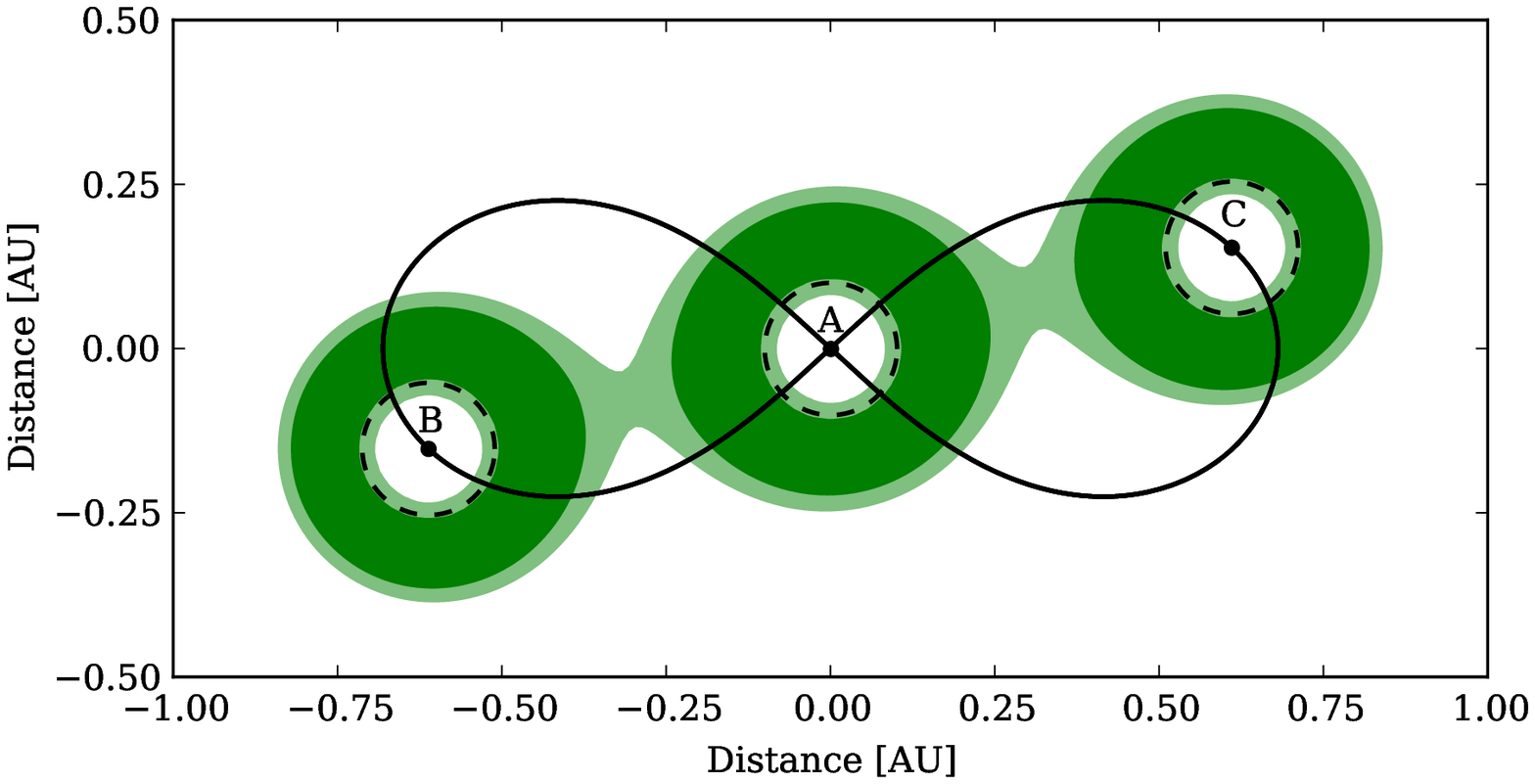}
\includegraphics[width=0.48\columnwidth]{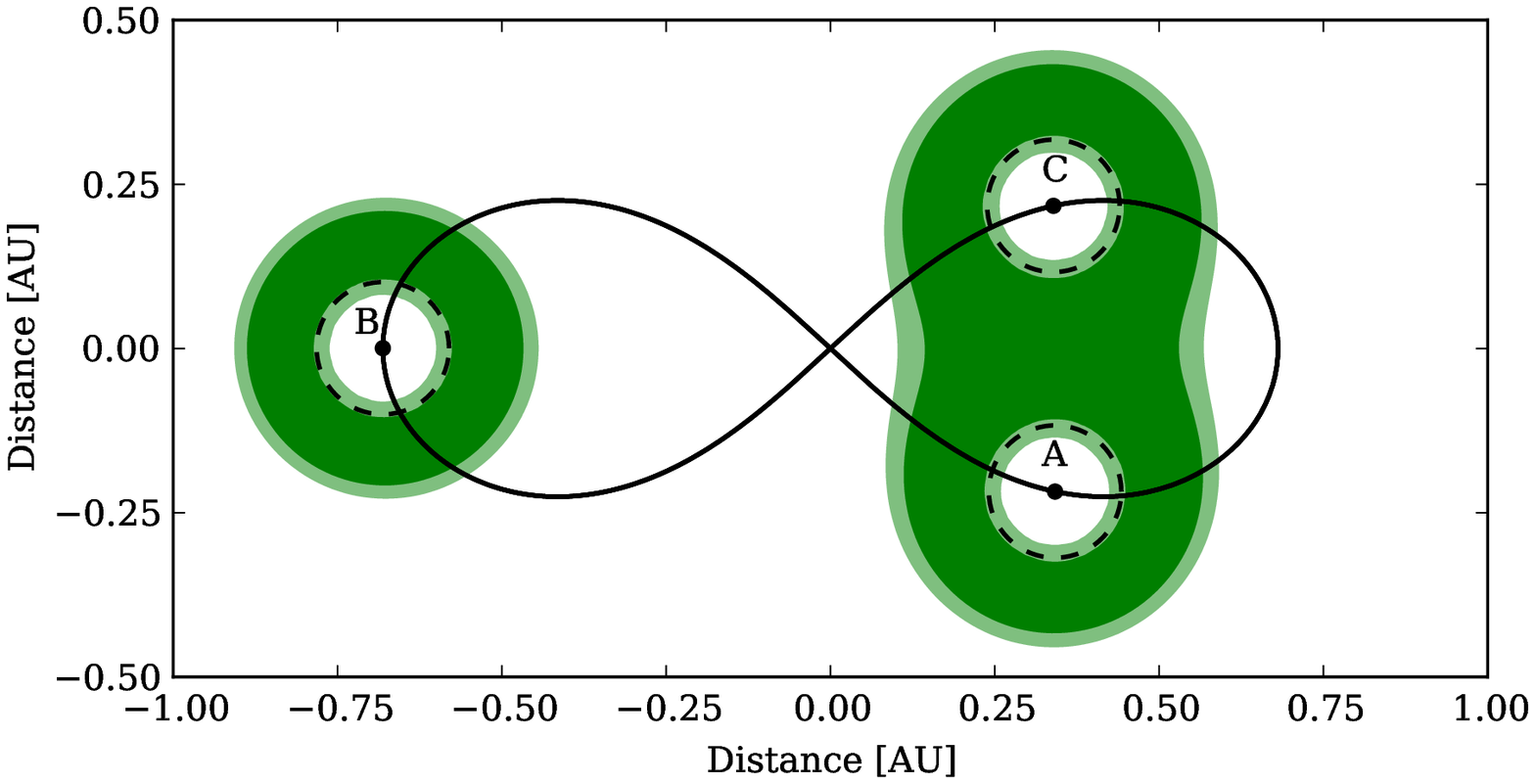}
\vskip 10pt
\caption{The HZ of a system of three 0.25 solar-mass M stars in a figure-eight orbit. From top-right panel and in a counter-clockwise
rotation, the figure show the evolution of the HZ of the system for one complete revolution around its center of mass. The initial
orbital elements and velocities of the stars are given in Table \ref{table5}. The dashed circles show the boundaries of the regions 
of planetary stability around each star ($\sim$0.101 AU). As shown here, Earth-like planets can have stable orbits in a small 
region in the empirical HZ around each star. A movie of the HZ of this system can be found at http://astro.twam.info/hz-multi.}
\label{fig:eight-Mstars}
\end{figure}

Although the HZ of the three-star system in Figure \ref{fig:eight-Sunlike} is dynamically unstable, the fact that small regions of stability
exist in close distances around each star motivated us to calculate the HZ of the system for cool and low-mass stars.
Figure \ref{fig:eight-Mstars} shows the results for three 0.25 solar-masses M dwarfs with luminosities of $ 0.0095\,L_{\sun} $ and 
temperatures of $ T = 2800\,\mathrm{K}$. From top-right panel and counter-clockwise, the figure corresponds to one complete revolution
of the system. The integrations of the motion of Earth-mass objects indicated that the planetary stability is limited to distances larger 
than 2.20 AU around the entire system, and smaller than 0.101 AU around each star (the dashed circles in Figure \ref{fig:eight-Mstars}).
As shown here, while during the motion of the stars, the HZ of the system extends to larger distances, 
only a small region of the empirical HZ around each star maintains planetary stability.
Movies of the time-evolution of the HZ of the system can be found at \url{http://astro.twam.info/hz-multi}.

\section{Summary}
\label{sec:summary}

We presented a general methodology for calculating the HZ of multiple star systems. We used the concept of spectral weight factor as
introduced by HK13 and KH13, and calculated the total flux received at the top of the atmosphere
of an Earth-like planet. By comparing this flux with that received at the top of Earth's atmosphere from the Sun, we determined
regions corresponding to narrow and empirical HZ in and around multiple stars systems. To demonstrate the applicability of our
methodology, we calculated the HZ of two triple star systems and studied the effects of high and low luminosity stars on the HZ around the other
stellar components. To streamline the calculations of HZ in binary and multiple star systems,
we developed an interactive website where by inputting the physical and dynamical properties of the stars, the HZ of the system
is obtained. 

We would like to note that the HZ, as calculated in this study is in fact an instantaneous HZ. In our calculations, we
did not consider the effect of the eccentricity of the stellar and planetary orbits. When the orbit of the planet and/or those of the stars
are eccentric, the close approaches of the stars to the planet will affect the total flux received at the top of the planet's
atmosphere, and therefore changes the locations where the planet can be habitable (i.e., the boundaries of the system's HZ). In an actual system with
an Earth-like planet, these effects have to be taken into consideration, and the region of the habitability of the planet
has to be determined by averaging the flux received by the planet over the longest orbital period of the system. Such calculations have to
be applied to actual systems in a case by case basis.

\begin{acknowledgements}
We would like to thank the anonymous referee for his/her constructive comments which have improved our manuscript.
Tobias M\"uller received financial support from the Carl-Zeiss-Stiftung. N.H. acknowledges 
support from the NASA ADAP grant NNX13AF20G, NASA Astrobiology Institute under Cooperative Agreement NNA09DA77 at the Institute for 
Astronomy, University of Hawaii, HST grant HST-GO-12548.06-A, and Alexander von Humboldt Foundation.
Support for program HST-GO-12548.06-A was provided by NASA through a grant from the Space Telescope Science 
Institute, which is operated by the Association of Universities for Research in Astronomy, Incorporated, 
under NASA contract NAS5-26555. N.H. is also thankful to the Computational Physics group at the Institute 
for Astronomy and Astrophysics, University of T\"ubingen for their kind hospitality during the course
of this project.
\end{acknowledgements}


\clearpage

\clearpage

\clearpage

\clearpage

\clearpage

\clearpage

\clearpage

\clearpage

\clearpage

\clearpage

\clearpage

\clearpage

\clearpage

\clearpage

\clearpage

\clearpage
\begin{deluxetable}{lcccc}
\tablecaption{Values of the coefficients of equation (\ref{eq:weightfactor}) from \citet{Kopparapu13b} \label{table1}}
\tablewidth{0pt}
\tablehead{\colhead{} & \multicolumn{2}{c}{Narrow HZ} & \multicolumn{2}{c}{Empirical HZ} \\
\hline
\colhead {} & \colhead{Runaway Greenhouse} & \colhead {Maximum Greenhouse} &  
\colhead{Recent Venus} & \colhead {Early Mars}
}
\startdata
$F_{\rm Sun}$ & 1.06  & 0.36  & 1.78 & 0.32 \\
$d_{\rm {Sun}}$ (AU) & 0.97  & 1.67  &  0.75  &   1.77   \\
$a$    & $1.2456 \times {10^{-4}}$   & $5.9578 \times {10^{-5}}$   & $1.4335 \times {10^{-4}}$   & $5.4471 \times {10^{-5}}$ \\
$b$    & $1.4612 \times {10^{-8}}$   & $1.6707 \times {10^{-9}}$   & $3.3954 \times {10^{-9}}$   & $1.5275 \times {10^{-9}}$ \\
$c$    & $-7.6345 \times {10^{-12}}$ & $-3.0058 \times {10^{-12}}$ & $-7.6364 \times {10^{-12}}$ & $-2.1709 \times {10^{-12}}$ \\
$d$    & $-1.7511 \times {10^{-15}}$ & $-5.1925 \times {10^{-16}}$ & $-1.1950 \times {10^{-15}}$ & $-3.8282 \times {10^{-16}}$ \\
\enddata
\end{deluxetable}

\begin{deluxetable}{lccccccccc}
\tablecaption{Values of the spectral weight factor \label{table2}}
\tablewidth{0pt}
\tablehead{\colhead{$T$ (K)} & \colhead{2800} & \colhead{3500} & \colhead{4500} & \colhead{5500} & \colhead{5636} & \colhead{5780} & 
\colhead{6407} &  \colhead{7400} & \colhead{8500}  }
\startdata
$W_{\rm in}$ (Narrow)     & 1.200  & 1.184 & 1.132 & 1.033 & 1.017 & 1.000 & 0.929 & 0.844 & 0.843  \\
$W_{\rm out}$ (Narrow)    & 1.529  & 1.417 & 1.237 & 1.048 & 1.024 & 1.000 & 0.906 & 0.809 & 0.807  \\
$W_{\rm in}$ (Empirical)  & 1.194  & 1.164 & 1.102 & 1.023 & 1.012 & 1.000 & 0.952 & 0.899 & 0.901  \\
$W_{\rm out}$ (Empirical) & 1.614  & 1.462 & 1.249 & 1.050 & 1.025 & 1.000 & 0.903 & 0.799 & 0.771  \\
\hline
\enddata
\end{deluxetable}

\clearpage
\begin{deluxetable}{lccc}
\tablecaption{The AB-C triplet in KID 5653126 system \label{table3}}
\tablewidth{0pt}
\tablehead{\colhead{Star} & \colhead{A} & \colhead{B} & \colhead{C}}
\startdata
Mass $(M_{\sun})$        & 1.043 & 1.528 & 0.4   \\
Luminosity $(L_{\sun})$  & 0.84  & 4.54  & 0.02  \\
Temperature (K)          & 5636  & 6407  & 3561  \\
\hline
\enddata
\end{deluxetable}

\begin{deluxetable}{lcccccccc}
\tablecaption{Values of the mass, luminosity, temperature, and the components of the position and velocity vectors of 
the equilateral three-star systems shown in Figures \ref{fig:equilateral-circle} and \ref{fig:equilateral-elliptical} \label{table4}}
\tablewidth{0pt}
\tablehead{\colhead{} & \multicolumn{3}{c}{Circular} & \colhead{}  & \multicolumn{3}{c}{Elliptical} \\
\hline
\colhead {} & \colhead {A} & \colhead {B} & \colhead {C} & \colhead {} & \colhead {A} & \colhead {B} & \colhead {C} 
}
 \startdata
$T\,({\rm K})$       & 3500   & 5500   & 7400   && 3500   & 5500   & 7400  \\
$L\,({L_\sun})$      &  0.028 &  0.063 &  6.6   &&  0.028 &  0.063 &  6.6  \\
$M\,({M_\sun})$      &  0.4   &  0.5   &  1.6   &&  0.4   &  0.5   &  1.6  \\
$a$ (AU)             &  5.32  &  5.13  &  2.19  &&  2.96  &  2.85  &  1.21 \\
$e$                  &  0     &  0     &  0     &&  0.8   &  0.8   &  0.8  \\
$X$ (AU)             &  3.64  & -0.14  & -3.64  &&  3.36  & -0.14  & -3.64 \\
$Y$ (AU)             & -3.88  &  2.18  & -3.88  && -3.88  &  2.18  & -3.88 \\
$V_X/2\pi$ (AU/day)  &  0.33  & -0.19  &  0.33  &&  0.13  & -0.08  &  0.15 \\
$V_Y/2\pi$ (AU/day)  &  0.29  & -0.01  & -0.31  &&  0.14  & -0.005 & -0.13 \\
\enddata
\end{deluxetable}

\clearpage
\begin{deluxetable}{lccccccc}
\tablecaption{Values of the mass, luminosity, temperature, and the components of the position and velocity vectors
of the three-star systems shown in Figures \ref{fig:eight-Sunlike} and \ref{fig:eight-Mstars} \label{table5}}
\tablewidth{0pt}
\tablehead{
\colhead{} & \multicolumn{3}{c}{Sun-like stars} & \colhead{}  & \multicolumn{3}{c}{M stars} \\
\hline
\colhead {} & \colhead {A} & \colhead {B} & \colhead {C} & \colhead {} & \colhead {A} & \colhead {B} & \colhead {C} 
}
\startdata
$T\,({\rm K})$       & 5780   & 5780   & 5780   && 2800    & 2800    & 2800    \\
$L\,({L_\sun})$      &  1.0   &  1.0   &  1.0   &&  0.0095 &  0.0095 &  0.0095 \\
$M\,({M_\sun})$      &  1.0   &  1.0   &  1.0   &&  0.25   &  0.25   &  0.25   \\
$X$ (AU)             & -0.97  &  0     &  0.97  && -0.61   &  0      &  0.61   \\
$Y$ (AU)             &  0.24  &  0     & -0.24  &&  0.15   &  0      & -0.15   \\
$V_X/2\pi$ (AU/day)  &  0.47  & -0.93  &  0.47  &&  0.29   & -0.59   &  0.29   \\
$V_Y/2\pi$ (AU/day)  &  0.43  & -0.86  &  0.43  &&  0.27   & -0.54   &  0.27   \\
\enddata
\end{deluxetable}

\end{document}